\documentclass[twocolumn]{emulateapj}
\slugcomment{{\sc ApJ 790:10, 2014, minor corrections after publication}}

\usepackage{amsmath}
\usepackage{natbib}
\usepackage{pifont}
\usepackage{longtable}

\shorttitle{H/H$_2$}
\shortauthors{Sternberg et al.}

\begin{document}
\title{H{\small I}-to-H$_2$ Transitions and H{\small I} Column Densities
in Galaxy Star-Forming Regions}

\author{Amiel Sternberg\altaffilmark{1}, Franck Le Petit\altaffilmark{2},
Evelyne Roueff,\altaffilmark{2}
and Jacques Le Bourlot\altaffilmark{2,3}}
\altaffiltext{1}
{Raymond and Beverly Sackler School of Physics \& Astronomy,
Tel Aviv University, Ramat Aviv 69978, Israel, amiel@astro.tau.ac.il}
\altaffiltext{2}
{LERMA, Observatoire de Paris, CNRS, 5 place 
Jules Janssen, 92190, Meudon, France}
\altaffiltext{3}
{Universit\'e Paris Diderot, 5 rue Thomas-Mann, 75205 Paris cedex 13, France}

\begin{abstract}
We present new analytic theory and radiative transfer computations for the
atomic to molecular (H{\small I}-to-H$_2$) transitions, and the build-up of 
atomic-hydrogen (H{\small I}) gas columns, in optically thick interstellar clouds, irradiated by 
far-ultraviolet photodissociating radiation fields.  
We derive analytic
expressions for the total H{\small I} column densities for (one-dimensional (1D)) planar slabs, for
beamed or isotropic radiation fields,
from the weak- to strong-field limits, for gradual or sharp atomic to molecular transitions,
and for arbitrary metallicity.
Our expressions may be used to evaluate the H{\small I} column densities as functions of the
radiation field intensity and the H$_2$-dust-limited dissociation flux, the hydrogen gas density, and the metallicity-dependent 
H$_2$ formation rate-coefficient and far-UV dust-grain
absorption cross-section. 
We make the distinction between
``H{\small I}-dust" and ``H$_2$-dust" opacity, and we present computations for the
``universal H$_2$-dust-limited effective dissociation bandwidth".
We validate our analytic formulae with {\it Meudon PDR code} computations 
for the H{\small I}-to-H$_2$ density profiles, and total H{\small I} column
densities. We show that our general 1D formulae 
predict H{\small I} columns and H$_2$ mass fractions that are essentially 
identical to those found in more complicated (and approximate) spherical (shell/core) models. 
We apply our theory to compute H$_2$ mass fractions and star-formation thresholds 
for individual clouds in
self-regulated galaxy disks, for a wide range of metallicities.
Our formulae for the H{\small I} columns and H$_2$ mass fractions may be incorporated into hydrodynamics simulations
for galaxy evolution.

\end{abstract}


\section{Introduction}

The atomic to molecular hydrogen (H{\small I}-to-H$_2$) transition
is of central importance for the evolution of the interstellar medium (ISM)
and for star-formation in galaxies, from local environments in the Milky Way 
to distant cold gas reservoirs in high-redshift systems.
Stars form in molecular gas, plausibly because H$_2$ formation enhances low-temperature 
cooling and cloud fragmentation, or perhaps simply because the molecular formation rates 
are elevated in the denser and more shielded components of the gravitationally collapsing regions.  
The atomic to molecular conversion is also the critical initiating step for the growth of chemical complexity 
in the ISM from large to small scales, e.g.~from diffuse clouds to dense star-forming cores to 
protoplanetary disks. Globally, the transition to H$_2$ appears to be associated with 
star-formation thresholds in galaxy-wide Kennicutt-Schmidt relations, and with the observed critical 
gas mass surface densities above which star-formation becomes probable.

In this paper, we revisit the theory of the H{\small I}-to-H$_2$ transition and the
build-up of atomic hydrogen gas layers in fully optically thick interstellar clouds irradiated by far-ultraviolet (FUV)
radiation fields. Atomic (H{\small I}) gas produced by rapid 
stellar far-ultraviolet ``Lyman-Werner" (LW) photodissociation
undergoes conversion to H$_2$ 
as the destructive radiation is absorbed.  In steady-state, a 
mass of H{\small I} is maintained in the outer FUV irradiated photon-dominated regions 
(PDRs) of the dense molecular clouds. Much of the (cold) H{\small I} gas in galaxies may reside 
in such cloud boundary layers and envelopes, interspersed with the recently formed FUV-emitting OB-type stars.    


The study of interstellar H{\small I}-to-H$_2$ conversion has had a long and venerable history.   
Early theoretical discussions (e.g., \citealt{Spitzer_48, Gould_63, Field_66,Stecher_67,deJong_72}; 
followed by \citealt{Aaronson_74,Glassgold_74,Jura_74,Black_77,Federman_79,vanDishoeck_1986})
focussed on the competing processes of (grain-surface) molecule formation, photodissociation, and shielding, in 
predominantly atomic gas - the classical warm and cold neutral medium (WNM and CNM) and diffuse gas - 
showing that significant concentrations of H$_2$ could be expected in the Galactic ISM,
especially in dusty dark clouds with high visual extinctions
 \citep{Meszaros_68,Hollenbach_71,Solomon_71,deJong_72}.
This was confirmed 
observationally with the first direct (far-UV LW absorption line) 
detections of interstellar molecular hydrogen in diffuse clouds and the 
correlation of the H$_2$ with $E(B-V)$ color excess and dust extinction 
\citep{Carruthers_70,Spitzer_73,Savage_77}, and with the discovery of fully 
molecular clouds via proxy millimeter-wave carbon monoxide (CO) emissions 
\citep{Wilson_70,Rank_71}.

Absorption line spectroscopy (Ly$\alpha$ for H{\small I}, and LW-band for H$_2$)
has been carried out for H{\small I}-to-H$_2$ along many Milky Way sight-lines,
through low-extinction diffuse-to-translucent gas in the disk,
and into the infrared cirrus and high-velocity gas in vertical directions
\citep{Savage_77,Bohlin_78,Richter_01,Liszt_02,Rachford_02,Rachford_09,Gillmon_06a,Gillmon_06b,
Wakker_06,Liszt_07,France_13, Fukui_14, Rohser_14}.
These studies probe systems in which the H$_2$ mass fractions range over many orders
of magnitude, from 
$\lesssim 10^{-5}$ up to $\sim 50\%$ in highly reddened systems.
Absorption line observations of damped- and sub-damped Ly$\alpha$ absorbers (DLAs)
at high redshifts also directly reveal the partial conversion of H{\small I} to H$_2$ in optically thin media
\citep{Levshakov_85,Foltz_88,Ge_97,Cui_05,Ledoux_06,Noterdaeme_10,Crighton_13, Albornoz_14}.
In the early Universe, the formation of the first stars (Population III) was enabled by
the partial conversion to H$_2$ via negative ion chemistry. The resulting H$_2$ 
rotational-line gas cooling rates were likely regulated  
by LW-photodissociation ``feedback" from the first stars and FUV sources
\citep{Palla_83, Lepp_1984, Haiman_96, Haiman_97, Abel_97, Ciardi_00, Glover_03, Yoshida_03, Wise_07, Dijkstra_08, Ahn_09, Bromm_09, Miyake_10, Wolcott_11, Fialkov_12, Holzbauer_12, Safranek_12, Visbal_14}.

In optically thick regions, 21~cm observations of very cold ($\lesssim 20$~K) narrow-line self-absorbed H{\small I}
(the Galactic ``HINSA"; \citealt{Li_03})
in combination with CO, OH, and dust mapping for 
locating the H$_2$ clouds, reveal the presence of trace atomic hydrogen inside dark dusty 
and predominantly molecular clouds 
\citep{Bok_55,Heiles_69,Knapp_74,Burton_78,McCutcheon_78,Liszt_79,Mebold_82,vanderwerf_88,Li_03,Goldsmith_05,Krco_10}.
Such clouds are fully
shielded against externally incident photodissociating FUV radiation, and the conversion to
H$_2$ is essentially complete. The residual atomic gas in the cloud cores is likely 
the product of impact-ionization by penetrating low-energy cosmic rays
\citep{Spitzer_68, Webber_98, Dalgarno_06}.
Somewhat warmer H{\small I} ($\sim 100$~K, so still ``cold") is also observed in dissociation zones surrounding 
Galactic H{\small II} regions associated with individual OB-type stars or clusters,
and/or as H{\small I} PDRs in molecular cloud envelopes exposed to ambient interstellar radiation
(e.g., \citealt{Sancisi_74,Myers_78,Read_81,Roger_81,Wannier_83,Elmegreen_87,vanderwerf_89,Wannier_91,Andersson_92,Gir_94,Reach_94,Williams_96,Gomez_98,Habart_03,Matthews_03,Roger_04,Lee_07}; Lee et al. 2012; \citealt{vanderwerf_13}).
In nearby galaxies, 
H{\small I} has been mapped in spiral arms showing that the atomic gas 
likely traces outer photodissociated layers in the star-forming giant molecular clouds
\citep{Allen_86,Shaya_87,Rand_92,Madden_93,Allen_97,Smith_00,Heyer_04,Knapen_06,Schuster_07,Heiner_09,Heiner_11}.
 
By the 1980's a conceptual switch had occurred with the recognition that much of the
hydrogen in galaxies is fully shielded H$_2$, and that in dense gas in
star-forming regions the H{\small I} is often a surface photodissociation ``product",
rather than being the dominant component within which some shielded H$_2$ may 
be present, as in the diffuse medium.  Over the decades
many model computations for the H{\small I}-to-H$_2$ transition in optically
thick media have been presented, with varying degrees of sophistication in
treating the critical roles of
FUV dust absorption and scattering, and H$_2$ absorption line self-shielding.  
These include one-dimensional (1D) plane-parallel (slab) models
assuming steady-state conditions with simplified (``isolated line") treatments of H$_2$ self-shielding 
(\citealt{Federman_79,deJong_80,Tielens_1985,Viala_86,Black_87,Sternberg_1988,Sternberg_1989,Burton_90,Spaans_94,Sternberg_99,Kaufman_99}),
models incorporating
``exact" radiative transfer for the combined effects of multiple H$_2$ absorption-line
overlap and dust absorption/scattering 
(\citealt{vanDishoeck_88,vanDishoeck_90,Viala_88,Abgrall_92,Draine_1996,Browning_03,Shaw_05,Goicoechea_2007,LePetit_2006}),
  spherically symmetric models  
(\citealt{Andersson_93,Diaz_98,Neufeld_96,Stoerzer_1996,SpaansNeufeld_97,Spaans_97,Krumholz_2008,Krumholz_2009,McKee_2010,Wolfire_10}),
 and also
time-dependent models for the H$_2$ formation and destruction
(\citealt{London_78,Roger_92,Goldshmidt_95,Hollenbach_95,Lee_96,Goldsmith_07}).
More recently, sophisticated multidimensional (2D and 3D) radiative transfer codes have been
developed for the atomic to molecular conversion,
also incorporating hydrodynamics 
(\citealt{Robertson_08,Gnedin_09,Glover_10,Bisbas_12,Christensen_12,MacLow_12,Dave_13,Offner_13,Thompson_14}),
 although the H$_2$ photodissociation rates and the implied H{\small I}/H$_2$ density ratios
are generally still estimated using 1D shielding prescriptions
for the individual hydrodynamic particles or cells.

In recent years interstellar H{\small I}-to-H$_2$ conversion has become an important issue in the study
of galaxy evolution on large scales, across entire galaxy disks, at both low- and high-redshifts,
and for varying metallicities
(e.g.,~\citealt{Wong_02,Boker_03,Blitz_04,Blitz_06,Bigiel_08,Leroy_08,Tacconi_10,Bolatto_11,Schruba_11,Welty_12,Genzel_12, Genzel_13, Tacconi_13}).
Galaxy mapping surveys suggest that on global scales the star-formation efficiencies are determined,
at least in part, by molecular gas fractions that may be sensitive to the varying mid-plane gas pressures
and/or metallicities
(e.g.,~\citealt{Hirashita_05,Fumagalli_10,Fu_10,Lagos_11,Feldmann_12, Kuhlen_13, Popping_13}).
Remarkably, the observations of disk galaxies on large scales (e.g.,~\citealt{Leroy_08}),
 and individual
Galactic molecular clouds on small scales (e.g.,~\citealt{Lee_12, Lee_14}),
indicate that for solar metallicity
the H{\small I}-to-H$_2$ conversion occurs for characteristic gas surface densities
of $\sim 10$~M$_\odot$~pc$^{-2}$ (for ``ambient" FUV radiation fields).  This surface density
corresponds to an FUV dust optical depth $\sim 1$, for typical grain properties and
dust-to-gas mass ratios, suggesting that dust absorption and hence metallicity is playing an essential role
in setting the critical gas surface densities.


An analytic theory for the H{\small I}-to-H$_2$ transition was
presented by \cite{Sternberg_1988} (hereafter S88) 
who derived a scaling-law for the growth of the H{\small I}
column density and the associated FUV-excited infrared H$_2$ vibrational emission intensities produced
in optically thick irradiated cloud surfaces, for application to Galactic emission-line sources
(see also \citealt{Jura_74,Hill_78}, and \citealt{Elmegreen_93}).
S88 included a general-purpose analytic formula for the total H{\small I} column density as a function of
the FUV radiation intensity, the cloud gas density, and the metallicity-dependent 
H$_2$ formation rate coefficient and FUV dust attenuation cross section.  
S88 also identified the fundamental dimensionless parameter that 
controls the H{\small I}-to-H$_2$ transitions and the build-up of the atomic hydrogen columns.

More recently, and motivated by the possible metallicity-dependence of molecular mass fractions
in galaxy disks,  \citealp{Krumholz_2008,Krumholz_2009} 
 and \cite{McKee_2010}
(hereafter KMT/MK10) presented new models for the H{\small I}-to-H$_2$ transition, 
and for the associated metallicity dependent H$_2$ mass fractions and star-formation surface density thresholds.
A novel feature of the KMT/MK10 study is their analytic focus on (idealized) spherical clouds 
embedded in ambient isotropic fields, as opposed to the
(also idealized) planar geometry and beamed fields adopted in much of the 
earlier PDR literature, including S88.


Our main goal and motivation in this paper is to reintroduce and extend
the S88 theory, for applications to global galaxy evolution studies. 
In \S2 we elaborate on S88 and present a detailed overview and discussion of the 
basic theoretical ingredients and parameters controlling the H{\small I}-to-H$_2$ 
transition in FUV irradiated clouds.  We rederive the fundamental S88 equation for the total
H{\small I} column density produced for beamed radiation into a (one-dimensional) optically thick slab.
We then extend the theory and consider irradiation by isotropic fields.  This will enable
our direct comparison to the more complicated (and more approximate) formalism for spheres.  
 In \S 3 we present detailed numerical ({\it Meudon PDR-code})
radiative transfer computations for the H{\small I}-to-H$_2$ transitions 
and integrated H{\small I} columns for a wide range of interstellar conditions.
The ratio of the free space FUV field intensity (or dissociation rate)
to the gas density (or H$_2$ formation rate) is an essential parameter,
as is the metallicity and dust-to-gas mass ratio.  We present numerical
computations for a verification of our analytic formulae for beamed
and isotropic irradiation from the weak- to strong-field limits
(gradual to sharp H{\small I}-to-H$_2$ transitions) and for low- to high-metallicity gas.
In \S 4 we compare our planar formulae to the KMT/MK10 theory for spheres.
This includes a discussion of the dimensionless parameters,
a comparison of the expressions for the total H{\small I} columns,
H$_2$ mass fractions, and star-formation thresholds,
as functions of the metallicity. An important application and comparison is
for ``self-regulated gas" in which the FUV-intensity to gas density
ratio is set by the condition of two-phased equilibrium for the H{\small I}.
We demonstrate that our
simpler, more general, and fully analytic 1D formulae,
predict H{\small I} columns and H$_2$ mass fractions
that are essentially identical to results for spheres
in the more restricted regime in which the spherical models are applicable
(intense fields, sharp transitions, low-metallicity).

This is a lengthy paper, and we develop the theory
and present our comparisons, step-by-step, in a pedagogical style.
In \S 5 we summarize and recap our basic analytic results for the H{\small I}
column densities and molecular mass fractions in FUV irradiated clouds,
including for self-regulated star-forming galaxies.
A glossary of symbols is in the Appendix.

\section{Analytic Overview}

In this section we present an analytic discussion and overview of the basic processes 
and quantities that control the H{\small I}-to-H$_2$ transitions and total H{\small I} columns 
in interstellar clouds exposed to photodissociating far-ultraviolet (FUV) 
radiation fields.  Our overview anticipates and also provides analytic 
representations for the detailed numerical results that we present in \S 3.
We focus on idealized static one-dimensional semi-infinite
uniform density isothermal and optically thick plane-parallel clouds that are 
irradiated by steady fluxes of FUV Lyman-Werner (LW) band photons.
We rederive the S88 formula for the steady-state H{\small I} column 
densities produced in slabs irradiated by normally incident 
uni-directional beamed fields, as appropriate for interstellar clouds exposed to
localized FUV sources.  We then show that this formula can be 
generalized to clouds embedded in isotropic radiation fields. Isotropic irradiation
may be more representative of global ambient conditions in galaxies.

We begin with our normalizations for the beamed and isotropic ultraviolet radiation fields,
and for the associated H$_2$ photodissociation rates (\S~2.1).
We then define the dissociation bandwidth and its derivative - the
 H$_2$ self-shielding function (\S~2.2.1).  We then describe our treatment of dust grains
 (\S~2.2.2). The grains provide FUV continuum opacity and are also the 
 H$_2$ formation sites. The metallicity of the gas then enters as an important parameter
because it controls the dust-to-gas mass ratio, and therefore also the associated 
H$_2$ formation efficiency and the FUV dust optical depth per gas column density.
We make the simplifying assumption
that the dust-to-gas mass ratio scales linearly with the metallicity.
If star-formation requires the conversion from H{\small I} to H$_2$, the metallicity will
be an essential parameter in controlling the star-formation thresholds.
 
We put the physical ingredients together and write down the 
depth-dependent steady-state H{\small I}/H$_2$ formation-destruction equation
for semi-infinite slabs exposed to beamed fields (\S~2.2.3).  Crucially, the differential
equation is separable, and this enables our definition of the 
 ``universal H$_2$-dust-limited LW dissociation bandwidth" (\S~2.2.4)
 and the ``effective dissociation flux" (\S~2.2.5).
 The H$_2$-dust-limited dissociation bandwidth is a fundamental quantity
 in the theory, and we present analytic expressions and numerical computations for it
 in this paper.
 
 We then integrate the H{\small I}/H$_2$ formation-destruction equation
to derive our analytic formula for the H{\small I} column density
for clouds irradiated by beamed fields (\S~2.2.6).
Our formula gives the H{\small I} column density
as a function of the physical variables, including gas density, FUV intensity, 
effective dissociation flux, 
H$_2$ formation rate coefficient, FUV dust-absorption cross-section, and metallicity.

As we discuss (in \S 2.2.6)
the H{\small I}-to-H$_2$ transition profiles and the total atomic column densities are controlled by
a single dimensionless parameter, ``$\alpha G$", first introduced by S88.
In a nutshell, $\alpha G$ determines the LW-band optical depth in the cloud due to the 
{\it {dust associated with the H{\small I} gas}}
(which we refer to in this paper as ``H{\small I}-dust"), and whether or not the H{\small I} is 
mixed with the H$_2$. The total H{\small I}-dust optical depth is a critical quantity
in the theory; for beamed fields our formula for it (as derived in \S 2.2.6) is
\begin{equation}
\tau_{1,\rm tot} \ = \ {\rm ln}[\frac{\alpha G}{2} + 1] \ \ \ .
\end{equation}
As we show (in \S~2.2.7) $\alpha G$ can be expressed in terms of the 
physical variables in several ways. Most simply,
$\alpha$ (also dimensionless) is the ratio of the 
free-space photodissociation rate to the H$_2$ formation rate, and
$G$ is a cloud-averaged H$_2$ self-shielding factor.
The product $\alpha G$ is then similar to the ionization parameter ``$U$"
for H{\small II} Str{\" o}mgren regions, where $U$ is proportional to the 
ratio of the photoionization rate to the H{\small I} formation rate via electron-proton recombination.
However, $\alpha G$ is also a measure of the dust-absorption efficiency of the H$_2$-dissociating photons. 
For the H{\small I}/H$_2$ density ratio in an optically thin ``free-space" radiation field, $\alpha G$ is the ratio of the 
H{\small I}-dust to H$_2$-line absorption rates of LW-band photons that are effectively available for H$_2$ dissociation. For sufficiently
large metallicities this {\it excludes} LW photons ``between the lines"
that are inevitably absorbed by dust associated with just the H$_2$
(``H$_2$-dust") in a predominantly molecular and dusty cloud.
As we will discuss, the mean shielding factor $G$ depends on the
competition between H$_2$ line-absorption and H$_2$-dust absorption.
Because of this competition, a metallicity dependence is introduced into $G$
and therefore also into our fundamental parameter $\alpha G$.


An important physical distinction occurs between the limits of small and large $\alpha G$ (\S~2.2.8).
Small $\alpha G$ is the ``weak-field-limit" for which H{\small I}-dust opacity 
is negligible and does not contribute to the absorption of the radiation,
although a substantial (observable) atomic column can nevertheless exist in this limit.
Large $\alpha G$ is the ``strong-field-limit", for which the atomic column becomes so large that
H{\small I}-dust dominates the attenuation of the radiation fields, reducing the fraction
of the incident radiation that is absorbed by the H$_2$. 

In general, $\alpha G$ is the dimensionless ``free parameter" in the problem,
with a value that is determined by local conditions
(density, radiation intensity, metallicity, etc.).  However, as invoked by KMT/MK10,
on global scales in star-forming galaxy disks the
gas density and radiation intensity may be correlated or self-regulated 
to conditions enabling a
two-phase equilibrium between cold and warm H{\small I} (CNM/WNM multiphase).
As we describe (in \S 2.2.9) this then drives $\alpha G$ to a narrow range,
of order unity, intermediate between the weak- and strong-field limits,
and only weakly dependent on the metallicity.

Finally, in \S~2.3 we extend our analysis to slabs
exposed to isotropic radiation fields. An angular integration 
over all photodissociating ray directions
is then also required in the computation of the H{\small I} column densities.
Our resulting analytic expressions for the atomic columns and associated
H{\small I}-dust opacities are similar to those for beamed fields, 
and with similar behavior in the weak- and strong-field limits.
Our analytic results for slabs irradiated by isotropic fields enable a direct comparison
to the KMT/MK10 results for spheres, as we discuss in depth in \S 4.
In \S~3 we verify our analytic results with detailed numerical model computations.


\subsection{Radiation Fields and H$_2$ Photodissociation Rate}  


We will consider static optically thick plane-parallel clouds (slabs)  exposed to either isotropic or corresponding 
beamed Lyman-Werner (LW) band radiation fields.  We define the spectral range of the LW band as 
912-1108~\AA~(11.3-13.6 eV) as appropriate for line absorptions occurring out of low-lying rotational
levels in the ground vibrational state.

Let $F_\nu\equiv 4\pi I_\nu$, where $I_\nu$ is the specific photon intensity 
(cm$^{-2}$  s$^{-1}$ Hz$^{-1}$ sr$^{-1}$) of an isotropic optically thin ``free-space" LW radiation field.
If an optically thick gas slab (or semi-infinite slab) is inserted, the flux density of the isotropic field at a cloud surface is 
equal to $\pi I_{\nu}=F_\nu/4$.  The corresponding unidirectional beamed field is defined such the 
LW photons are normally incident on the cloud surface with
flux density $2\pi I_{\nu}=F_{\nu}/2$. The surface flux of the isotropic field
is half that of the corresponding beamed field, but the energy densities are equal.
The energy densities at the cloud surfaces are half that in the (full $4\pi$) free-space radiation field.

In this paper we adopt the standard \citet{Draine_1978, Draine_2011} expression
\begin{equation}
\label{Draine}
{\cal I}_\nu^{ISM} = \frac{1}{4\pi}\biggl\lbrace \frac{1.068\times 10^{-3}}{\lambda} - \frac{1.719\times 10^0}{\lambda^2} 
+ \frac{6.853\times 10^2}{\lambda^3}\biggr\rbrace \\
\end{equation} 
for the specific intensity (photons~s$^{-1}$~cm$^{-2}$~Hz$^{-1}$~sr$^{-1}$)
of the isotropic free-space far-ultraviolet (5-13.6 eV FUV) radiation field
in the Galactic interstellar medium. 
In Equation~(\ref{Draine}),
 $\lambda$ is the photon wavelength in Angstroms. For this spectrum, 
 the specific intensity varies by a factor of eight across the LW band.
 At~1000~\AA,~${\cal I}_\nu^{ISM}=2.73\times 10^{-9}$~photons~s$^{-1}$~cm$^{-2}$~Hz$^{-1}$~sr$^{-1}$,
 and the energy density $4\pi h \nu^2 {\cal I}_\nu / c = 6.8\times 10^{-14}$~erg~cm$^{-3}$
 (where $\nu$ is the photon frequency in Hz).
The total photon density in the 912-1108~\AA~LW band is $6.9\times 10^{-4}$~cm$^{-3}$.

To consider radiation fields with greater or lesser intensities we multiply
by an overall field-strength scaling factor $I_{\rm UV}$,
such that $I_{\rm UV}=1$ corresponds to the unit free-space Draine field 
given by Equation~(\ref{Draine}).
In this paper we do not consider radiation fields with alternate spectral shapes.
For completeness we recall that the Draine energy density
is 1.7 times larger than the \citet{Habing_1968} estimate for the LW energy 
density at 1000~\AA
\footnote{We note that in the classical ``PDR literature" the FUV field-strength is designated
variously as $G_0$, $I_{\rm UV}$, or $\chi$ 
(e.g., \citealt{Hollenbach_71,vanDishoeck_1986,Sternberg_1989,Draine_1996}).
In this paper we adopt $I_{\rm UV}$ for the field-strength
 to avoid confusion with the
``$\alpha G$-factor" defined below (in \S 2.2.5) or with the KMT/MK10 ``$\chi$" which, as we 
discuss (in \S 4) is equivalent to our $\alpha G$ in the low metallicity limit.  
The adopted normalizations can be confusing.
For example, in the classic \citet{Tielens_1985} paper, $G_0=1$
refers to an FUV field for which the energy density at the surface of an optically thick cloud 
is equal to the energy density in the free-space (all $4\pi$) \citet{Habing_1968} field. 
That is, $G_0=0.5$ for an optically thick slab embedded in a unit (isotropic) Habing field. 
With our definitions,
$I_{\rm UV}=1$ (not $0.5$)
for a cloud inserted into a unit isotropic Draine field with $I_\nu=I_\nu^{ISM}$, or
for a cloud illuminated by a corresponding beamed field with surface 
flux density $F_\nu=2\pi I_\nu^{ISM}$. For $I_{\rm UV}=1$, 
the H$_2$ photodissociation rates at the cloud surfaces are equal to
half the full-$4\pi$ ``free-space" rate in a unit Draine 
field (see also Equation~[\ref{D0}]).
}.

For the free-space fields, we define the LW-band photon flux-integral
\begin{equation}
\label{F0}
F_0 \equiv \int_{{\nu_1}}^{{\nu_2}} F_\nu d\nu
\end{equation}
where $\nu_1$ to $\nu_2$ is the frequency range of the LW band, and where
$F_\nu\equiv 4\pi {\cal I}_\nu^{ISM} I_{\rm UV}$.
For the Draine spectrum, $F_\nu = 3.4\times 10^{-8} I_{\rm UV}$~photons~s$^{-1}$~cm$^{-2}$~Hz$^{-1}$
at 1000~\AA, and 
\begin{equation}
\label{F0norm}
F_0 \ = \ 2.07 \times 10^7\  I_{\rm UV} \ \ \ \ {\rm photons} \ {\rm cm}^{-2}  \ {\rm s}^{-1} \ \ \ .
\end{equation}
At the cloud surfaces, the 1000~\AA~flux densities are then
$F_\nu/4 =  8.6\times 10^{-9} I_{\rm UV}$ and $F_\nu/2 = 1.7\times 10^{-8} I_{\rm UV}~$photons~s$^{-1}$~cm$^{-2}$~Hz$^{-1}$
for the isotropic
and corresponding beamed fields. The total LW band surface fluxes are 
$F_0/4 = 5.18 \times 10^6 I_{\rm UV}$
and $F_0/2= 1.03 \times 10^7 I_{\rm UV}$~photons~cm$^{-2}$~s$^{-1}$.
For a given $I_{\rm UV}$, the energy densities
of the isotropic and corresponding beamed fields are equal at the cloud surfaces.

The photodissociation of H$_2$ occurs via line absorption of LW photons
in allowed transitions from the ground electronic $X ^1\Sigma_g^+$ state to the excited 
$B ^1\Sigma_u$ or $C ^1\Pi_u$  states.  These are followed by rapid decays to
either bound ro-vibrational levels or to the continuum of the ground $X$ state. 
Decays to the continuum lead to dissociation 
(P.~M.~Solomon, private communication in \citealt{Field_66, Stecher_67,
Stephens_72, Abgrall_92}).
The $B$-$X$ and $C$-$X$ bound-bound transitions are mainly to excited vibrational
levels followed by a near-infrared quadrupole radiative cascade
(\citealt{GouldHarwit_63,Black_76,Shull_1978,Black_87}; S88;
 \citealt{Sternberg_1989, Draine_1996, Neufeld_96}).   Thus, all
of the LW-band photons absorbed by the H$_2$ are removed, but only a
fraction of these absorptions ($\sim 10\%$) lead to photodissociation.

The H$_2$ photodissociation rate in the ISM is a fundamental quantity,
and we recompute it in \S~3 assuming the Draine spectrum,
for a range of assumed gas temperatures, densities, and field intensities.
We find that for dissociation out of the 912-1108~\AA~LW band, the
optically thin (full-$4\pi$) free-space photodissociation rate is
\begin{equation}
\label{D0}
D_0=5.8\times 10^{-11} \ I_{\rm UV} \ \ \ {\rm s}^{-1} \ \ \ .
\end{equation}
At a cloud surface the dissociation rate, $D(0)\equiv D_0/2$, is half the free-space rate and
\begin{equation}
D(0) = 2.9 \times 10^{-11} I_{\rm UV} \ \ \ {\rm s}^{-1} \ \ \ .
\end{equation}
Because the dissociation rate is proportional to the radiation 
energy density, the dissociation rates for isotropic and corresponding beamed fields 
are identical at a cloud surface.

The photodissociation rate diminishes with cloud depth due
to the combination of H$_2$-line and dust absorptions.
The attenuation of the LW radiation field is crucial in 
determining the depth dependence of the atomic and molecular densities,
the shapes of the H{\small I}-to-H$_2$ transition profiles, and the resulting
atomic hydrogen column densities. The depth-dependent attenuation 
depends on the assumed field geometry, and we analyze the behavior for both
beamed and isotropic fields, starting with beamed fields which are simpler.


\subsection{Beamed Fields}

We consider the H{\small I}-to-H$_2$ transition and total column density of atomic hydrogen
on one side of an optically thick plane-parallel slab of gas (or semi-infinite slab) 
that is exposed to a steady flux 
of Lyman-Werner band photons normally incident on the cloud surface as uni-directional 
beamed radiation.

\subsubsection{Dissociation Bandwidth and Self-Shielding Function}

Let $N_2$ be the H$_2$ column density (cm$^{-2}$) at some depth
normal to the cloud surface. Then, neglecting dust-absorption of the LW photons, and 
for beamed radiation, the
photodissociation rate (s$^{-1}$) for a single LW absorption line $\ell$ may be written as
\begin{equation}
\label{DN2}
D_{\ell}(N_2) \ = \ \frac{1}{2}  \int_0^\infty F_\nu \ \sigma_{\nu,d} \ {\rm e}^{-\sigma_\nu N_2} \ d\nu 
\ = \ \frac{1}{2} F_\nu \ \frac{dW_{\ell,d}}{dN_2}
\end{equation}
where 
\begin{equation}
\label{Wd}
W_{\ell,d}(N_2) \equiv \int_0^\infty [1 - \frac{\sigma_{\nu,d}}{\sigma_\nu}{\rm e}^{-\sigma_\nu N_2}] \ d\nu \ \ \ .
\end{equation}
In these expressions, $F_\nu/2=2\pi {\cal I}_\nu^{ISM}I_{\rm UV}$ is the incident beamed 
flux-density (photons cm$^{-2}$~s$^{-1}$~Hz$^{-1}$) at the cloud surface,
$\sigma_{\nu,d}$ is the cross-section (cm$^2$) for absorptions that lead to molecular dissociation
\footnote{
In our notation, the subscript ``d" refers to H$_2$ photodissociation.
Thus, $\sigma_{\nu,d}$ is the cross-section for line-absorption followed by dissociation.
The subscript ``g" refers to dust grains. Thus, $\sigma _g$ is the far-UV dust grain
absorption cross section (\S\ 2.2.2).},
and $\sigma_{\nu}$ is the cross section for all photon absorptions (not just
those that are followed by dissociation).  
The dissociation probabilities,
$f_{\rm diss}\equiv \sigma_{\nu,d}/\sigma_\nu$,
range from $\sim$~0 to more than 0.5 for individual LW transitions depending on 
the rotational quantum number in the excited B or C states.
The mean (typical) dissociation probability averaged over all lines is 
$\langle f_{\rm diss}\rangle =0.12$.
For a single absorption line, 
\begin{equation} 
\begin{split}
\sigma_d  & \equiv \int \sigma_{\nu,d} \ d\nu \\
& = \ f_{\rm diss} \ \frac{\pi e^2}{m_e c} \ f_{osc} \ \simeq  \ 2.7\times 10^{-5} \ \ \ {\rm cm}^2 \ {\rm Hz}
\end{split}
\end{equation}
for a typical LW band oscillator strength $f_{osc}\approx 0.01$, 
and dissociation probability $f_{\rm diss}\approx 0.1$
(and where $e$ and $m_e$ are the electron charge and mass, and $c$ is the speed-of-light).

In Equation~(\ref{DN2}), we pull $F_\nu$ out of the integral because we assume that the flux-density 
varies very slowly over the narrow line profile (as represented by $\sigma_\nu$). 
In Equation~({\ref{Wd}), $W_{\ell,d}(N_2)$ is defined as the 
``equivalent bandwidth" (Hz) of radiation absorbed in H$_2$
dissociations via absorption line $\ell$, up to {\it molecular} column $N_2$. 
The dissociation rate $D_{\ell}(N_2)$ decreases with the molecular column $N_2$
as the absorption line become optically thick, and
the ratio
\begin{equation}
\label{shieldS}
f_{\ell, shield}(N_2) \ \equiv \ \frac{D_{\ell}(N_2)}{D_{\ell}(0)}= \frac{1}{\sigma_d} \ \frac{dW_{\ell,d}}{dN_2} \ \ \ ,
\end{equation}
is the individual H$_2$-line ``self-shielding" function. 
It quantifies the reduction of the line dissociation rate, where
$D_{\ell}(0)$ is the line dissociation rate at the cloud surface.
By definition, the self-shielding function is proportional to the derivative of the
dissociation bandwidth $W_{\ell,d}$ (\citealt{Federman_79, vanDishoeck_1986}; S88; \citealt{Draine_1996}).

For the full multi-line LW band system, the dissociation rate (again neglecting dust absorption)
may be written compactly as
\begin{equation}
\label{DN2p}
D(N_2) \ = \  \frac{1}{2}{\bar F_\nu} \ \frac{dW_d}{dN_2} \ \ \ ,
\end{equation}
where $W_d(N_2)$ is the equivalent dissociation bandwidth 
{\it summed} over all of the (possibly overlapping) absorption lines,
and ${\bar F_\nu}$ is a mean flux density. A plot of
$W_d$ versus $N_2$ is the  effective ``curve-of-growth"  for the dissociating LW radiation
bandwidth in a dust free cloud. We present computations for $W_d(N_2)$ in \S 3
(the blue curve in Figure~\ref{Fig:WGtot_N2}).

The mean flux density in Equation~(\ref{DN2p}) is given by
\begin{equation}
\label{defbarF}
{\bar F_\nu} \  \equiv \ 4\pi \frac{\sum I_{{\nu_{ij}}}x_i \sigma_d^{ij}}{\sum x_i \sigma_d^{ij}}\ \ \  .
\end{equation}  
Here $x_i$ are the fractional
populations of H$_2$ molecules in ro-vibrational levels $i$ of the ground electronic $X$-state, 
$I_{{\nu_{ij}}}$ is the (full $4\pi$) free-space specific intensity at frequencies 
$\nu_{ij}$ of LW band transitions between level $i$ and levels $j$ in the excited $B$ or $C$ states,
and $\sigma_d^{ij}$ are the absorption-line dissociation cross sections (cm$^2$~Hz).
The mean flux density ${\bar F_\nu}$ is weighted by the relative strengths of the dissociation transitions.
For the Draine spectrum,
${\bar F_\nu}=2.46\times 10^{-8}  I_{\rm UV}$ photons cm$^{-2}$ s$^{-1}$ Hz$^{-1}$.

The denominator in expression~(\ref{defbarF}),
 is the total (frequency integrated) H$_2$ dissociation cross-section 
 (cm$^2$~Hz) summed over all absorption lines
 \begin{equation}
 \label{sigdtot}
 \sigma_d^{\rm tot}\equiv\sum x_i \sigma_d^{ij} \ \ \ .
 \end{equation}
Most of the H$_2$ line absorptions occur out of the lowest few rotational levels,
and the total effective dissociation cross section is insensitive to the fractional
populations $x_i$. We find that 
\begin{equation}
\sigma_d^{\rm tot} \ = \ 2.36\times 10^{-3} \ \ \ {\rm cm}^{2} \ {\rm Hz} \ \ \ .
\end{equation} 
The ratio, $\sigma_d^{\rm tot}/\sigma_d\simeq 80$ is then the 
approximate number of strong LW 
absorption lines involved in the multi-line H$_2$ photodissociation process
(see also Figure~\ref{Fig:spectrum} in \S~3).

With these definitions, the free-space photodissociation rate may be expressed as
\begin{equation}
\label{D0F}
D_0 = {\bar F_\nu} \ \sigma_d^{\rm tot} \ \ \ .
\end{equation}
At a cloud surface
the photodissociation rate is $D(0)\equiv D_0/2=\frac{1}{2}{\bar F_\nu} \ \sigma_d^{\rm tot}$.
The ratio
\begin{equation}
\label{shield}
f_{shield}(N_2) \ \equiv \ \frac{D(N_2)}{D(0)}= \frac{1}{\sigma_d^{\rm tot}} \ \frac{dW_d}{dN_2} \ \ \ ,
\end{equation}
is then the complete multi-line H$_2$ ``self-shielding" function.
It quantifies the reduction of the total dissociation rate due opacity in all of the absorption lines.
 
For a single line the self-shielding function varies as $N_2^{-1/2}$ for large $N_2$
because absorptions can always occur far out on the Lorentzian damping wings
\footnote{This can be seen directly from Equation~(\ref{DN2}). For damped lines, $\sigma_\nu$ and
$\sigma_{\nu,d}$ are proportional to $\nu^{-2}$, and the integral over frequency is then proportional
to $N_2^{-1/2}$.}. Therefore, for a single absorption line,  $W_{\ell,d}$ as given by Equation~(\ref{Wd}) diverges as $N_2^{1/2}$.
For strong lines (with $f_{osc}\sim 0.01$) 
this ``square-root" part of the curve-of-growth begins when $N_2\gtrsim 10^{17}$~cm$^{-2}$.
For the realistic multi-line system the absorption lines will
overlap for sufficiently large ($\gtrsim 5\times 10^{20}$~cm$^{-2}$) molecular columns,
and $W_d$ does not diverge as does $W_{\ell, d}$. For the multiline system
the total dissociation bandwidth 
\begin{equation}
W_{d,{\rm tot}} \equiv \int_0^\infty \frac{dW_d}{dN_2} \ dN_2
\end{equation}
is limited to a finite maximal value (even in the absence of dust).
We find that for the Draine spectrum $W_{d,{\rm tot}} = 9.1\times 10^{13}$~Hz 
(as computed in \S~3.1.2).
In \S 3.1.3 we present our computations for the multi-line self-shielding function (Figure \ref{Fig:fsh_DB}).
At the cloud surface $f_{shield}=1$. As the Doppler cores
become optically thick at $N_2\gtrsim 10^{14}$~cm$^{-2}$, $f_{shield}$ becomes small and the molecules 
are then said to self-shield against the dissociating radiation.  The decline is more gradual
at intermediate columns, $10^{17}$ to $10^{22}$~cm$^{-2}$, for which most of the
absorption is out of the line-wings.  Finally, 
as line overlap occurs and the dissociating radiation is fully absorbed, 
$f_{shield}$ becomes vanishingly small.

In the limit of complete-line-overlap, every photon in the (912-1108 \AA) LW-band
is absorbed in H$_2$-lines.  The product $\frac{1}{2}{\bar F}_\nu W_{d,{\rm tot}}$ is 
then the LW ``dissociation flux" (photons~cm$^{-2}$~s$^{-1}$)
at the cloud surface. In the absence of dust absorption 
this flux is equal to the H$_2$ dissociation rate per unit area. 
For complete absorption of the LW-band radiation, the mean dissociation probability is
\begin{equation}
\label{fdissdef}
{\bar f}_{\rm diss} = \frac{{\bar F}_\nu W_{d,{\rm tot}}}{F_0}
\end{equation}
where $F_0/2$ is the total incident LW-band flux.
For the radiative transfer computations we present in \S~3 we find that ${\bar f}_{\rm diss}=0.12$,
and essentially equal to the simple average over the individual line dissociation probabilities
for the matrix of $X$-$B$, and $X$-$C$ transitions.
Our result for ${\bar f}_{\rm diss}$ is consistent with many previous calculations
(e.g. \citealt{Black_87,Draine_1996,Browning_03}).


\subsubsection{Dust and Metallicity}

In addition to the H$_2$ line absorptions, the LW band photons are also absorbed
by dust grains, further reducing
the photodissociation rate.  We assume that the dust is mixed uniformly with the gas
with a dust-to-gas mass ratio that depends linearly on the metallicity of the cloud.
For the grain-photon interaction we assume pure absorption and no scattering
(or equivalently only forward scattering). With the 
inclusion of dust, the local dissociation rate, $D$, at any cloud depth is then
\begin{equation}
\label{Dfull}
D \ = \  \frac{1}{2}D_0 \ f_{shield}(N_2) \ {\rm e}^{-\tau_g} \ \ \ ,
\end{equation}
where $D_0$ is the free-space dissociation rate (Equation~[\ref{D0}]).
In this expression, $\tau_g \equiv \sigma_g N$ is the dust continuum optical depth, where
$N\equiv N_1+2N_2$ is the column density of hydrogen nuclei, in molecules {\it plus} atoms.
Here, $\sigma_g$ is the dust-grain LW-photon absorption cross section (cm$^{2}$)
per hydrogen nucleon. 
With the inclusion of dust attenuation, the dissociation rate given by Equation~(\ref{Dfull})
depends on both $N_2$ {\it and} the column density of atomic hydrogen $N_1$.

For simplicity we also assume that $\sigma_g$
is independent of photon frequency over
the narrow LW band.  For a standard interstellar extinction curve, with
a ratio of total-to-selective extinction
$R_V\equiv A_V/E(B-V)=3.1$, a
1000~\AA~grain albedo $\approx 0.3$, and a
scattering asymmetry factor $\langle cos\theta\rangle \approx 0.6$, the effective
absorption cross section per hydrogen nucleus $\sigma_g=1.9\times 10^{-21}$~cm$^2$ \citep{Draine_2003}.
$R_V=3.1$ is for diffuse gas (with densities $n\sim 10^2$~cm$^{-3}$). 
For $R_V=3.1$, $A_V/N=5.35\times 10^{-22}$~mag~cm$^2$.
In dense regions
($n\gtrsim 10^3$~cm$^{-3}$) $R_V$ can be larger (up to $\sim 5.8$),
but with an extinction curve that is less steep towards the ultraviolet,
with $\sigma_g\approx 8\times 10^{-22}$~cm$^{2}$
\citep{Cardelli_1989, Fitzpatrick_1999, Draine_1996, Draine_2011}.
With the assumption that the dust-to-gas mass ratio
is linearly proportional to the metallicity $Z'$ of the gas, we therefore set
\begin{equation}
\label{Sstand}
\sigma_g = 1.9\times 10^{-21} \phi_g \ Z' \ \ \ {\rm cm}^2
\end{equation}
where $Z'=1$ corresponds to the solar photospheric abundances of the
heavy-elements (``solar metallicity")
and where $\phi_g$ of order unity depends on the grain composition and size distribution.

Dust grains are also essential for H$_2$ formation \citep{Hollenbach_71, Jura_74, Barlow_76, LeitchDevlin_85, Pirronello_97, Takahashi_99, Cazaux_02, Habart_04}.
We assume that per hydrogen nucleon the rate-coefficient 
for H$_2$ formation on grains is given by
\begin{equation}
\label{Rstand}
R = 3\times 10^{-17}  \ \Bigr(\frac{T}{100 \ {\rm K}}\Bigl)^{1/2} \ Z' \ \ \ {\rm cm}^3 \ {\rm s}^{-1}
\end{equation}
where $T$ is the gas temperature in $^\circ$K. 
Our standard value is then $R=3\times 10^{-17}$~cm$^{3}$~s$^{-1}$,
for $T=100$~K and $Z'=1$. 


\subsubsection{H{\small I}/H$_2$ Formation-Destruction Equation}

For a steady-state in which molecular photodissociation is balanced everywhere by
grain surface H$_2$ formation,  
the H{\small I}/H$_2$ formation-destruction equation may be written as
\begin{equation}
\begin{split}
\label{formdes}
Rn \ n_1 \  & = \ \frac{1}{2}{\bar F_\nu} \ \frac{dW_d}{dN_2} \ {\rm e}^{-\tau_g} \  n_2 \\
\ & = \ \frac{1}{2}D_0 \ f_{shield}(N_2) \ {\rm e}^{-\tau_g} \  n_2 \ \ \  .
\end{split}
\end{equation}
In this equation, $n_1$ and $n_2$ are the local volume densities (cm$^{-3}$) 
of the H{\small I} atoms and H$_2$ molecules, and 
\begin{equation}
\label{ntot}
n\equiv n_1+2n_2
\end{equation}
is the total volume density of hydrogen nuclei. 
The right-hand-side of Equation~(\ref{formdes}) is the H$_2$
photodissociation rate per unit volume (s$^{-1}$~cm$^{-3}$)
at some cloud depth where the rate is 
reduced by the combined effects of self-shielding and dust attenuation.
The left-hand-side 
is the rate per unit volume
of H$_2$ formation on dust grains.
We ignore all other formation or destruction processes
(such as formation in the gas phase; or destruction by X-ray or cosmic-ray ionization).

Given the free-parameters, $n$, $D_0$ (or equivalently $I_{\rm UV}$), 
$R$, and $\sigma_g$
(or given $Z'$ which determines $R$ and $\sigma_g$)
Equation~(\ref{formdes}) together with particle conservation 
Equation~(\ref{ntot}) can be solved for the local atomic and molecular
densities $n_1$ and $n_2$, and for the integrated atomic and molecular columns
$N_1$ and $N_2$.  If it is assumed that
the incident LW radiation is fully absorbed, then (as we show below)
Equation~(\ref{formdes}) gives our fundamental formula for  
the total atomic column that is maintained in the cloud.

The density ratio $n_1/n_2 = dN_1/dN_2$, and
Equation~(\ref{formdes}) may be written as the separable differential equation 
(S88; see also 
\citealt{Jura_74,Hill_78}).
\begin{equation}
\label{sep}
Rn \ {\rm e}^{\sigma_g N_1} \ dN_1 = \frac{1}{2} {\bar F_\nu} \ \frac{dW_d}{dN_2} \ {\rm e}^{-2\sigma_g N_2}\ dN_2 \ \ \ .
\end{equation}
In writing the formation-destruction equation this way, a key insight is that the dust opacities 
associated with the atomic and molecular columns can be considered {\it separately}.
We will refer to ``H$_2$-dust" or ``H{\small I}-dust" as the dust opacities associated with
either just the H$_2$ or the  H{\small I} gas respectively,
and {\it whether or not the H{\small I} gas is mixed with the H$_2$}.

Integrating this expression, and assuming that $R$ and $n$ are constants 
(i.e., do not vary with cloud depth) gives
\begin{equation}
\label{formdesint}
Rn \int_0^{N_1} {\rm e}^{\sigma_g N'_1} \ dN'_ 1 
\ = \  \frac{1}{2}{\bar F_\nu}  \int_0^{N_2}   \ \frac{dW_d}{dN'_2} \ {\rm e}^{-2\sigma_g N'_2} \ dN'_2 \ \ \ ,
\end{equation}
which is a functional relationship, $N_1(N_2)$, between the atomic and  
molecular column densities. We note that the independent variable 
parameterizing the cloud depth is here chosen to be $N_2$
rather than the total gas column density $N$.  Choosing $N_2$
as the independent variable is essential for our analysis, even though it is $N$ that
is proportional to the visual extinction $A_V$, or to
the length scale $z\equiv N/n$.

\subsubsection{H$_2$-Dust-Limited Dissociation Bandwidth}

Most importantly, for a given value of $\sigma_g$ the integral on the right-hand-side 
of Equation~(\ref{formdesint}),
\begin{equation}
\label{ddb}
W_g(N_2) \equiv \int_0^{N_2}   \ \frac{dW_d}{dN'_2} \ {\rm e}^{-2\sigma_g N'_2} \ dN'_2 
\end{equation}
is a function of the molecular column $N_2$ only. This is because the exponential cut-off factor
in the integrand is due to H$_2$-dust opacity only - it excludes H{\small I}-dust -
and because $W_d$ itself depends only on $N_2$.   Furthermore,  $W_g(N_2)$ is only
very weakly dependent on $n$, $R$, or $D_0$, and is essentially independent of these
parameters (see \S~3). So, for a given $\sigma_g$, the effective equivalent width 
$W_g(N_2)$ is a quantity  that can be calculated {\it in advance} as a ``universal dust-limited 
curve-of-growth" for the H$_2$ line absorption of LW radiation, independent of the
other parameters, $n$, $R$, and $D_0$, that together with $\sigma_g$
determine the depth-dependent H{\small I}/H$_2$ density ratios and the H{\small I}-to-H$_2$ transition profiles.

Thus, $W_g(N_2)$ (as opposed to $W_d[N_2]$) is the effective bandwidth 
of dissociating LW radiation 
in a {\it dusty} H$_2$ cloud, 
where now this bandwidth is limited by {\it H$_2$-dust} absorption of
LW photons that would otherwise be available for H$_2$ photodissociation in a dust-free cloud.
H$_2$-dust opacity is important if it becomes 
large before the H$_2$ absorption lines can fully overlap.
For sufficiently small $\sigma_g$, the lines do overlap completely, and then $W_g(N_2)=W_d(N_2)$.
For sufficiently large $\sigma_g$, the H$_2$-dust provides a cut-off, and then $W_g(N_2)<W_d(N_2)$
at large $N_2$.

It is a remarkable physical coincidence that the H$_2$ column density at which the 
H$_2$ lines begin to overlap - a column that depends on the internal
molecular oscillator strengths and energy level spacings - is comparable to the H$_2$ column
at which the H$_2$-dust opacity $2\sigma_g N_2 \gtrsim 1$
for standard interstellar dust absorption cross sections.  Thus, both regimes
of ``small-$\sigma_g$" and ``large-$\sigma_g$" for the dissociation bandwidth $W_g(N_2)$ are relevant 
for the range of interstellar dust properties and metallicities in galaxies.
In \S 3.1.2, we present computations of $W_g(N_2)$ for a wide range of $\sigma_g$
encompassing these regimes.

As $N_2 \rightarrow \infty$, the equivalent width $W_g(N_2)$ converges to a finite limit
\begin{equation}
\label{Wtot}
W_{g,{\rm tot}}(\sigma_g) \ \equiv  \ \int_0^{\infty}   \  \frac{dW_d}{dN_2} \ {\rm e}^{-2\sigma_g N_2} \ dN_2  \ \ \ ,
\end{equation}
either because the exponential H$_2$-dust attenuation factor 
cuts off the integrand (for large-$\sigma_g$),
or because $dW_d/dN_2$ itself vanishes as the lines overlap (for small-$\sigma_g$).
Thus, $W_{g,{\rm tot}}(\sigma_g)$ is the {\it total H$_2$-dust-limited effective dissociation bandwidth}.
For large-$\sigma_g$, $W_{g,{\rm tot}} < W_{d,{\rm tot}}$, and for 
small-$\sigma_g$, $W_{g,{\rm tot}}=W_{d,{\rm tot}}$.
For large-$\sigma_g$ the absorption lines remain separated but are nevertheless highly damped for 
most of the integration range up to the H$_2$-dust cut-off.  

For a single absorption line
$dW_d/dN_2 \propto N_2^{-1/2}$ in the damped regime, and
it follows from Equation~(\ref{Wtot}) that $W_{g,{\rm tot}}(\sigma_g)$ scales as  $\sigma_g^{-1/2}$.
Our numerical computations (\S~3.1.2) show that for large-$\sigma_g$ this scaling behavior is 
maintained in the full multi-line problem to a good approximation. In \S~3, we find that the simple formula
\begin{equation}
\label{Wtotfit}
W_{g,{\rm tot}}(\sigma_g) \ \simeq \ \frac{9.9 \times 10^{13}} {1 + (\sigma_g/ \ 7.2\times 10^{-22} \ {\rm cm}^{2})^{1/2}} \ \ \ {\rm Hz}
\end{equation}
is an excellent fit to our numerical radiative transfer results.
The normalized H$_2$-dust limited dissociation bandwidth
\begin{equation}
\begin{split}
\label{defw}
w \ \equiv \ \frac{W_{g,{\rm tot}}}{W_{d,{\rm tot}}} &\ \simeq \ \frac{1} {1 + (\sigma_g/ \ 7.2\times 10^{-22} \ {\rm cm}^{2})^{1/2}} \\
 & \ = \ \frac{1}{1 + (2.64\phi_gZ')^{1/2}} \ \ \ ,
\end{split}
\end{equation}
where in the last equality we have assumed $\sigma_g=1.9\times 10^{-21}\phi_gZ'$ (Equation[\ref{Sstand}]).
In these expressions, $W_{g,{\rm tot}}\rightarrow W_{d,{\rm tot}}$ and $w\rightarrow 1$ for small $\sigma_g$ (low metallicity), and
decrease as $\sigma_g^{-1/2}$ for large $\sigma_g$ (high metallicity).
The normalized bandwidth $w$ decreases from $0.9$ to $0.2$
for $\sigma_g$ ranging from $1\times 10^{-23}$~cm$^{2}$ (small) to 
$\sim 6\times 10^{-21}$~cm$^{2}$ (large),
or for $Z'$ ranging from $\sim 0.01$ to 3 (assuming $\sigma_g \propto Z'$) which is the relevant range
for galaxies.


\subsubsection{Effective Dissociation Flux and Dissociation Probability}

Given Equation~(\ref{fdissdef}) we may now write
\begin{equation}
{\bar F_\nu}W_{g,{\rm tot}} = w{\bar f}_{\rm diss}F_0 \ \ \ ,
\end{equation}
and define the effective dissociation probability
\begin{equation}
\label{pdiss}
{\bar p}_{\rm diss} \equiv \frac{{\bar F_\nu} W_{g,{\rm tot}}}{F_0} \ 
\equiv \  w{\bar f}_{\rm diss} \ \ \ .
\end{equation}
The product $\frac{1}{2}{\bar F_\nu}W_{g,{\rm tot}}$ 
is the ``effective dissociation flux" 
for dusty clouds in which H$_2$-dust may absorb some of the 
incident LW radiation.  The effective dissociation flux
depends on the competition between H$_2$-line and H$_2$-dust absorption
as given by the dependence of $W_{g,{\rm tot}}$ on $\sigma_g$.
The effective dissociation flux is the H$_2$ photodissociation rate per unit surface area 
for a dusty and optically thick molecular slab in which H{\small I}-dust opacity is negligible. 

When H$_2$-dust is negligible, $w=1$, and ${\bar p}_{\rm diss}={\bar f}_{\rm diss}$.
When H$_2$-dust opacity is significant, $w<1$, and ${\bar p}_{\rm diss}<{\bar f}_{\rm diss}$.
The effective dissociation probability
${\bar p}_{\rm diss}$ is the fraction of the total 912-1108 
\AA~LW-band flux that is absorbed in
H$_2$ photodissociation events in a dusty optically thick and predominantly molecular slab
(with vanishing H{\small I}-dust opacity). For low $Z'$,  ${\bar p}_{\rm diss}=0.12$ is a constant.
For high-$Z'$, ${\bar p}_{\rm diss}$ 
decreases as $Z'^{-1/2}$.


\subsubsection{Formula for the H~{\small I} Column Density}

Returning now to Equation~(\ref{formdesint}), it follows that
\begin{equation}
\label{intsep}
Rn \int_0^{N_{1}}{\rm e}^{\sigma_g N'_1} dN'_ 1 \ = \
\frac{1}{\sigma_g}Rn[{\rm e}^{\sigma_g N_1}-1]
\ = \ \frac{1}{2}{\bar F_\nu} W_g(N_2) \ \ \ ,
\end{equation}
or,
\begin{equation}
\label{S88pProf}
N_1(N_2) \ = \ \frac{1}{\sigma_g} \ {\rm ln}\Bigl[\frac{1}{2}\frac{\sigma_g {\bar F}_\nu W_g(N_2)}{Rn} + 1\Bigr] \ \ \ .
\end{equation}
Following S88 we now define the dimensionless parameter
\begin{equation}
\label{defalpha}
\alpha \ \equiv \ \frac{D_0}{Rn} 
\ = \ \frac{\sigma_d^{\rm tot}{\bar F}_\nu}{Rn}
\end{equation}
where $D_0$ is the {\it free-space} dissociation rate
\footnote{In S88 $\alpha$ was defined with the
surface dissociation rate $D(0)=D_0/2$ in the numerator.
In this paper we adjust the definition and instead use $D_0$,
for a clear comparison of planar and spherical geometries
as presented in \S 4.  With this adjustment factors of $1/2$
appear in our formulae.
},
and we define the dimensionless ``$G$-integral"
\begin{equation}
\begin{split}
\label{defG}
G(N_2) \   & \equiv \ \sigma_g \int_0^{N_2} f_{shield}(N'_2) \ {\rm e}^{-2\sigma_g N'_2} \ dN'_2 \\
\  &= \ \frac{\sigma_g}{\sigma_d^{\rm tot}} W_g(N_2) \ \ \ .
\end{split}
\end{equation}
We can then write
\begin{equation}
\label{S88Prof}
N_1(N_2) \ =  \ \frac{1}{\sigma_g}{\rm ln}[\frac{\alpha G(N_2)}{2} + 1] \ \ \ 
\end{equation}
where 
\begin{equation}
\label{aG}
\alpha G(N_2) \ =  \ \frac{\sigma_g{\bar F_\nu} W_g(N_2)}{Rn} \ \ \ .
\end{equation}
Given the ``universal" dust-limited curve of growth $W_g(N_2)$ - which can be computed
``in advance" for any $\sigma_g$ - the atomic column is then given by Equation~(\ref{S88Prof}) (or Equation~[\ref{S88pProf}])
for any surface dissociation rate $D_0$ (or $I_{\rm UV}$), rate cofficient $R$, and density $n$. 

We refer to Equation~(\ref{S88Prof}) as the ``semi-analytic integral H{\small I}-to-H$_2$ profile".
It shows that the H{\small I}
column at any depth depends on only two quantities -  $\sigma_g$ and $\alpha G(N_2)$.
In dimensionless form
\begin{equation}
\label{tau1oftau2}
\tau_1(\tau_2) \ = \ {\rm ln}[\frac{\alpha G(\tau_2)}{2}+ 1]
\end{equation}
where $\tau_2\equiv 2\sigma_gN_2$ is the dust optical depth associated with
the molecular column, and
$\tau_1(\tau_2)\equiv \sigma_g N_1$ is the H{\small I}-dust optical depth 
at $\tau_2$.

 
As $N_2\rightarrow \infty$, $W_g(N_2)\rightarrow W_{g,{\rm tot}}$. The
{\it total} atomic column density is therefore finite, and is given by
\begin{equation}
\begin{split}
\label{S88p}
N_{1,{\rm tot}}\ & = \ \frac{1}{\sigma_g} \ {\rm ln}[\frac{1}{2}\frac{\sigma_g{\bar F_\nu} W_{g,{\rm tot}}}{Rn} +1] \\
\ & = \ \frac{1}{\sigma_g} \ {\rm ln}[\frac{1}{2}{\bar f}_{\rm diss}\frac{\sigma_g wF_0}{Rn} + 1]
\end{split}
\end{equation}
or,
\begin{equation}
\label{S88}
N_{1,{\rm tot}} \ = \ \frac{1}{\sigma_g}{\rm ln}[\frac{\alpha G}{2} + 1] \ \ \ .
\end{equation}
Here, the dimensionless parameter
\begin{equation}
\label{defGparam}
G(\sigma_g) \  \equiv \  \frac{\sigma_g}{\sigma_d^{\rm tot}} W_{g,{\rm tot}}(\sigma_g) \ \ \ 
\end{equation}
is the limit of $G(N_2)$ as $N_2\rightarrow \infty$, so that 
\begin{equation}
\alpha G \ = \ \frac{D_0G}{Rn} 
\ = \ \frac{\sigma_g{\bar F}_\nu W_{g,{\rm tot}}}{Rn} 
\ = \ {\bar f}_{\rm diss}\frac{\sigma_g wF_0}{Rn}  \ \ \ .
\end{equation}
We discuss these and additional expressions for $\alpha G$ in \S 2.2.6 below.

Equation~(\ref{S88}) for the total H{\small I} column on one side of an optically thick cloud
was first derived by S88
(see Equation~[9] of that paper), and it is the fundamental relation in 
our analysis\footnote{The main goal and result of S88 was an analytic
formula for the intensity of UV excited (fluorescent) IR H$_2$ emission lines from 
photon-dominated regions (PDRs).
Because the FUV-pumped H$_2$ vibrational excitation rate is proportional to the dissociation rate,
the IR intensity
is proportional to the H{\small I} column density, as expressed in Equation~[10] of that paper.
}.
The basic assumption is that all of the dissociating LW-band radiation is absorbed, as
in a classical ``ionization-bounded" H{\small II} region or layer \citep{Stromgren_39}.
However, because of the three-way competition between H{\small I}-dust, H$_2$-dust,
and H$_2$-lines, the behavior for H{\small I} is more complicated than for H{\small II}.
For a steady (dust-free) photoionized planar Str{\" o}mgren layer the H{\small II} column equals the ratio
of the Lyman continuum flux to the recombination rate, independent of the photoionization cross section. 
Similarly, in our Equations~(\ref{S88p}) or (\ref{S88}) the H$_2$ line absorption cross section 
does not appear explicitly (although it is implicit in our definition of the effective dissociation flux). 
The H{\small I} column depends on the ratio of the effective dissociation flux to the H$_2$ formation rate,
but this ratio is multiplied by the dust absorption cross section and appears inside
a logarithm.  We discuss this behavior, and the connection to Str{\" o}mgren relations, in 
our description of the weak- and strong-field limits in \S~2.2.8.

In dimensionless form,
the total H{\small I}-dust optical depth associated
with the total {\it atomic} column is
\begin{equation}
\label{S88tau}
\tau_{1,{\rm tot}} = {\rm ln}[\frac{\alpha G}{2} + 1] \ \ \ .
\end{equation}
The total H{\small I}-dust optical depth
depends on the single dimensionless parameter $\alpha G$
constructed from the cloud variables $D_0$, (or $F_0$, or $I_{\rm UV}$) $n$, $R$, and $\sigma_g$
(or $Z'$ which determines $R$ and $\sigma_g$).

The total gas column $N\equiv N_1(N_2)+2N_2$.  It therefore follows from
Equation (36) that the atomic column as a function of the total (atomic
plus molecular) column, $N_1(N)$, also depends on just $\sigma_g$ and
$\alpha G(N_2)$.  Similarly, for $\tau_g\equiv \sigma_g N$ the HI-dust optical depth
$\tau_1(\tau_g)$ depends on $\alpha G(\tau_2)$. Then, since $n_1/n_2=dN_1/dN_2$
it follows that the {\it shapes} of the HI-to-H$_2$ transition profiles are
for any $\sigma_g$ {\it invariant} for identical $\alpha G$.

  
In \S~3.1.4 we present detailed numerical computations for the H{\small I}-to-H$_2$ transition profiles.
We show that the transitions are ``gradual" when $\alpha G \ll 1$, and are ``sharp"
when $\alpha G \gg 1$.  An essential feature of our derivation and analytic expression for the 
total H{\small I} column is that no assumptions need to be made on the 
shape of the H{\small I}-to-H$_2$ transition profile. Our Equations~(\ref{S88p}) or~(\ref{S88})
are universally valid for all profile shapes, gradual or sharp. 

\subsubsection{$\alpha G$}

It is useful to consider the physical meaning of the dimensionless parameters
$\alpha$ and $G$, and their product $\alpha G$.

First, $\alpha$ is the ratio of the unattenuated free-space H$_2$ photodissociation rate to the H$_2$ formation rate,
and can be expressed as 
\begin{equation}
\begin{split}
\label{alphaDef}
\alpha \ = \ & 1.93\times 10^4 \  
\biggl(\frac{D_0}{5.8\times 10^{-11} \ {\rm s}^{-1}}\biggr) \\
& \times
\biggl(\frac{3\times 10^{-17} \ {\rm cm}^3 \ {\rm s}^{-1}}{R}\biggr) 
\biggl(\frac{100 \ {\rm cm}^{-3}}{n}\biggr) \ \ \ .
\end{split}
\end{equation} 
Thus, $\alpha$ is just the free-space atomic to molecular density ratio 
$n_1/n_2$, 
and $\alpha/2$ is the density ratio at the surface of an optically thick slab.
For a characteristic interstellar cloud gas density $n\sim 10^2$~cm$^{-3}$,
and with $D_0=5.8\times 10^{-11}I_{\rm UV}$~s$^{-1}$,
the atomic to molecular
density ratio $n_1/n_2 \gg 1$ at the cloud edge in the absence of shielding, unless $I_{\rm UV}$ is
unrealistically small. Conversion to the molecular phase in the ISM generally requires
significant attenuation of the ambient and destructive radiation fields.

Now consider the parameter $G$. 
Defining the H$_2$-dust opacity $\tau_2\equiv 2\sigma_g N_2$, (see Equation~[\ref{defG}])
\begin{equation}
\begin{split}
\label{Gshield}
G = \ & \frac{1}{2}\int_0^\infty f_{shield}(\tau_2){\rm e}^{-\tau_2} \ d\tau_2
 \equiv \frac{1}{2} \langle{f_{shield}}\rangle \\
\approx & \ \frac{1}{2}\int_0^1 f_{shield}(\tau_2) \ d\tau_2  \ \ \ .
\end{split}
\end{equation}
Thus, $G$ is the {\it average} H$_2$ self-shielding factor. The average is over
an H$_2$ column for which the H$_2$-dust opacity $\tau_2 \approx 1$.
Thus,  $\frac{1}{2}D_0 G$ is the characteristic photodissociation rate for self-shielded H$_2$
in a fully molecular cloud,
prior to the onset of any H$_2$-dust attenuation.  Because $f_{shield}$ 
is already $\ll 1$ for $\tau_2 \ll 1$, the H$_2$ molecules are very self-shielded for an H$_2$-dust 
opacity $\tau_2 \sim 1$, and $G$ is generally very small.

The product $\alpha G/2=\frac{1}{2}D_0G/Rn$ is then the atomic to molecular density ratio $n_1/n_2$
for the average {\it shielded} H$_2$ dissociation rate. This is our first interpretation for $\alpha G$
(and as adopted in S88).
If $n_1>2n_2$ for the shielded dissociation rate, then H{\small I}-dust must also contribute to the attenuation 
of the LW-radiation since then $\tau_1>\tau_2$ within the H$_2$-dust attenuation column. 
If $n_1\ll 2n_2$, then H{\small I}-dust attenuation is negligible.

Alternatively,  $G\equiv\sigma_gW_{g,{\rm tot}}/\sigma_d^{\rm tot}$ (Equation~[\ref{defGparam}])
is the ratio of the UV continuum dust absorption cross section (cm$^2$) to the total H$_2$ line
dissociation cross section (cm$^2$~Hz) averaged over the effective dissociation bandwidth (Hz).
Again, $G$ is generally very small because H$_2$ line absorption is
so much more efficient than dust absorption. 
Given $\sigma_g=1.9\times 10^{-21}\phi_gZ'$~cm$^2$ (Equation~[\ref{Sstand}]), and
our expression (\ref{Wtotfit}) for $W_{g,{\rm tot}}$, and with $\sigma_d^{\rm tot}=2.36 \times 10^{-3}$~cm$^2$~Hz,
we have that
\begin{equation}
\label{Gfit}
G \ \simeq \ \frac{7.97\times 10^{-5} Z' \phi_g}{1 \ + \ (2.64Z'\phi_g)^{1/2}} \ \ \ .
\end{equation}
For example, for standard values, $Z'=1$ and $\phi_g=1$, Equation~(\ref{Gfit}) gives $G=3.0\times 10^{-5}$,
and the shielded H$_2$ dissociation rate $\frac{1}{2}D_0G=8.7\times 10^{-16} I_{\rm UV}$~s$^{-1}$.
For low-metallicity, small-$\sigma_g$, it follows that $G \propto Z'$ (or $G\propto \sigma_g$).
For high-metallicity, large-$\sigma_g$, $G\propto Z'^{1/2}$, (or $G\propto \sigma_g^{1/2}$).

Since $\alpha$ is the free-space atomic-to-molecular density ratio we have that
\begin{equation}
\begin{split}
\label{defGalt}
\alpha G \ =  \ \frac{\sigma_gW_{g,{\rm tot}}}{\sigma_d^{\rm tot}} \frac{n_1}{n_2}  \ \Biggr\rvert_{\rm free space}\\
\ = \ \frac{\sigma_g {\bar F}_\nu W_{g,{\rm tot}} n_1}{D_0n_2} \ \Biggr\rvert_{\rm free space} \\
\  = \ \frac{{\bar f}_{\rm diss} w F_0  \sigma_g n_1}{D_0n_2} \ \Biggr\rvert_{\rm free space}
\end{split}
\end{equation}
where in the second and third equalities we have used the relations $D_0\equiv \sigma_d^{\rm tot}{\bar F}_\nu$
and ${\bar F}_\nu W_{g,{\rm tot}}\equiv {\bar f}_{\rm diss}w F_0$.
This gives our second interpretation for $\alpha G$. It is the free-space ratio
of the H{\small I}-dust to H$_2$-line absorption rates of LW photons in the 
H$_2$-dust-limited dissociation band.

Third, since $D_0G={\bar f}_{\rm diss}\sigma_g wF_0$ (and ${\bar p}_{\rm diss}\equiv w{\bar f}_{\rm diss}$)
we have that
\begin{equation}
\label{aG2}
\alpha G
\ = \ {\bar f}_{\rm diss}\frac{\sigma_g wF_0}{Rn} 
 \ = \ {\bar p}_{\rm diss}\frac{\sigma_g F_0}{Rn} \ \ \ .
\end{equation}
Thus, $\alpha G$ is the free-space ratio of the dust absorption rate 
of the effective dissociation flux - {\it per hydrogen atom} - 
to the molecular formation rate per atom.



When $\alpha G \ll 1$, H{\small I}-dust
plays no role anywhere in the cloud,
since it is negligible even for the free-space field where the
atomic density is largest.  However, when $\alpha G \gg 1$, H{\small I}-dust becomes important 
and absorbs an increasing fraction of the LW photons that are otherwise available
for H$_2$ photodissociation.

Thus, with Equations~(\ref{alphaDef}) and (\ref{Gfit}),
\begin{equation}
\begin{split}
\label{aG_A}
\alpha G \ = & \ 1.54 \ \biggl(\frac{D_0}{5.8\times 10^{-11} \ {\rm s}^{-1}}\biggr) 
 \biggl(\frac{3\times 10^{-17} \ {\rm cm}^3 \ {\rm s}^{-1}}{R}\biggr)\\
 & \times
\biggl(\frac{100 \ {\rm cm}^{-3}}{n}\biggr) \ \frac{\phi_gZ'}{1+(2.64\phi_gZ')^{1/2}} \ \ \ .
\end{split}
\end{equation}  
Or, with Equation(\ref{aG2}),
\begin{equation}
\label{aG_B}
\begin{split}
\alpha G \ = \ 1.54 \ \biggl(\frac{\sigma_g}{1.9\times 10^{-21} \ {\rm cm}^2}\biggr)
\biggl(\frac{F_0}{2.07\times 10^7 \ {\rm cm}^{-2} \ {\rm s}^{-1}}\biggr)\\
\times \biggl(\frac{3\times 10^{-17} \ {\rm cm}^3 \ {\rm s}^{-1}}{R}\biggr)
\biggl(\frac{100 \ {\rm cm}^{-3}}{n}\biggr)\ \frac{1}{1+(2.64\phi_gZ')^{1/2}}
\end{split}
\end{equation}
With Equations~(\ref{F0norm}) or (\ref{D0}) for $F_0$ or $D_0$,
and Equations~(\ref{Sstand}) and (\ref{Rstand}) for $\sigma_g$ and $R$ we then have
\begin{equation}
\label{aGfin}
\alpha G \ = \ 1.54 \ \frac{I_{\rm UV}}{(n/100~{\rm cm}^{-3})} \ \frac{\phi_g}{1+(2.64\phi_gZ')^{1/2}} \ \ \ .
\end{equation}
For $\sigma_g$ and $R$ varying the same way with $Z'$ there is a cancellation, 
but a metallicity dependence still remains via the
dissociation bandwidth $w$ (Equation~[\ref{defw}]) and its dependence on the 
competition between H$_2$-dust and H$_2$-line absorption.
For low $Z'$ (complete line overlap limit, $w=1$),
$\alpha G$ is independent of the metallicity, but 
for high $Z'$,
$w\sim \sigma_g^{-1/2}$, and $\alpha G \sim Z'^{-1/2}$.


\subsubsection{Weak- and Strong-Field Limits}


Expression~(\ref{aGfin}) also shows that the regimes of large and small $\alpha G$ are both
relevant for the widely varying conditions in localized environments of the ISM in galaxies. 
For example, $I_{\rm UV}$ can range from $\sim 1$ at ``average" locations to $\gtrsim 10^5$ 
near hot stars, whereas $n$ can range from $\sim 10$ cm$^{-3}$ in diffuse gas to 
$\gtrsim 10^6$~cm$^{-3}$ in dense molecular clouds. 
On global scales in star-forming galaxy disks $I_{\rm UV}$ and $n$ may be
correlated (see \S~2.2.9).  But on 
small scales, enhanced radiation fields will not necessarily be 
offset by higher gas densities, and $I_{\rm UV}/n$ may span a wide range,
from ``large" to ``small".

For $\alpha G/2 \ll 1$ the absorption of the LW radiation is dominated by
the combination of H$_2$-line and H$_2$-dust absorption, and H{\small I}-dust is negligible
\footnote{This limit is often referred to in the literature as the ``self-shielding" limit.
However, this is potentially confusing if self-shielding is properly understood to be
associated with just H$_2$-line absorption.  As we have been emphasizing, 
H$_2$-dust also absorbs and cannot be ignored for small $\alpha G$, 
unless the metallicity is very low.  So the weak-field limit may be referred
to as the ``H$_2$-line plus H$_2$-dust" shielding limit.
}.
For $\alpha G/2~\gg~1$ H{\small I}-dust absorption dominates the attenuation of the radiation field.
We refer to $\alpha G/2 \ll 1$ as the ``weak-field-limit" and to $\alpha G/2 \gg 1$ as the ``strong-field-limit".

We now consider the behavior of $N_{1,{\rm tot}}$ in these two limits.

It follows from Equations~(\ref{S88}) and (\ref{S88p}) that for $\alpha G/2 \ll 1$,
\begin{equation}
\begin{split}
\label{N1low}
N_{1,{\rm tot}}   &   \ = \frac{1}{\sigma_g}\frac{\alpha G}{2} \ = \ \frac{1}{2}\frac{1}{\sigma_g}\frac{D_0}{Rn} G \\
& \ = \ \frac{1}{2} \frac{{\bar F_\nu} W_{g,{\rm tot}}}{Rn}
\ = \  \frac{1}{2}{\bar f}_{\rm diss}\frac{wF_0}{Rn} \ \ \ .
\end{split}
\end{equation}
So for weak fields, the total H{\small I} column density is equal to the ratio of the
effective LW dissociation flux to the H$_2$ formation rate (which is the removal rate of    
for the H{\small I}). The atomic column is proportional to the surface 
dissociation rate $D_0$, (or to the field-strength $I_{\rm UV}$)
and inversely proportional to the cloud gas density $n$.

In the weak-field limit 
the H{\small I}-dust opacity associated with the total atomic column 
$\tau_{1,{\rm tot}}\equiv \sigma_g N_{1,{\rm tot}} \ll 1$. We again see that H{\small I}-dust
plays no role in attenuating the LW radiation field in this limit. 
For weak fields, the total H{\small I} 
column depends on $\sigma_g$ only via $W_{g,{\rm tot}}$,
through the possible competition between H$_2$-line and
H$_2$-dust absorption.  For small-$\sigma_g$
where H$_2$-dust is negligible, dust absorption plays no role whatsoever, and
$N_{1,{\rm tot}}$ is completely independent of $\sigma_g$.
In the small-$\sigma_g$ limit, $G\propto \sigma_g$, and the dust-absorption cross section
cancels out completely.
For $R\propto Z'$, we then have that $N_{1,{\rm tot}} \propto 1/Z'$,
and the metallicity dependence enters entirely via the H$_2$ formation rate coefficient.
For large-$\sigma_g$, H$_2$-dust absorption is non-negligible,  $W_{g,{\rm tot}} \propto \sigma_g^{-1/2}$
(and $G\propto \sigma_g^{1/2}$) 
so $N_{1,{\rm tot}}$ scales as $\sigma_g^{-1/2}$.
Then with $\sigma_g \propto Z'$ and $R \propto Z'$ we have that
$N_{1,{\rm tot}} \propto Z'^{-3/2}$.

Expression~(\ref{N1low}) can be written as the simple
``Str{\" o}mgren relation"
\footnote{
This is analogous 
to the Str{\" o}mgren expression, $\alpha_B n_e N_{{H^+}} = F_{\rm Lyc}$,
for the column density, $N_{{H^+}}$, of ionized hydrogen in a slab 
that fully absorbs a flux, $F_{\rm Lyc}$, of Lyman continuum photons. Here $n_e$ is the 
electron density, and $\alpha_B$ is the electron-proton recombination coefficient.} 
\begin{equation}
Rn N_{1,{\rm tot}} \ = \ \frac{1}{2}{\bar F}_\nu W_{g,{\rm tot}}   \ \ \ .
\end{equation}
The effective dissociation flux on the right-hand-side is the rate per unit area at which 
dissociating photons (those absorbed in H$_2$ lines but not by H$_2$-dust) 
penetrate the cloud surface. By definition 
these photons are fully absorbed by the H$_2$ when H{\small I}-dust is negligible, so this is also
the photodissociation rate per unit surface area. In
steady-state this must equal the total H$_2$ formation rate per unit area,
which is the left-hand-side. 


For the strong-field limit $\alpha G/2 \gg 1$ it follows from Equation~(\ref{S88}) that
\begin{equation}
\begin{split}
\label{N1high}
N_{1,{\rm tot}} & \ = \frac{1}{\sigma_g}{\rm ln}[\frac{\alpha G}{2}] 
\ =  \ \frac{1}{\sigma_g}{\rm ln}[\frac{1}{2}\frac{D_0 G}{Rn}] \\
& \ = \ \frac{1}{\sigma_g}{\rm ln}\Bigl[\frac{1}{2}\frac{\sigma_g{\bar F_\nu} W_{g,{\rm tot}}}{Rn}\Bigr]\\
& \ = \  \frac{1}{\sigma_g}{\rm ln}\Bigl[\frac{1}{2}{\bar f}_{\rm diss}\frac{\sigma_g wF_0}{Rn}\Bigr]  \ \ \ . 
\end{split}
\end{equation}
For strong fields, $\tau_{1,{\rm tot}}=\sigma_g N_{1,{\rm tot}} \gtrsim 1$, and the 
H{\small I}-dust opacity associated with the total
atomic column contributes significantly to the attenuation of the incident LW flux.
This leads to a saturation, and logarithmic dependence of the atomic column
on the cloud parameters. For example, increasing the atomic column
by increasing the LW-band flux also leads to more effective absorption
of the LW photons by the larger H{\small I}-dust column.
A decreasing fraction of the LW photons is then absorbed by the H$_2$, and the 
growth of the atomic column is limited.
Similarly, increasing the H$_2$ formation rate reduces the atomic column,
but the H{\small I}-dust opacity is then also reduced, which increases the LW fraction 
available for photodissociation, thereby moderating the reduction of the H{\small I} column.

If we neglect the logarithmic factor, then up to a factor of order unity, we have that
\begin{equation}
N_{1,{\rm tot}} \ \approx \ 1/\sigma_g 
\end{equation}
for intense fields.
Indeed, if the attenuation is dominated by H{\small I}-dust 
the atomic column must approach a value such that 
$\tau_{1,\rm{tot}}\equiv \sigma_g N_{1,{\rm tot}} \gtrsim 1$.
Then, if $\sigma_g  \propto Z'$ the H{\small I} column
$N_{1,{\rm tot}}\propto 1/Z'$ (neglecting the logarithmic factor). 
In the strong-field limit the metallicity dependence enters
via the grain absorption cross section.

Finally, Equation~(\ref{N1high}) for the strong-field limit
may also be expressed as the Str{\" o}mgren relation
\footnote{To see this, note that
${\rm e}^{-\tau_{1,{\rm tot}}} = 1/\alpha G$ when $\alpha G \gg 1$, .}
\begin{equation}
\label{strom2}
Rn N_{1,{\rm tot}} \ = \ \frac{1}{2} {\bar F}_\nu W_{g,{\rm tot}} \  u \ \ \ .
\end{equation}
Here
\begin{equation}
u \equiv  \tau_{1,{\rm tot}} \ {\rm e}^{-\tau_{1,{\rm tot}}} 
\end{equation}
is a reduction factor 
that accounts for H{\small I}-dust attenuation of the effective dissociation flux.
The right-hand-side of Equation~(\ref{strom2}), including the factor $u$,
is the H$_2$ photodissociation rate per unit surface area, and this
equals the total H$_2$ formation rate per unit area, which is the left-hand side.


\subsubsection{$(\alpha G)_{\rm CNM}$ for Two-Phase Equilibria}

Galaxy disks may be self-regulated such that the thermal pressures in the H{\small I} gas 
enable a cold/warm (CNM/WNM) two-phased mixture, 
with cold-neutral-H{\small I} (CNM) accumulating in the UV illuminated PDRs 
of the star-forming molecular clouds
(\citealt{Ostriker_10,Faucher_13,Kim_13}).  As invoked by KMT/MK10, 
the ratio $I_{\rm UV}/n$ may then be restricted to a narrow range.
This then gives rise to a characteristic $\alpha G$ for self regulated disks, as follows.

For a given heating rate,
two-phased equilibrium occurs for a narrow range of thermal pressures and
associated CNM and WNM densities, $n_{\rm CNM}$ and $n_{\rm WNM}$, 
as controlled by the combined action of 
Ly$\alpha$ and C{\small II} fine-structure emission line cooling, with a
metallicity dependence via the abundance of the gas-phase carbon ions
(\cite{Field_69},\citealt{Wolfire_2003}).
Given the FUV heating rates and 
the metallicity dependent emission-line cooling rates,
the characteristic CNM density for two-phase equilibrium will be close to the minimum
gas density for which CNM is possible. \citet{Wolfire_2003}
developed the analytic formula
\begin{equation}
\label{ncnm}
n_{\rm CNM} =\frac{31\phi_{\rm CNM}}{1 + 3.1Z'^{0.365}}I_{\rm UV} \ \ \ {\rm cm}^{-3} \ \ \ ,
\end{equation}
for the characteristic CNM density, assuming that FUV grain photoelectric emission 
is the dominant heating mechanism for the gas. In this expression,
$I_{\rm UV}$ is again the FUV intensity (normalized to the Draine field),
$Z'$ is the metallicity, and $\phi_{\rm CNM}$ is a factor of order unity.
Following KMT/MK10 we set $\phi_{\rm CNM}=3$. This gives
$n_{\rm CNM}=23$~cm$^{-3}$ for $I_{\rm UV}=1$, at $Z'= 1$.

Most importantly, $n_{\rm CNM}$ is proportional to $I_{\rm UV}$.  If the gas density $n$
in the FUV illuminated gas is set equal to $n_{\rm CNM}$, 
the ratio $I_{\rm UV}/n$ then depends on the metallicity only, as given by
Equation~(\ref{ncnm}).  With Equation~(\ref{aGfin}) we then have
\begin{equation}
\label{aGcnm}
(\alpha G)_{\rm CNM} \ = \ 6.78 \ \Bigl(\frac{1 + 3.1 Z'^{0.365}}{4.1}\Bigr)\frac{\phi_g}{1 + (2.64 \phi_gZ')^{1/2}}
\end{equation}
for self-regulated systems.  In Figure \ref{Fig:aGCNM_Z} the solid curve is
$(\alpha G)_{\rm CNM}$ versus $Z'$ as given by Equation~(\ref{aGcnm}) (assuming $\phi_g=1$).  
The dashed curve excludes the $(2.64 \phi_gZ')^{1/2}$ term that accounts for
H$_2$-dust reduction of the effective dissociation bandwidth (not considered
by KMT/MK10 as discussed in \S 4). Remarkably, the enhanced $n_{\rm CNM}$
associated with reduced cooling efficiency at low metallicity (Equation~(\ref{ncnm}) 
is offset by the increased dissociation bandwidth at low $Z'$.  
Thus, so long as grain photoelectric heating dominates we have that 
\begin{equation}
\frac{(\alpha G)_{\rm CNM}}{2}\ \approx  \ 1
\end{equation} 
for self-regulated systems, independent of $Z'$. The expected H{\small I} columns are
then midway between the weak- and strong-field limits.

\begin{figure}[h!]
\centerline{\includegraphics[width=1.0 \columnwidth]{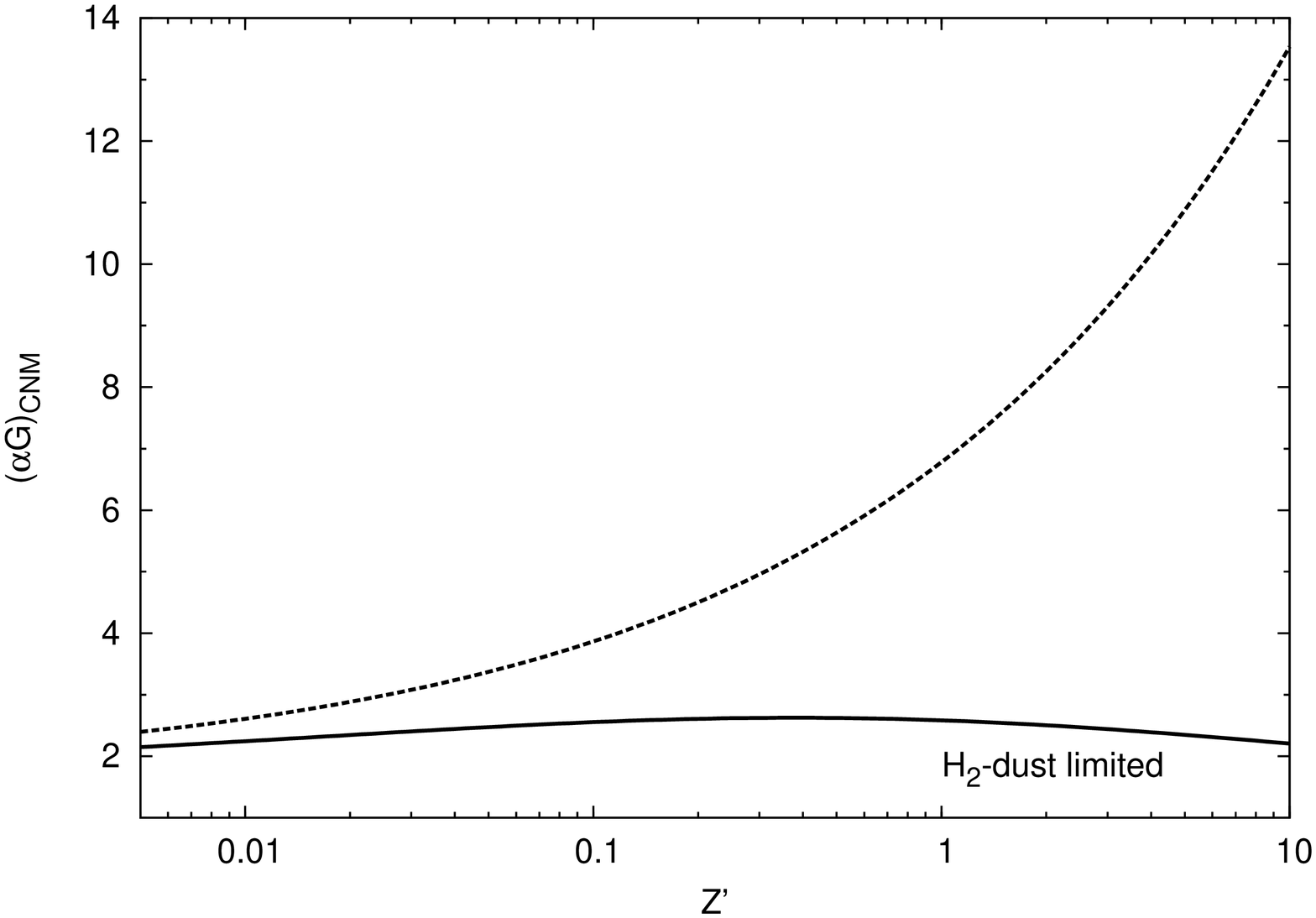}}
\caption{Metallicity dependence of $(\alpha G)_{\rm CNM}$ for two-phase equilibrium as given by Equation~(\ref{aGcnm}),
with and without (solid and dashed curves) H$_2$-dust reduction of the effective dissociation bandwidth.}
\label{Fig:aGCNM_Z}
\end{figure}

~\\
~\\
~\\

\subsection{Isotropic Fields}

We now consider clouds illuminated by isotropic radiation fields.

Let $I_\nu$ be the specific LW photon intensity of an isotropic field, and let $\mu\equiv cos\theta$, 
where $\theta$ is the angle of an incident ray relative to the cloud normal.
Again, we are assuming plane-parallel clouds.
In the absence of dust, the photodissociation rate at a 
molecular column $N_2$ normal to the cloud surface may be written as
\begin{equation}
\begin{split}
\label{Diso}
D(N_2)  &  \ = \ \frac{D_0}{2} \int_0^1 f_{shield}(N_2/\mu)\  d\mu \\
&  \ = \ 2\pi {\bar I}_\nu \int_0^1 \frac{dW_d(N_2/\mu)}{dN_2} \ \mu \ d\mu
\end{split}
\end{equation}
where $f_{shield}(N_2)$ is the uni-directional self-shielding function defined by 
Equation~(\ref{shield}), and $W_d(N_2)$ is the {\it same} effective ``multi-line" curve-of-growth for 
the (dust-free) dissociating bandwidth that appears in Equation~(\ref{DN2p}) for beamed fields.
In Equation~(\ref{Diso}), the contribution to the angular-integrated dissociation rate is reduced by the
self-shielding factor $f_{shield}(N_2/\mu)$ along each ray. Here,
 ${\bar I}_\nu\equiv {\bar F}_\nu/4\pi$ is the free-space specific intensity averaged over the dissociating transitions, as given by Equation~(\ref{defbarF}).


With the inclusion of dust absorption along each ray, the H{\small I}/H$_2$ formation-destruction equation
at a molecular column $N_2$ is
\begin{equation}
\begin{split}
Rn\ & dN_1 \ = \ 2\pi {\bar I}_\nu  \\
&\times \int_0^1 \frac{dW_d(N_2/\mu)}{dN_2}
{\rm e}^{-2\sigma_gN_2/\mu} {\rm e}^{-\sigma_gN_1/\mu} \ \mu  d\mu \ dN_2 \ \ \ .
\end{split}
\end{equation}
Like Equation~(\ref{sep}) for beamed fields, this is a differential equation for $N_1(N_2)$.
However, because of the angular integration, 
the exponential term ${\rm exp}(-\sigma_g N_1/\mu)$ for the H{\small I}-dust attenuation along a ray
cannot be moved to the left-hand-side as was done in Equation~(\ref{sep}).
To enable a separation of variables nevertheless, we replace the variable $\mu$ with some constant average 
$\langle\mu\rangle$ in the H{\small I}-dust term, and we {\it then} move this term to the left-hand-side.  
For H$_2$-dust attenuation on the right-hand-side we keep $\mu$ as a variable in the exponential term.

Making this approximation and integrating over the atomic and molecular columns gives,
\begin{equation}
\begin{split}
\label{isostep}
Rn \ \int_0^{N_{1,{\rm tot}}}{\rm e}^{\sigma_g N'_1/\langle\mu\rangle} \ dN'_1 
& \ = \ 2\pi {\bar I}_\nu \ W_{g,{\rm tot}} \int_0^1 \mu d\mu \\
& \  =  \ \frac{1}{4}\ {\bar F}_\nu \ W_{g,{\rm tot}}
\end{split}
\end{equation}
where
\begin{equation}
\label{Wtotmu}
W_{g,{\rm tot}} \equiv \int_0^\infty \mu \frac{dW_d(N_2/\mu)}{dN_2} \ {\rm e}^{-2\sigma_gN_2/\mu}
\ d\bigl(\frac{N_2}{\mu}\bigr) \ \ \ .
\end{equation}
For $W_{g,{\rm tot}}$ as defined by
Equation~(\ref{Wtotmu}), the integration variable is $N_2/\mu$, and the integration is
from 0 to $\infty$. Thus,
the $W_{g,{\rm tot}}$ that appears here is the {\it same}
effective total H$_2$-dust-limited dissociation bandwidth that we defined for beamed radiation in Equation~(\ref{Wtot}) 
(and represented analytically by Equation~[\ref{Wtotfit}]). The effective bandwidths are {\it identical} for
isotropic and beamed fields because the relative fractions of LW photons absorbed by H$_2$-lines versus H$_2$-dust 
along a ray do not depend on the ray's orientation. 
The product ${\bar F}_\nu W_{g,{\rm tot}}/4$ is then the effective dissociating flux
for the isotropic field.

It follows from Equation~(\ref{isostep}) that the total H{\small I} column is given by
\begin{equation}
\begin{split}
\label{S88iso}
N_{1,{\rm tot}} & \ = \ \frac{\langle\mu\rangle}{\sigma_g} \ 
{\rm ln}\bigl[\frac{1}{4}\frac{\sigma_g}{\langle\mu\rangle}\frac{{\bar F_\nu} W_{g,{\rm tot}}}{Rn} \ + \ 1\bigr]\\
& \ = \ \frac{\langle\mu\rangle}{\sigma_g} \
{\rm ln}\bigl[\frac{1}{4}\frac{\sigma_g}{\langle\mu\rangle}{\bar f}_{\rm diss}\frac{wF_0}{Rn} \ + \ 1\bigr]
\end{split}
\end{equation}
or
\begin{equation}
\label{S88piso}
N_{1,{\rm tot}}  \ = \ \frac{\langle\mu\rangle}{\sigma_g} \ {\rm ln}\bigl[\frac{1}{\langle\mu\rangle}\frac{\alpha G}{4} \ + \ 1\bigr]
\end{equation}
where again $\alpha\equiv D_0/Rn$,
and $G\equiv \sigma_g W_{g,{\rm tot}}/\sigma_d^{\rm tot}$.
The total H{\small I}-dust optical depth in the normal direction is then
\begin{equation}
\label{S88tauiso}
\tau_{1,\rm tot} \ = \ \langle\mu \rangle \ {\rm ln}\bigl[\frac{1}{\langle\mu\rangle}\frac{\alpha G}{4} \ + \ 1\bigr]
\end{equation}
Equations~(\ref{S88iso}), (\ref{S88piso}), and (\ref{S88tauiso}) for isotropic fields
are very similar to Equations~(\ref{S88p}), (\ref{S88}), and (\ref{S88tau}) for beamed fields.
The values of $\alpha$ are {\it equal} for corresponding beamed and
isotropic fields (i.e., fields with the same $I_{\rm UV}$). Furthermore, $G$ is independent of the field geometry.
A geometrical factor of $1/4$ appears for slabs irradiated by isotropic fields
(as opposed to $1/2$ for beamed fields).

The average angle ${\langle\mu\rangle}$ appears in Equations~(\ref{S88iso}), (\ref{S88piso}), 
and (\ref{S88tauiso}) because the radiation fraction absorbed by H{\small I}-dust {\it does} depend on the field geometry,
and is larger for isotropic fields for which the relative H{\small I}-dust attenuation is increased along inclined rays.
In \S~3.1.5 we calculate ${\langle\mu\rangle}$ by fitting these analytic expressions to the results of our
numerical computations for the atomic columns for isotropic fields. We find that universally and
to an excellent approximation
${\langle\mu\rangle}=0.8$, independent of $\alpha G$ or $Z'$, i.e.
independent of the cloud parameters $n$, $R$, $D_0$ or $\sigma_g$.

We again consider the weak- and strong-field limits.

For weak fields ($\alpha G/4 \ll 1$),
\begin{equation}
\begin{split}
\label{N1lowiso}
N_{1,{\rm tot}} & \ = \ \frac{1}{\sigma_g}\frac{\alpha G}{4} = \frac{1}{\sigma_g}
\frac{1}{4}\frac{D_0}{Rn} G \\
& \ = \ \frac{1}{4}\frac{{\bar F_\nu} W_{g,{\rm tot}}}{Rn}
\ = \  \frac{1}{4}{\bar f}_{\rm diss}\frac{wF_0}{Rn}   \ \ \ ,
\end{split}
\end{equation}
and the H{\small I}-dust opacity $\sim \sigma_g N_{1,{\rm tot}}/{\langle\mu\rangle}$
is negligible and plays no role in attenuating the LW flux.
The total atomic column depends on $\sigma_g$ only via the dissociation bandwidth $W_{g,{\rm tot}}$,
i.e.~via the competition between H$_2$-line and H$_2$-dust absorption.
As for beamed radiation, Equation~(\ref{N1lowiso}) is a simple Str{\" o}mgren relation, and
$N_{1,{\rm tot}}$ is equal to the ratio of the 
effective dissociation flux (or dissociation rate per unit surface area)
to the H$_2$ formation rate.
In the weak-field-limit, and
for a given $\alpha G$ and $\sigma_g$, i.e., for a given $D_0$ (or $F_0$, or $I_{\rm UV}$),
$n$, $R$, and $\sigma_g$ (or $Z'$) 
the atomic column for isotropic radiation is
equal to half that produced by a corresponding beamed field. This is simply
due to the factor-of-two difference in the LW photon fluxes for corresponding isotropic versus beamed 
fields for a given field strength $I_{\rm UV}$.

In the strong-field limit, $\alpha G/4 \gg 1$, 
\begin{equation}
\begin{split}
\label{N1highiso}
N_{1,{\rm tot}} & \ = \ \frac{{\langle\mu\rangle}}{\sigma_g}
{\rm ln}\Bigl[\frac{\alpha G}{4}\Bigr]
\ =  \ \frac{{\langle\mu\rangle}}{\sigma_g}{\rm ln}\Bigl[\frac{1}{4}\frac{D_0 G}{Rn}\Bigr] \\
& \ = \ \frac{{\langle\mu\rangle}}{\sigma_g}
{\rm ln}\Bigl[\frac{1}{4}\frac{\sigma_g}{\langle\mu\rangle}\frac{{\bar F_\nu} W_{g,{\rm tot}}(\sigma_g)}{Rn}\Bigr] \ \ \ .
\end{split}
\end{equation}
For strong fields, $\sigma_g N_{1,{\rm tot}} \gtrsim 1$
and the H{\small I}-dust opacity is significant, and dominates the attenuation
of the radiation. As for beamed fields, the total atomic column saturates, and
$N_{1,{\rm tot}}$ is insensitive to the cloud parameters except for $\sigma_g$.
Up to the logarithmic factor of order-unity we then have
\begin{equation}
N_{1,{\rm tot}} \ \approx \ \frac{\langle\mu\rangle}{\sigma_g}
\end{equation}
Because ${\langle\mu\rangle}=0.8$
the saturation columns are only slightly smaller for isotropic versus beamed fields
for large $\alpha G$, and they are not very different.


\section{Numerical Model Computations}

With our analytic results (\S~2) in mind we now present detailed numerical calculations for 
the H{\small I}-to-H$_2$ transition profiles and the associated build-up of the atomic-hydrogen columns,
for planar clouds illuminated by either beamed or isotropic far-ultraviolet LW-band radiation fields. 
For this purpose we use the {\it Meudon PDR code}\footnote{Publically available at http://pdr.obspm.fr 
}
 \citep{LePetit_2006} for the computation of the UV radiative transfer and depth-dependent 
photodissociation rates, and for the steady-state atomic and molecular hydrogen gas densities.
The code implements the ``extended spherical harmonics" method \citep{Flannery_1980, Goicoechea_2007} for an exact numerical solution of the coupled H$_2$-line 
and dust scattering and absorption radiative transfer.  An adaptive frequency grid is employed
with sufficient resolution ($\Delta\nu/\nu \sim 10^{-5}$) to capture the contributions of 
the narrow H$_2$-line Doppler cores and broad wings to the the total ultraviolet opacities. The
competition between dust absorption and scattering and H$_2$-line-absorptions and the important effects 
of H$_2$-line overlap are included in the calculation of the local radiation field intensities and 
H$_2$ photodissociation rates. 

A principal feature of the {\it Meudon PDR code} is that the radiative transfer can be
calculated for either beamed or isotropic fields.  We assume the Draine
spectrum (Equation~[\ref{Draine}]) and we calculate models for 
beamed and isotropic configurations. As defined in \S 2.2, for corresponding beamed and
isotropic fields the radiation energy densities at the cloud surfaces are identical (\S2.1).
We compare our numerical results to the analytic formulae discussed above.

One of our main goals is the accurate computation of the effective H$_2$-dust-limited dissociation band-width
$W_{g, {\rm tot}}(\sigma_g)$ (Equation~[\ref{Wtot}]) summed over all of the LW-band absorption lines,
for a wide range of FUV-absorption dust cross sections $\sigma_g$ (as set by the metallicity $Z'$).  
We also calculate H{\small I}-to-H$_2$ transition profiles and the resulting
atomic-hydrogen columns from the weak- to strong-field limits (small to large $\alpha G$), for both beamed 
and isotropic fields, and for a wide range of metallicities.  

For computational efficiency in our large parameter-space and for direct comparisons
to our analytic formulae, we have made several simplifying assumptions and modifications to 
the standard {\it Meudon PDR code}. First, instead of considering a full range of grain 
sizes and wavelength dependent absorption/scattering properties we assume 
a single representative $\sigma_g$ for each metallicity (as given by Equation~[\ref{Sstand}])
independent of photon frequency within the narrow LW-band. We assume pure forward scattering 
by the grains, and neglect the (small) effects of back-scattering discussed by \citet{Goicoechea_2007}.
Thus, our $\sigma_g$ enters as a simple effective absorption cross section.  We have verified 
by spot checks within our parameter-space that our assumption of a constant $\sigma_g$ alters the
results for the computed H{\small I} columns by no more than 10\% for any metallicity
compared to computations incorporating a standard grain-size distribution.

Second, we decouple the H{\small I}/H$_2$ formation-destruction equation from the complex
gas and grain networks that govern the heavy-element chemistry. We assume
that the H$_2$ is formed on grain-surfaces only with a rate-coefficient $R$ 
as given by Equation~(\ref{Rstand}), and is destroyed only by (depth-dependent) LW photodissociation.  
We exclude gas-phase formation, e.g., via the sequence $H+e \rightarrow H^- + \nu$, $H^- + H \rightarrow H_2 + e$
\footnote{
With our assumption that the grain surface H$_2$ formation rate coefficient 
$R$ is linearly proportional to the metallicity, the gas phase formation
routes become important for $Z'\lesssim 5\times 10^{-3}$, depending on the temperature
and fractional ionization, $x_e$, of the gas. For an estimate of the effective rate coefficient, $R_-$, for 
H$_2$ formation via the negative-ion $H^-$ intermediary, see
e.g., Eq.~A7 of MK10, $R_-=8\times 10^{-19}x_{e,-3}T_3^{0.88}$~cm$^3$~s$^{-1}$.
},
or destruction by cosmic-ray or X-ray secondary-ionization
(and we set the ionization rate $\zeta$ equal to zero).
Thus, at depths where the LW radiation field is fully absorbed
the atomic density vanishes, and the entire H{\small I} column 
is maintained by photodissociation.

Third, we consider isothermal clouds rather than solving a heating-cooling equation
for the gas temperature $T$.  Our standard is $T=100$~K for which
the H$_2$ formation rate-coefficient $R=3\times 10^{-17}$~cm$^3$~s$^{-1}$ for $Z'=1$. 
And fourth, we assume that the total hydrogen nucleon density, $n=n_1+2n_2$,
is a constant independent of cloud depth, so that the local H$_2$ formation rate
$Rn$ (s$^{-1}$) is also constant for any given model.  

Fifth, we ignore absorption by neutral atomic carbon (C{\small I})
in the H$_2$ photodissociation layers.  The carbon (continuum) photoionization band 
1100-912 \AA~coincides almost exactly with the LW-band for H$_2$ photodissociation,
and C{\small I}, H$_2$, and dust compete for the same photons. The carbon 
photoionization cross section is $1.6\times 10^{-17}$~cm$^{2}$  (\citealt{vanDishoeck_06}),
so that the C{\small I} opacity $\tau_{{\rm CI}}=1.6\times 10^{-17}AZ'Nx_{{\rm CI}}\approx 1.6\times 10^{-21}Z'Nx_{{\rm CI}}$,
where $N$ is the hydrogen gas column, $A\approx 10^{-4}$ is the gas phase carbon abundance
for solar ($Z'=1$) metallicity, and $x_{{\rm CI}}$ is the fraction of carbon present in atomic form.
For $x_{{\rm CI}}\approx 1$, $\tau_{{\rm CI}}$ is competitive with the dust opacity 
$\tau_g=\sigma_gN\approx 1.9\times 10^{-21}N$.  However, the carbon is primarily C$^+$ and
$x_{{\rm CI}}$ is generally very small,
in the dust-limited or H$_2$-line-limited absorption layers in which the H{\small I} columns are built up, as follows.
In the absence of dust or H$_2$ absorption, 
and assuming that C{\small I} absorbs the entire LW flux in maintaining an 
outer and optically thick C$^+$ zone, then
$F_0 \simeq  \alpha n^2_{{\rm C^+}} {\ell}=  \alpha (AZ^\prime)^2 nN$,
where $F_0=2\times 10^7 I_{\rm UV}$ is the ionizing photon flux (Equation~[\ref{F0}]),
$\alpha\approx 2\times 10^{-11}$~cm$^3$~s$^{-1}$ is the electron-carbon recombination
rate-coefficient (\citealt{Wolfire_08}), $n_{{C^+}}=AZ^\prime n$ is the volume density of C$^+$ ions,
$\ell$ is the length-scale of the optically thick C$^+$ Str{\" o}mgren
layer, and $N$ is the {\it hydrogen} gas column 
associated with the C$^+$ layer.  We are assuming that the
carbon ions are neutralized by recombination with free electrons,
and that $n_e\approx n_{{C^+}}$.
However, if the dust opacity associated with this gas column is large, i.e., if
$\sigma_g N \gtrsim 1$, then
the C$^+$ layer is limited by dust and C{\small I} absorption may be ignored.  
The condition 
$\sigma_g N \gtrsim 1$ may be expressed as
$\sigma_g F_0/[(AZ')^2\alpha n] \gtrsim 1$, 
or $I_{\rm UV}/n \gtrsim 5\times 10^{-6}Z'$~cm$^{3}$, or $\alpha G \gtrsim 10^{-3}Z'$.
Thus, unless $\alpha G$ is unusually small, C{\small I} absorption is negligible.
Using our PDR code, and turning on the effects of additional C$^+$ neutralization
processes such as dust-assisted recombination and chemical removal processes,
we find that even for $Z'=1$ the C{\small I} absorption is less than a 10\% effect 
for $\alpha G=0.01$, and is negligible for larger $\alpha G$.

Sixth, we ignore H{\small I} Lyman-series line absorption of the LW-band photons.
These atomic lines
(beginning with $Ly\beta$ at 1026~\AA) do appear within the molecular LW 
absorption band (e.g., \citealt{Draine_1996}), and the {\it Meudon PDR code} includes them.
 However, very large atomic columns are required for the
atomic-line equivalent widths to contribute significantly to the absorption.
We find that an H{\small I} column 
of $\sim 10^{24}$~cm$^{-2}$ is required for the summed equivalent widths of 
the $Ly$-series lines to equal half the LW bandwidth (most of this
absorption is due to just $Ly\beta$).  Such large H{\small I} columns 
are produced only for very small values of $\sigma_g$ and $Z'$, even in the
strong-field large $\alpha G$ limit.  For example,  for $\alpha G\sim 100$ it follows
from Equation~(\ref{S88}) that $N_{1,{\rm tot}}\gtrsim 10^{24}$~cm$^{-2}$ requires 
$\sigma_g \lesssim 3.9\times 10^{-24}$~cm$^2$, or $Z'\lesssim 2 \times 10^{-3}$.
Thus, for the relevant range of metallicities, the atomic line absorptions can be
ignored, and we exclude them in the radiative transfer.

With the above assumptions, the basic inputs to the code are the intensity, $I_{\rm UV}$, and spectral shape
of the radiation field (we assume the Draine representation, Eq.~[\ref{Draine}]), in either a 
beamed or isotropic configuration,
the total gas density $n$, the associated H$_2$ formation rate $Rn$ (temperature 
and metallicity dependent), and the dust-grain absorption cross section $\sigma_g$ (metallicity dependent).
The dimensionless parameter $\alpha G$ is formed from these cloud variables
as described in \S 2.2.5 and 2.2.6.

For any set of parameters the critical numerical computation is for the depth-dependent
line-plus-continuum absorptions and the attenuation of the H$_2$ photodissociation rate. 
For the H$_2$, our code includes
all 302 ro-vibrational $vj$ levels in the $X~^1\Sigma_g^+$ ground electronic state, and the entire matrix
of Lyman and Werner transitions to discrete $v'j'$ levels in the excited $B~^1\Sigma_u$ and
$C~^1\Pi_u$ states.  We exclude transitions 
with energies greater than the hydrogen ionization energy of 13.59~eV.
(As is standard, we assume that any ionizing photons 
are always absorbed in adjacent H{\small II} regions outside of the PDRs).
Our code includes transitions out of excited
$v$-levels of the $X$-state. However, for almost all conditions of interest and
throughout most of the H{\small I}-to-H$_2$ transition zones
the dominating line absorptions are from the lowest few rotational levels
($j$=0 to 5) levels of the $v=0$ level 
\footnote{Photodissociation out of excited vibrational levels becomes significant when
$I_{\rm UV}\gtrsim 10^5$ and the UV excitation rates becomes comparable to the 
quadrupole vibrational decay rates \citep{Shull_1978}.  However, even for such intense fields, the
excitation rates become small as the lines rapidly become optically thick,
and for most of the HI layer absorption out of excited vibrational states
is negligible (S88).
}.
Thus, the relevant UV transitions lie between 1108 and 912 \AA, and this is the wavelength 
range of our ``standard LW-band".

The fractional populations, $x_{vj}$, of the $vj$ levels in the $X$-state are computed assuming 
population and depopulation by the upward and downward $X$-$B$ and $X$-$C$ transitions, 
quadrupole radiative (cascade) transitions between the $X$-state $vj$ levels, 
and excitations and deexcitations in collisions with He, H$^+$, H, and other H$_2$ molecules.    
We use the \citet{Abgrall_1993_a} and \citet{Abgrall_1993_b} 
Lyman- and Werner-band oscillator strengths and transition wave-numbers,
and the \citet{Abgrall_2000} probabilities for spontaneous radiative dissociations from the 
individual ro-vibrational levels in the 
excited $B$ and $C$ states.  For the radiative quadrupole transitions we use the Einstein-A values 
computed by \citet{Wolniewicz_1998}. For the collisional processes we use the 
ro-vibrational state-to-state rate coefficients reported in \cite{LeBourlot_99} and \cite{Wrathmall07} 
in their study of H$_2$ excitation in astrophysical media.

As an example of our radiative transfer computations we show in Figure~\ref{Fig:spectrum} the
energy density (erg cm$^{-3}$ \AA$^{-1}$) at a cloud column depth 
$N=3.74\times10^{20}$~cm$^{-2}$, or $\tau_g=0.7$
($A_V=0.2$), for $\sigma_g=1.9\times 10^{-21}$~cm$^2$
($Z'=1$) for a model computation with $I_{\rm UV}=35.5$ (beamed), and $n=1000$~cm$^{-3}$,
or $\alpha G/2=1$.  At this depth the cores of the individual absorption line are
very optically thick, but the continuum between the lines has not yet been significantly
attenuated by overlapping line wings. Around 80 strong lines are visible
in Figure~\ref{Fig:spectrum}, consistent with our analytic estimate 
$\sigma_d^{\rm tot}/\sigma_d \simeq 80$ for the number of lines involved in
the photodissociation process, as discussed in \S~2.2.1.

\begin{figure}[h!]
\centerline{\includegraphics[width=1.0 \columnwidth]{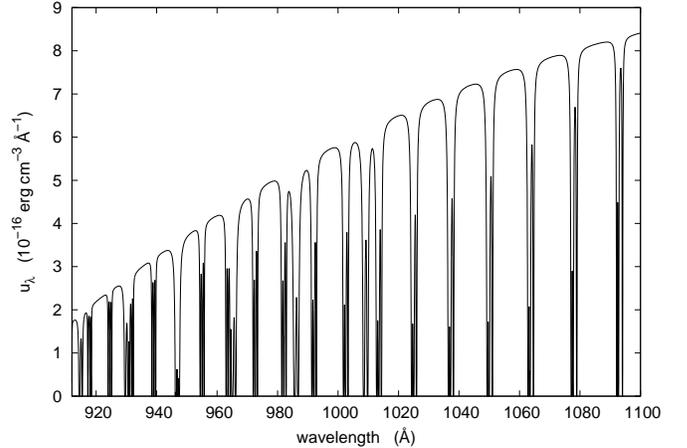}}
\caption{Absorbed far-UV spectrum showing partially overlapping Lyman-Werner band
absorption lines, for beamed radiation into a cloud, at a total hydrogen gas column density of 
$3.74\times 10^{20}$~cm$^{-2}$, for a free-space radiation intensity $I_{\rm UV}=35.5$, gas density
$n=10^3$~cm$^{-3}$, and metallicity $Z'=1$ ($\alpha G/2$ = 1).}
\label{Fig:spectrum}
\end{figure}

\subsection{Results}

We now present our numerical code results for (a) the unattenuated free-space 
H$_2$ photodissociation rate; (b) the curves-of-growth for the H$_2$-dust-limited dissociation 
bandwidth $W_g(N_2)$; (c) the total dissociation bandwidth $W_{g,{\rm tot}}(\sigma_g)$;
(d) the self-shielding function $f_{shield}(N_2)$; and (e) the mean self-shielding factor $G(Z')$.
We then present H{\small I}-to-H$_2$ transition profiles and total integrated H{\small I} columns,
for beamed and isotropic fields,
for a range of $\alpha G$ and metallicities $Z'$. 

\subsubsection{Free-Space H$_2$ Photodissociation Rate}

The optically thin (full $4\pi$) free-space H$_2$ photodissociation rate, $D_0$,  is a fundamental parameter
for ISM and galaxy evolution studies, and we have recalculated it here for the Draine FUV spectrum
(Equation~[\ref{Draine}]).  In Table 1, and for $I_{\rm UV}=1$, we list the free-space UV-excitation rates,
$P_{vj}$ (s$^{-1}$)
out of the 14 lowest-lying H$_2$ $(v,j)$ ro-vibrational levels. Each rate 
is summed over all upward LW transitions. We also list the mean dissociation probabilities
$\langle f_{\rm diss}\rangle_{vj}$, averaged over all of  the transitions, and the
resulting dissociation rates, $D_{vj}$ (s$^{-1}$) 
out of each $(v,j)$ level.  Our numbers are consistent 
with \citet{Draine_1996} (see their Table~2) who used our basic input molecular data sets.

\begin{table}[h]
\begin{center}
\begin{tabular}{ccccc}
\hline
     ($v,j$)  & $E_{i}$ (cm$^{-1}$) & $P_{vj}$ (s$^{-1}$)& $\langle f_{{\rm diss}}\rangle_{vj}$ & $D_{vj}$ (s$^{-1}$)\\ \hline
\hline
(0,0)  & 0.000     & 4.71(-10) & 0.117   & 5.51(-11)  \\
(0,1)  & 118.505   & 4.75(-10) & 0.119   & 5.65(-11)  \\
(0,2)  & 354.363   & 4.83(-10) & 0.123   & 5.94(-11)  \\
(0,3)  & 705.567   & 4.95(-10) & 0.130   & 6.44(-11)  \\
(0,4)  & 1168.825  & 5.11(-10) & 0.145   & 7.41(-11)  \\
(0,5)  & 1740.277  & 5.30(-10) & 0.141   & 7.47(-11)  \\
(0,6)  & 2414.852  & 5.57(-10) & 0.160   & 8.91(-11)  \\
(0,7)  & 3187.691  & 5.86(-10) & 0.160   & 9.38(-11)  \\
(0,8)  & 4051.884  & 6.19(-10) & 0.175   & 1.08(-10)  \\
(1,0)  & 4161.259  & 7.14(-10) & 0.051   & 3.64(-11)  \\
(1,1)  & 4273.913  & 7.21(-10) & 0.055   & 3.97(-11)  \\
(1,2)  & 4497.992  & 7.31(-10) & 0.057   & 4.17(-11)  \\
(1,3)  & 4831.595  & 7.43(-10) & 0.057   & 4.24(-11)  \\
(0,9)  & 5002.162  & 6.58(-10) & 0.197   & 1.30(-10)  \\
\hline
\end{tabular}
\caption{H$_2$ Excitation and Dissociation Rates}
\label{tab1}
\tablecomments{The H$_2$ excitation rates, $P_{vj}$, mean dissociation fractions, $\langle f_{\rm diss}\rangle_{vj}$,
and dissociation rates, $D_{vj}$; for the free-space (optically thin) 
Draine radiation field, out of the lowest 14 ro-vibrational H$_2$ $(v,j)$ levels in order of
energy, $E_i$ (in cm$^{-1}$), relative to the ground (0,0) level.}
\end{center}
\end{table}

The {\it total} dissociation rate is weighted by the population fractions $x_{vj}$, but is
insensitive to the gas temperature or density when the 
excitation is mainly out of the lowest few $j$-levels. For
$T$ between 10 to $10^3$~K, and for $n$ ranging from 10 to $10^6$~cm$^{-3}$,
we find that to within at most a $2\%$ variation $D_0=5.8\times 10^{-11}$~s$^{-1}$
for $I_{\rm UV}=1$.  The dissociation rate is essentially proportional to the field intensity.
For very intense fields the rate is increased by enhanced excitation of the
rotational states and photodissociation out of these states by photons 
longward of 1108~\AA, outside our nominal LW band.
For $T=100$~K, and $n=100$~cm$^{-3}$, we find that for $I_{\rm UV}$ from $1$ to $10^3$,
$D_0=5.8\times 10^{-11} \phi_{\rm ex} I_{\rm UV}$~s$^{-1}$,
where the ``rotational excitation factor" $\phi_{\rm ex}$ increases from 1 to 1.5
for this range of field intensities.  We have also computed the mean flux density in the free-space radiation field, 
as defined by Equation~(\ref{defbarF}). For the Draine spectrum 
we find that ${\bar F}_\nu=2.46\times 10^{-8}\phi_{\rm ex}I_{\rm UV}$~photons~cm$^{-2}$~s$^{-1}$~Hz$^{-1}$, for 
the same range of $T$, $n$, and $I_{\rm UV}$. For the analysis we present in \S~2, we assume $\phi_{\rm ex}=1$.

\subsubsection{$W_g(N_2)$ and $W_{g,{\rm tot}}(\sigma_g)$}

As discussed in \S~2.2.4, the ``H$_2$-dust-limited dissociation bandwidth" $W_g(N_2)$
(Equation~[\ref{ddb}]) is a fundamental quantity for the H{\small I}-to-H$_2$ transitions
and the build-up of the H{\small I} column densities. 

In Figure~\ref{Fig:WGtot_N2} we plot our 
curve-of-growth computations for $W_g(N_2)$ integrated over all of the LW-band absorption lines,  
for $\sigma_g$ ranging from $1.9\times 10^{-20}$ to $1.9\times 10^{-23}$~cm$^{2}$,
corresponding to metallicities $Z'$ from 0.01 to 10.  We set the 
Doppler-$b$ parameters for all of the lines equal to a typical ISM cloud value of $2$~km~s$^{-1}$.
Our results are insensitive to the precise choice for $b$ because the dominant absorption lines
are very highly damped.
We extract the $W_g(N_2)$ curves from our numerical radiative transfer computations
for the radiation flux absorbed in the lines, self-consistently accounting for the 
flux reduction due to the presence of H$_2$-dust. For any $\sigma_g$, the curve-of-growth
$W_g(N_2)$ depends primarily on the internal molecular
oscillator strengths, line-profile cross-sections, and dissociation probabilities for the excited states.
We have verified by explicit computations that $W_g(N_2)$ is indeed
very insensitive to external cloud parameters such as the field intensity $I_{\rm UV}$ 
and/or gas density $n$, or temperature $T$. The curves-of-growth are also insensitive
to the rotational-level distributions and ortho-to-para H$_2$ ratio. 
The specific curves displayed in Figure~\ref{Fig:WGtot_N2} were 
extracted from model runs with $I_{\rm UV}=4$, $n=10^5$~cm$^{-3}$, and $T=100$~K,
and for a constant ortho-to-para ratio set equal to 3 (with H$^+$--H$_2$ proton exchange
reactions turned off). 

\begin{figure*}
\centerline{\includegraphics[width=0.9 \textwidth]{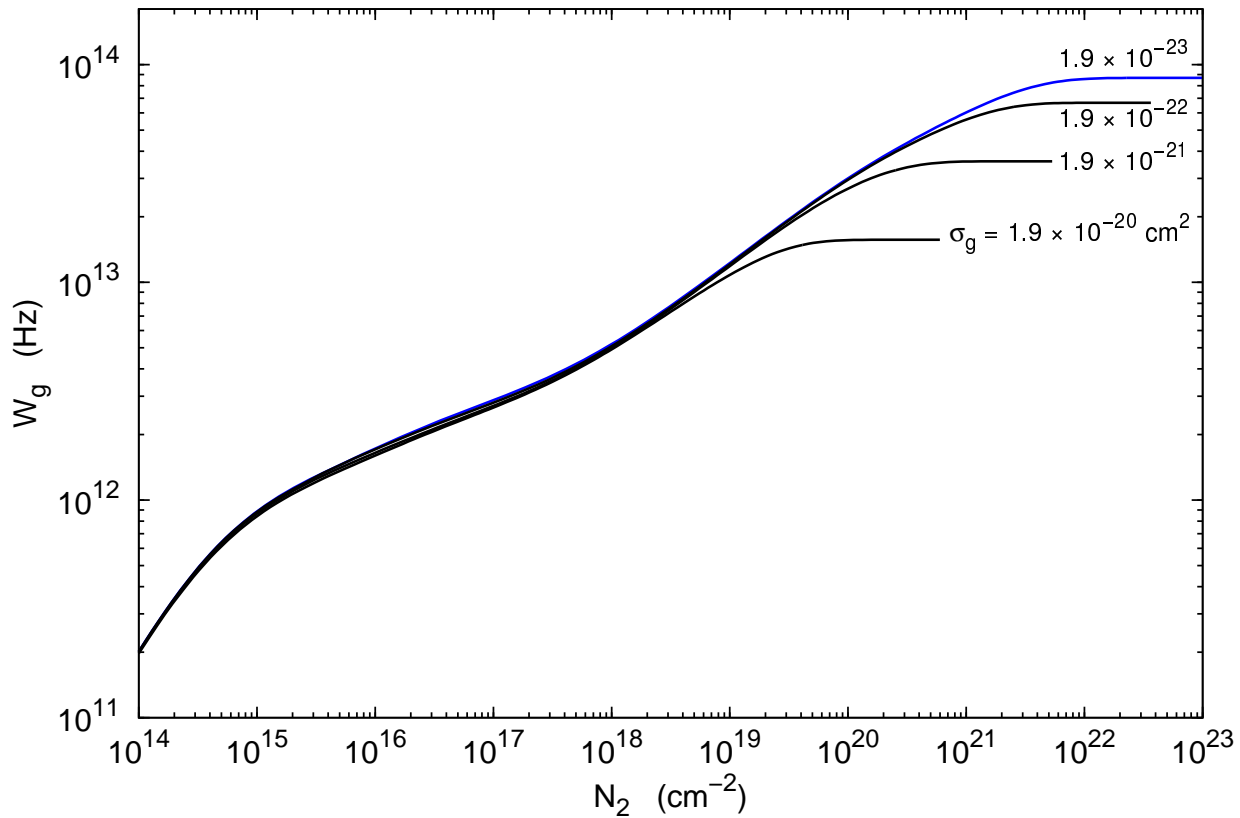}}
\caption{Curves-of-growth, $W_g(N_2)$, for the H$_2$-dust-limited dissociation bandwidth.  
The FUV dust absorption cross sections $\sigma_g$ range from $1.9\times 10^{-20}$
to $1.9\times 10^{-23}$~cm$^{2}$.  For $\sigma_g=1.9\times 10^{-23}$~cm$^{2}$ 
(blue curve) the H$_2$ lines fully overlap, and $W_g(N_2)=W_d(N_2)$ (see text).}
\label{Fig:WGtot_N2}
\end{figure*}

For $N_2 \lesssim 10^{14}$~cm$^{-2}$, all of the lines are optically thin, and $W_g$
increases linearly with $N_2$.
Between $10^{15}$ and $10^{17}$~cm$^{-2}$
the growth is logarithmic as the Doppler cores become optically thick.
At larger columns, $W_g$ increases more rapidly again as absorptions start occurring out of the line wings. 
For $N_2$ between $10^{18}$ and $10^{20}$~cm$^{-2}$
we find that $W_g$ grows as $N_2^{3/8}$, a bit more slowly than for a single damped line
(for which it would be $N_2^{1/2}$). 

$W_g$ saturates at sufficiently large H$_2$ columns. When H$_2$-dust is negligible 
the entire LW-band is absorbed in fully overlapping lines, and $W_g$ reaches a maximal value of
$9\times 10^{13}$~Hz, for $N_2\gtrsim 10^{22}$~cm$^{-2}$.  
In Figure~\ref{Fig:WGtot_N2} the absorption is essentially dust-free
for $\sigma_g=1.9\times 10^{-23}$~cm$^2$ since the lines overlap before the
H$_2$-dust opacity becomes significant, 
and for that (blue) curve $W_g(N_2)=W_d(N_2)$ (see \S~2.2.1 and 2.2.4).
For larger $\sigma_g$, and for $N_2 \gtrsim 1/(2\sigma_g)$, the asymptotic $W_g$ is limited by H$_2$-dust opacity.

Figure~\ref{Fig:Fit_Wtot} shows our results for 
the ``total dust-limited bandwidth" $W_{g,{\rm tot}}(\sigma_g)$  
(Equation[\ref{Wtot}]) 
for $\sigma_g$ from $10^{-24}$ to $10^{-20}$~cm$^2$.
The points are our numerical results, and the solid
curve is our analytic representation
\begin{equation}
\label{WtotfitB}
W_{g,{\rm tot}}(\sigma_g) \ \simeq \ \frac{9.9 \times 10^{13}} {1 + (\sigma_g/ \ 7.2\times 10^{-22} \ {\rm cm}^{2})^{1/2}} \ \ \ {\rm Hz} \ \ \ ,
\end{equation}
as already introduced in \S 2. This expression
is accurate to within 4\% compared to the numerical results.  The transition from the line-overlap
to H$_2$-dust limited regimes (small- to large-$\sigma_g$) occurs at $\sigma_g \sim 7.2\times 10^{-22}$~cm$^{2}$.
Line overlap is just starting to become important for solar ($Z'\sim 1$) metallicities. 
For a fully molecular slab, most of the LW-band radiation is absorbed by H$_2$-dust for $Z' \gtrsim 0.5$.
For $Z'  \lesssim 0.5$ most of the radiation is absorbed in H$_2$-lines.
Our results show that the regimes of small- and large-$\sigma_g$ are both
relevant for the realistic range of metallicities in galaxies.

\begin{figure}[h!]
\centerline{\includegraphics[width=1.0 \columnwidth]{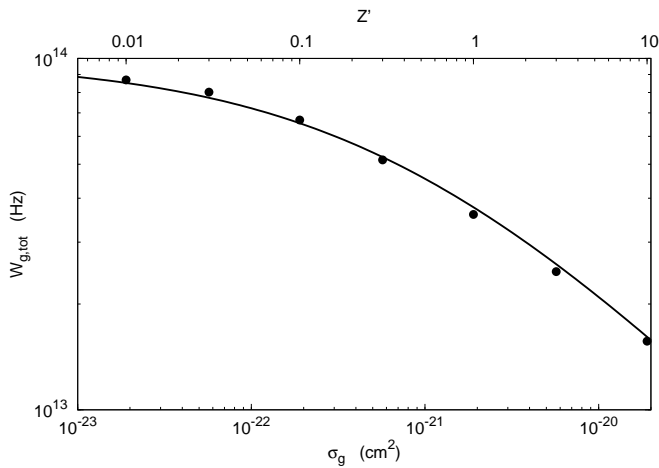}}
\caption{The total dust-limited dissociation bandwidth $W_{g,{\rm tot}}(\sigma_g)$.
The points are the results of our radiative transfer computations. The solid curve
is our fitting formula Equation~(\ref{Wtotfit}).}
\label{Fig:Fit_Wtot}
\end{figure}

To a good approximation  $W_{g,{\rm tot}}\propto \sigma_g^{-1/2}$ for large $\sigma_g$ as
indicated by our numerical results and Equation~(\ref{Wtotfit}).
This is the expected scaling for a
{\it single} (effective) damped absorption line in competition with H$_2$-dust
(see \S~2.2.4). 
We adopt $W_{g,{\rm tot}}\propto \sigma_g^{-1/2}$ for large $\sigma_g$
for our analytic scaling relations in \S~2, although 
$W_g(N_2)$ grows somewhat more slowly with $N_2$
than for a single line.

\subsubsection{Self-Shielding Function and G(Z')}

By definition, the H$_2$ ``self-shielding function" 
$f_{shield}(N_2)\equiv (1/\sigma_d^{\rm tot})dW_d/dN_2$ 
where $W_d(N_2)$ is the dissociation bandwidth for vanishing $\sigma_g$ (see Equation~[\ref{shield}]).

In Figure~\ref{Fig:fsh_DB} we plot (solid curve) our numerically computed derivative $f_{shield}(N_2)$.
We also plot (dashed) the \cite{Draine_1996} fit (their Equation [37]) for the shielding function, given by
\begin{equation}
\begin{split}
\label{DB96fit}
f_{shield}(N_2) = & \frac{0.965}{(1+x/b_5)^2} +\frac{0.035}{(1+x)^{0.5}} \\
& \times {\rm exp}[-8.5\times 10^{-4}(1+x)^{0.5}]
\end{split}
\end{equation}
where $x={N_2}/{5 \times 10^{14}~{\rm cm}^{-2}}$ and $b_5 = b /10^5$~cm~s$^{-1}$
(we assume $b_5=2$).

\begin{figure}[h!]
\centerline{\includegraphics[width=1.0 \columnwidth]{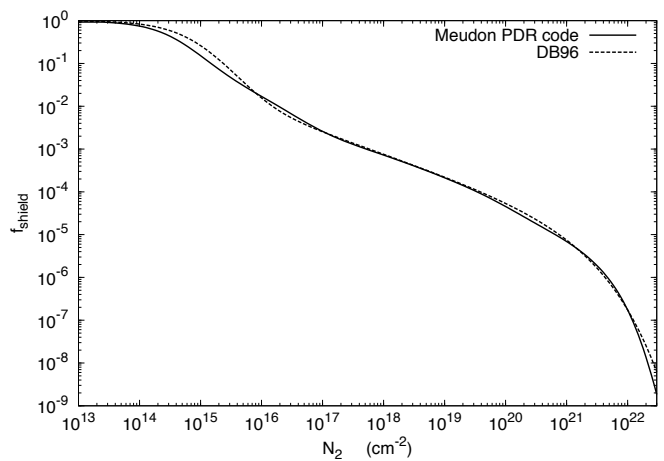}}
\caption{Our computed H$_2$ self-shielding function (solid curve) and the \cite{Draine_1996}
fitting formula (dashed curve) as given by Equation~(\ref{DB96fit}) for a Doppler parameter 
$b=2$~km~s$^{-1}$.}
\label{Fig:fsh_DB}
\end{figure}

It is evident that our computed shielding function is in excellent agreement with 
this formula.

At low $N_2$ the lines are optically thin and $f_{shield}~=~1$. As the line cores
become optically thick for $N_2~\gtrsim~10^{14}$~cm$^{-2}$ the
molecules ``self-shield" and $f_{shield}$ decreases. By  $10^{18}$~cm$^{-2}$, $f_{shield}=5\times 10^{-4}$. 
Between $10^{18}$ and $10^{20}$~cm$^{-2}$, the shielding function declines as $N_2^{-5/8}$,
as expected given the (integral) behavior of $W_d(N_2)$ in this range.
Finally, at larger columns $f_{shield}$ drops sharply as the lines fully overlap.
As found by \cite{Draine_1996} the reduction due to line overlap sets in at 
a column of $3\times 10^{20}$~cm$^{-2}$.  We again see that attenuation due to line overlap 
becomes important for $Z' \lesssim 1$. For such metallicities H$_2$-dust opacity becomes
significant only at columns
$1/(2\sigma_g)~\gtrsim~3\times 10^{20}$~cm$^{-2}$ at which point
the lines have already overlapped and the LW photons fully absorbed.
 
In Figure~\ref{Fig:fsh_GvsZ} we plot the ``average self-shielding factor", $G ~\equiv~(\sigma_g/\sigma_d^{\rm tot})W_{g,{\rm tot}}(\sigma_g)$
(Equations~[\ref{defGparam}] or [\ref{Gshield}]), as a function of the metallicity, assuming $\sigma_g=1.9\times10^{-21}Z'$~cm$^{2}$ ($\phi_g=1$).
The points are our numerical results, and the curve  
is our analytic fitting formula Equation~(\ref{Gfit}).
As expected, for low-metallicity (full overlap) $G\propto Z'$, but for high metallicity $G\propto Z'^{1/2}$
due to the H$_2$-dust cutoff.  For $Z'=10$, 1, 0.1, and 0.01, we find that $G$ equals
$1.3\times 10^{-4}$, $2.8\times 10^{-5}$, $5.4\times 10^{-6}$, and $7.1\times 10^{-7}$.  For these metallicities the average
self-shielded dissociation rates are
$D_0G/2=3.8\times 10^{-15}$, $8.1\times 10^{-16}$, $1.6\times 10^{-16}$, and $2.1\times 10^{-17}$~s$^{-1}$.
As discussed in \S 2.2.6 the average is over an H$_2$-dust optical depth $\tau_2\sim 1$.
For low-metallicities $D_0G/2$ becomes very small because the LW radiation is fully
absorbed in lines at very low H$_2$-dust optical depths.

\begin{figure}[h!]
\centerline{\includegraphics[width=1.0 \columnwidth]{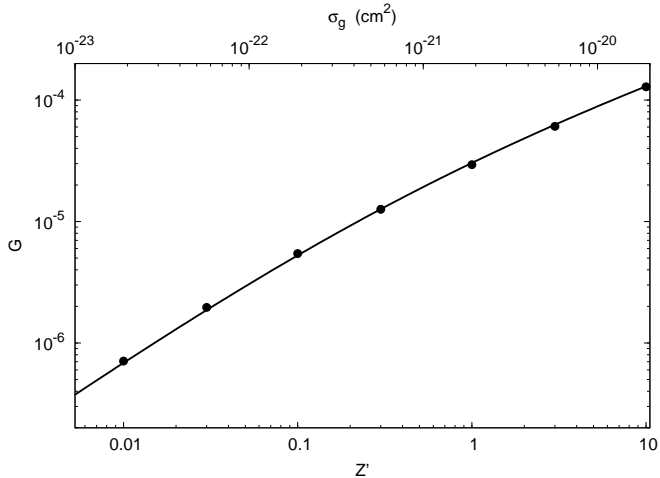}}
\caption{The average self-shielding factor $G$ as a function of metallicity $Z'$
or dust absorption cross section $\sigma_g$.  The points are our numerical results, and
the curve is our analytic expression Equation~(\ref{Gfit}).}
\label{Fig:fsh_GvsZ}
\end{figure}

~\\

\subsubsection{H{\small I}-to-H$_2$ Transition Profiles}

We now present illustrative computations for the H{\small I}-to-H$_2$ transition profiles
for a range of $\alpha G$
spanning the weak- to strong-field limits, and for metallicities $Z'$ from high to low,
for beamed and isotropic fields. 


As discussed in \S~2.2.5, for a given $\sigma_g$ the profile shapes, i.e.~the density ratios $n_1/n$ and $2n_2/n$
as functions of the total gas column $N$ (where $N\equiv N_1+2N_2$),
depend on just the single dimensionless parameter $\alpha G\equiv D_0G/Rn$.  
This includes the locations of the H{\small I}-to-H$_2$ transition points. We define the transition point as the cloud depth where
$n_1/n=2n_2/n=0.5$. For both beamed and isotropic fields we
compute five transition profiles
for $\alpha G/2 = 0.01$, 0.1, 1, 10 and 10$^2$, for $Z'=1$ and
$\sigma_g=1.9\times 10^{-21}$~cm$^{2}$ ($\phi_g=1$).
These sequences illustrate the change in profile shapes, from ``gradual" to ``sharp", 
from the weak-field (H{\small I}-dust negligible) to strong-field limits (H{\small I}-dust dominant).
In these computations $G=2.8\times 10^{-5}$ as appropriate for $Z'=1$, and
we set $D_0=2\times 10^{-9}$~s$^{-1}$ ($I_{\rm UV}\approx 35$),
and $R=3\times 10^{-17}$~cm$^3$~s$^{-1}$ ($T=100$~K). 
To alter $\alpha$ and $\alpha G$ we vary $n$ from $10$ to $10^{5}$~cm$^{-3}$.

For each $\alpha G/2$ we plot the atomic and
molecular fractions, $n_1/n$ and $2n_2/n$, as functions of the total
(atomic plus molecular) column density, $N$.
For fully atomic gas $n_1/n=1$, and
for fully molecular gas $2n_2/n=1$.  For each model we also plot the {\it normalized}
atomic column, ${\tilde N_1}\equiv N_1/N_{1,{\rm tot}}$, also as a function of $N$.
Thus, ${\tilde N_1}\rightarrow 1$ at sufficiently large cloud depths where the LW radiation is fully absorbed.
The curves for ${\tilde N_1}$  show how and where the atomic 
column is built up relative to the atomic-to-molecular transition points.
 
Figure~\ref{Fig:transition_HH2}, left panels, display the H{\small I}-to-H$_2$ transition profiles 
for the five beamed-field models with varying $\alpha G$.

\begin{figure*}
\centerline{\includegraphics[width=0.9 \textwidth]{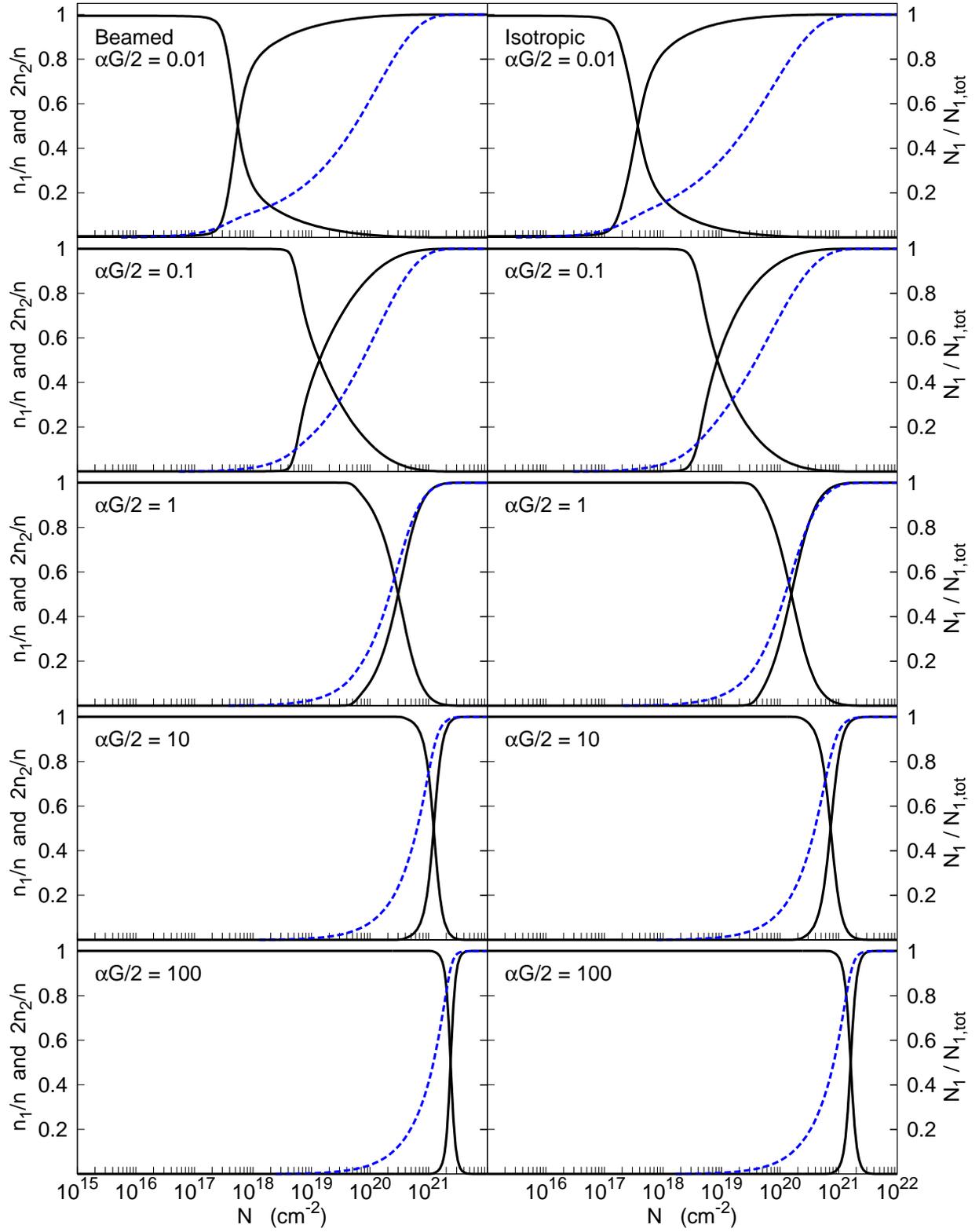}}
\caption{H{\small I}/H$_2$ transitions for beamed (left panels) and isotropic (right panels) radiation fields,
for $\alpha G/2$ ranging from 0.01 (top panels) to 100 (bottom panels) for $Z'$ = 1 and $\sigma_g = 1.9 \times 10^{-21}$ cm$^{2}$. The solid curves show the
atomic and molecular gas fractions, $n_1/n$ and $2n_2/n$, as functions of the total gas column
density $N$ into the cloud.  The (blue) dashed curves are the normalized atomic columns,
${\tilde N}_1\equiv N_1/N_{1,{\rm tot}}$ as functions of cloud depth.}
\label{Fig:transition_HH2}
\end{figure*}

Several important features can be seen in these plots.  
First, the atomic to molecular transition points 
move deeper into the cloud with increasing $\alpha G$.
For $\alpha G/2 = 0.01$, 0.1, 1, 10, and 10$^2$, the total gas columns
$N_{{1\rightarrow 2}}$ at the transition points equal $5.6\times 10^{17}$, 
$1.4\times 10^{19}$, $3.0\times 10^{20}$, $1.2\times 10^{21}$, and $2.4\times 10^{21}$~cm$^{-2}$.
For the assumed $\sigma_g=1.9\times 10^{-21}$~cm$^{2}$ these columns correspond to
total dust optical depths $\tau_g=1.05\times 10^{-3}$, $2.6\times 10^{-2}$, $0.58$, $2.3$, and $4.5$.
The H{\small I}-dust optical depths at the transition points are $9.0\times 10^{-4}$, 
$2.0\times 10^{-2}$, 0.45, 2.1, and 4.2.
In the weak-field limit, the transition depths increase rapidly with $\alpha G$.
In the strong field limit the UV penetration is moderated
by H{\small I}-dust absorption and the transition depths increase slowly.

Second, the profile shapes vary with $\alpha G$.  In the weak-field limit the atomic to molecular 
conversion is controlled by H$_2$-line self-shielding, but significant H{\small I} exists beyond 
the transition point up to the H$_2$-dust cut-off.
In the weak-field limit most of the H{\small I} column is built up past the transition point
where the gas is {\it predominantly 
molecular}.
In the strong-field limit the transition point is controlled by (exponential) H{\small I}-dust absorption,
and most of the H{\small I} is built up in an outer full atomic layer. Thus, in the weak-field limit
the transitions are {\it gradual}, and in the strong-field limit the transitions are {\it sharp}.
For example, at the H{\small I}-to-H$_2$ transition points
the normalized atomic column ${\tilde N_1}=0.08$,
$0.19$, $0.63$, $0.85$ and $0.92$, from small to large $\alpha G$.
For $\alpha G/2=0.01$, 92\% of the total atomic column is built up past
the transition point inside the molecular zone. For $\alpha G/2=100$, only 8\%
of the atomic column is produced past the transition point.

Figure~\ref{Fig:transition_HH2}, right panels, show the transition profiles for {\it isotropic} fields for
$\alpha G/2$ ranging from 0.01 to 100, for $Z'=1$ and $\sigma_g=1.9\times 10^{-21}$~cm$^2$.
As for the beamed-field models the shapes of the profiles vary from gradual
to sharp from the weak- to strong-field limits.  The transition
gas columns are $N_{1\rightarrow 2}=3.7\times 10^{17}$, $8.4\times 10^{18}$, 
$1.6\times 10^{20}$, $7.0\times 10^{20}$, and $1.6\times 10^{21}$~cm$^{-2}$,
or $\tau_g=5.8\times 10^{-4}$, $1.6\times 10^{-2}$, 
0.30, 1.4, and 3.0.
The normalized atomic columns at the transition points are
${\tilde N}_1=0.10$, 0.23, 0.59, 0.82, and 0.90. The H{\small I}-dust 
opacities are $5.8\times 10^{-4}$, $1.3\times 10^{-2}$,
0.24, 1.2, and 2.8.

For fixed $\alpha G \gtrsim 1$ but varying $Z'$ and $\sigma_g$ we expect the profile
shapes to be weakly dependent on the dust optical depth $\tau_g$, and for the UV penetration
scale-lengths and transition gas columns, $N_{1 \rightarrow 2}$,  to vary simply
as $1/Z'$ or $1/\sigma_g$.  This behavior is illustrated in 
the upper panel of Figure~\ref{Fig:transition_aGCNM} showing four transition profiles for
$Z'=10$, 1, 0.1, and 0.01 (again for beamed fields). 
For clarity we only plot the atomic $n_1/n$ curves.
For these models we set $\alpha G/2=(\alpha G)_{\rm CNM}(Z')/2=
1.11$, 1.30, 1.28, and 1.12, as given by Equation~(\ref{aGcnm}) for self-regulated multi-phased gas.
Because $\alpha G$ is about the same for these models 
the profile shapes are indeed very similar, and the transition columns
grow inversely with $Z'$.
For these four models, $N_{{1\rightarrow 2}}= 3.0\times10^{19}$, $3.7\times10^{20}$, $4.0\times10^{21}$, and $4.2\times10^{22}$~cm$^{-2}$,
corresponding to dust opacities $\tau_g=0.57$, 0.70, 0.76, and 0.78.
The H{\small I}-dust opacities at the transition points are 0.46, 0.56, 0.63, and 0.70. 
This behavior is in excellent agreement with the 
analytic theory presented in \S~2. For two-phased equilibrium 
$\alpha G/2\sim 1$ for all metallicities, and is intermediate between the weak- and strong field-limits
for which H{\small I}-dust opacity is just becoming significant.

Figure~\ref{Fig:transition_aGCNM}, lower panel, shows the $n_1/n$ curves for $Z'=10$, 1, 0.1, and 1,
for the isotropic-field models, with $\alpha G/2=(\alpha G)_{\rm CNM}(Z')/2$
as for the beamed-field models. Again, because $\alpha G$
is about the same for all four models the profile shapes are very similar.
The transition point columns are $N_{1\rightarrow 2}=1.6\times 10^{19}$, 
$1.9\times 10^{20}$, $2.1\times 10^{20}$, and $2.2\times 10^{22}$~cm$^{-2}$,
or $\tau_g=0.31$, 0.36, 0.40, and 0.42.
The H{\small I}-dust optical depths at the transition points are 0.25, 0.29, 0.33, and 0.37.

\begin{figure}[h!]
\centerline{\includegraphics[width=1.0 \columnwidth]{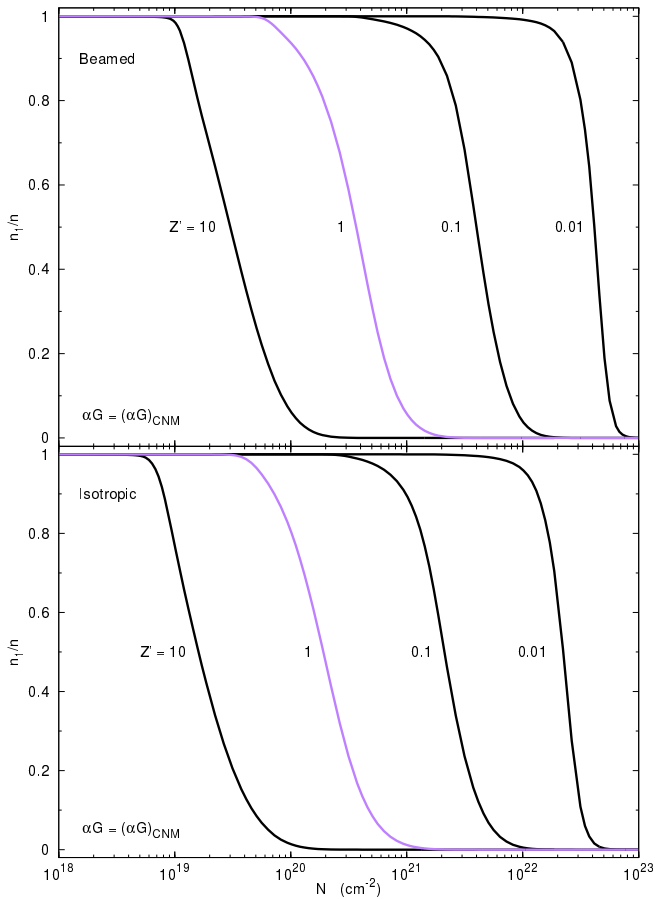}}
\caption{H{\small I} gas fractions, $n_1/n$, for beamed (upper panel)
and isotropic (lower panel) radiation, for $\alpha G= (\alpha G)_{\rm CNM}$
for metallicities $Z'$ from 0.01 to 10 times solar. }
\label{Fig:transition_aGCNM}
\end{figure}

The transition columns and optical depths for all of the isotropic-field models are
smaller than for the beamed-field models because of the factor-of-two reductions in
the incident fluxes for corresponding isotropic fields with the same $\alpha G$.


~\\
~\\

\subsubsection{Total H{\small I} Columns}

A key goal is the computation of the total atomic column densities, $N_{1,{\rm tot}}$, for
beamed and isotropic fields for a comparison to our analytic formulae.

In Figure~\ref{Fig:N1_aG}, upper panel, we display $N_{1,{\rm tot}}$ for beamed fields,
as a function of $\alpha G/2$ spanning the range from 0.01 to 100, for $Z'=10$, 1, 0.1, and 0.01. 

\begin{figure}[h!]
\centerline{\includegraphics[width=1.0 \columnwidth]{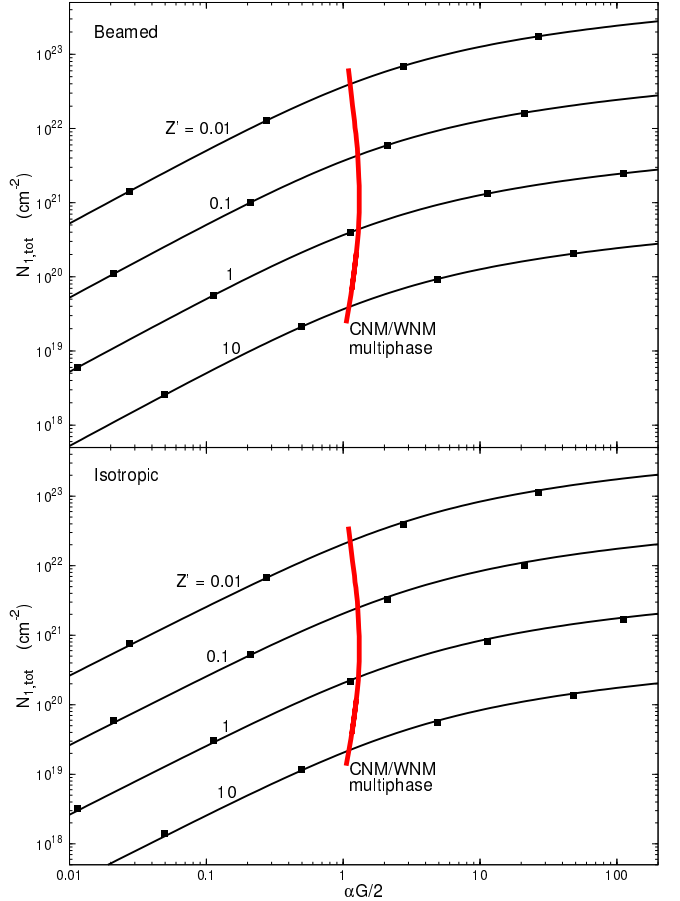}}
\caption{The total H{\small I} column as a function of $\alpha G/2$ for
beamed (upper panel) and isotropic (lower panel) radiation, for
metallicities $Z'$ from 0.01 to 10 times solar.  The square points
are the numerical results, and the solid (black) curves are as
given by our analytic formulae (Equations~(\ref{S88}) and \ref{S88iso}).
The red curves show $(\alpha G)_{\rm CNM}$ as functions of $Z'$ for
self-regulated two-phased equilibrium.}
\label{Fig:N1_aG}
\end{figure}

For our numerical models we again set $D_0=2\times 10^{-9}$~s$^{-1}$,
$R=3\times 10^{-17} Z'$~cm$^3$~s$^{-1}$ and vary the gas density $n$
(to select $\alpha\equiv D_0/Rn$) and we set $\sigma_g=1.9\times 10^{-21}Z'$~cm$^2$.
For the given input parameters
$D_0$, $R$, $n$, and $\sigma_g$, we compute the total integrated 
atomic column density, and also calculate $W_{g,{\rm tot}}$ 
and the associated average self-shielding factor 
$G=(\sigma_g/\sigma_d^{\rm tot})W_{g,{\rm tot}}$.  For each
model we compute $N_{1,{\rm tot}}$ for the corresponding $\alpha G$ and $\sigma_g$.
The points in Figure~\ref{Fig:N1_aG} are our model results for selected $\alpha G$ and $Z'$.
Again, for the four metallicities, $G=1.3\times 10^{-4}$, $2.8\times 10^{-5}$, $5.4\times 10^{-6}$, and $7.1\times 10^{-7}$.
The curves in Figure~\ref{Fig:N1_aG} are as given by our fundamental analytic formula 
for beamed fields
\begin{equation}
\label{S88B}
N_{1,{\rm tot}}\ = \ \frac{1}{\sigma_g}{\rm ln}\bigl[\frac{\alpha G}{2}+1\bigr] \ \ \ ,
\end{equation}
where $1/\sigma_g = 5.3\times 10^{20}/Z'$~cm$^{-2}$ as for the numerical models.  

Our analytic and numerical results are in excellent agreement
from the weak- to strong-field regimes.  
For weak fields (high gas densities) 
$N_{1,{\rm tot}}=(1/\sigma_g)\alpha G/2=(1/2){\bar F}_\nu W_{g,{\rm tot}}/Rn$ 
(Equation~[\ref{N1low}]).
In this regime the total atomic column 
is proportional to the field intensity (dissociation flux) and varies inversely with the gas density.
For strong fields (low densities) the dependence is logarithmic and
$N_{1,{\rm tot}}=(1/\sigma_g){\rm ln}(\alpha G)$ (Equation~[\ref{N1high}]).
In this regime the atomic column is weakly dependent on $I_{\rm UV}/n$, and is limited to values close to $1/\sigma_g$
due to the dominating H{\small I}-dust absorption.  For example, for $Z'=1$ and 
$\alpha G/2=0.01$, 0.1, 1, 10, and 100, our model computations give
$N_{1,{\rm tot}}=5.6\times 10^{18}$, $5.3\times 10^{19}$, $3.8\times 10^{20}$,
$1.2\times 10^{21}$, and $2.4\times 10^{21}$~cm$^{-2}$, as also given
by our analytic formula.   

Expressed as a function of $\alpha G$, $N_{1,{\rm tot}}\propto 1/Z'$. 
However, we recall that 
for large-$\sigma_g$ ($Z'\gtrsim 1$) $\alpha G$ is itself
metallicity dependent and $\alpha G\propto Z'^{1/2}$ (see Equation~[\ref{aGfin}]).  
Thus, for high metallicities in the weak-field limit $N_{1,{\rm tot}}\propto Z'^{-3/2}$.
In this regime the column decreases by one power of $Z'$ due to the
enhanced H$_2$ formation efficiency, and by an additional factor of $Z^{1/2}$
due the reduction of the effective dissociation flux by the H$_2$-dust.
For low metallicities in the weak-field limit the dissociation flux is maximal
and $N_{1,{\rm tot}}\propto Z'^{-1}$.  In the strong field limit
the atomic column is only weakly (logarithmically) dependent on $\alpha G$,
and the metallicity dependence is mainly via the H{\small I}-dust opacity,
and $N_{1,{\rm tot}}\propto Z'^{-1}$ for all metallicities.
These metallicity scalings, discussed in \S~2.2.8, are validated by our
numerical model results for the atomic column.

In Figure~\ref{Fig:N1_aG} we also plot a (red) curve for $(\alpha G)_{\rm CNM}(Z')$
as given by Equation~(\ref{aGcnm}) for self-regulated multi-phased H{\small I}. 
As is again illustrated in Figure~\ref{Fig:N1_aG}, $(\alpha G)_{\rm CNM}(Z')/2\sim 1$.
Given the assumptions underlying Equation~(\ref{aGcnm}), the H{\small I} columns for
self-regulated multi-phased gas are intermediate between the linear (weak-field) and logarithmic
(strong-field) regimes for all metallicities.

Our computational results show that our analytic expression for $N_{1,{\rm tot}}$ is
valid for {\it all} transition profile shapes, whether gradual (weak fields) or sharp (strong fields), and
no matter where the atomic column is built up, whether in the molecular zone or in the
outer fully atomic layer.
Indeed, in our analytic derivation we did not make any assumptions on the profile shape.
Our expression is universal and valid for all regimes, weak- and strong-fields, high-
and low-metallicities, and for all transition profile shapes, gradual and sharp.
We emphasize this point again in 
our alternate derivation via the 
KMT/MK10 ``transfer-dissociation-equation" (\S~4.1.1).

In Figure~\ref{Fig:N1_aG}, lower panel, we plot the total atomic columns for our isotropic-field models. 
For these models we select $D_0$ and $R$, and vary $n$ to set $\alpha$, and we compute the total H{\small I} column.  
As discussed in \S3.2, the total dust-limited dissociation bandwidth, $W_{\rm tot}$, 
and the average self-shielding factor, $G$, are identical for beamed and isotropic fields.  
In Figure~12 we plot $N_{1,{\rm tot}}$ as a function of $\alpha G$, where
for each $Z'$ we use the $G(Z')$ computed using beamed fields.  
Again, $\alpha G$ ranges from 0.01 to 100, and we present results for $Z'$ from 10 to 0.01.
The $(\alpha G)_{\rm CNM}(Z')$ curve is also displayed.
In Figure~\ref{Fig:N1_aG}, the points are our numerical results, and the curves are 
as given by our fundamental analytic formula for isotropic fields (\S~2.3)
\begin{equation}
\label{S88pisoB}
N^{\rm i}_{1,{\rm tot}}  \ = \ \frac{\langle\mu\rangle}{\sigma_g} \ {\rm ln}\bigl[\frac{1}{\langle\mu\rangle}\frac{\alpha G}{4} \ + \ 1\bigr]
\end{equation}
where again $1/\sigma_g = 5.3\times 10^{20}/Z'$~cm$^{-2}$.  
To obtain the match between the numerical results and analytic formula 
shown in Figure~\ref{Fig:N1_aG} we have fit for
the average-angle-factor ${\langle\mu\rangle}$.
We find that
\begin{equation}
{\langle\mu\rangle}\ = \ 0.8
\end{equation}
provides an excellent fit for all $\alpha G$ and for all $\sigma_g\propto Z'$.

For isotropic fields in the weak-field limit 
$N_{1,{\rm tot}}=(1/\sigma_g)\alpha G/4=(1/4){\bar F}_\nu W_{g,{\rm tot}}/Rn$
(Equation~[\ref{N1lowiso}]). As discussed in \S~2.3, this is
half the column for the corresponding beamed field due to the factor-of-two
difference in the incident dissociation fluxes. For weak-fields, $N_{1,{\rm tot}}$
is independent of ${\langle\mu\rangle}$.
In the strong-field limit, $N_{1,{\rm tot}}=({\langle\mu\rangle}/\sigma_g){\rm ln}(\alpha G/4{\langle\mu\rangle})$
(Equation~[\ref{N1highiso}]) and the column is reduced by the factor
${\langle\mu\rangle}=0.8$ compared to beamed-field models
(neglecting the small difference in the logarithms).
For example, for $Z'=1$ and $\alpha G/2=0.01$ 0.1, 1, 10, and 100, 
our model results for the total atomic column are
$3.1\times 10^{18}$, $2.9\times 10^{19}$, $2.1\times 10^{20}$, $7.7\times 10^{20}$,
and $1.6\times 10^{21}$~cm$^{-2}$,
smaller in the expected way compared to beamed-field models.
The linear and logarithmic behaviors for small and large $\alpha G \propto I_{\rm UV}/n$
and the scaling with $Z'$ are identical for isotropic and beamed fields.

Our expression for the total column for isotropic fields is also valid for all profile shapes, from
gradual to sharp.

\section{Comparison to KMT/MK10}

In a series of papers, \citet{Krumholz_2008, Krumholz_2009} and \citet{McKee_2010},
(hereafter KMT/MK10) considered the interstellar atomic-to-molecular transition in the context
of galaxy-wide conditions for star-formation.  For this purpose, KMT/MK10 developed analytic 
models for the H{\small I}-to-H$_2$ transitions in {\it spherical} clouds irradiated by isotropic radiation 
fields.  They also considered planar models. KMT/MK10 were particularly interested 
in exploring the role of metallicity in setting star-formation thresholds, and in MK10 they applied their 
spherical models to compute metallicity-dependent H$_2$ mass fractions in star-forming clouds
as functions of the gas-mass surface densities in galaxy disks. 
The integral H$_2$ mass fraction, $f_{{H_2}}$,  may be a
critical parameter in galaxy-wide Kennicutt-Schmidt (KS) relations, especially
if conversion to H$_2$ is required for star-formation\footnote{For example, in the expression
${\dot \Sigma}_{SFR} = \epsilon f_{H_2} \Sigma_{\rm gas}/\tau_{\rm dyn}$ the molecular
fraction $f_{H_2}$ enters as a threshold, and the star-formation rate is quenched
as $f_{H_2}$ becomes small. In this form of the KS relation
${\dot \Sigma}_{SFR}$ is the star-formation rate per unit mass surface density,
$\Sigma_{\rm gas}$ is the total gas mass surface density, $\tau_{\rm dyn}$ is
the dynamical time, and $\epsilon$ is the star-formation efficiency per-dynamical time.}.
For spheres, $f_{{H_2}}\equiv M_{{H_2}}/M_{gas}$,
where $M_{{H_2}}$ and $M_{gas}$ are the molecular and total gas masses
within the volumes.
For slabs $f_{{H_2}}\equiv \Sigma_{{H_2}}/\Sigma_{gas}$
where $\Sigma_{{H_2}}$ and $\Sigma_{gas}$ are the molecular and total gas mass
surface densities in the normal directions.  


The assumption of spherical versus plane-parallel geometry complicates 
the analysis and computation of the H{\small I}-to-H$_2$ transition because for spheres oblique rays 
cross through the clouds even when optically thick H$_2$ cores are present.
Optically thick slabs are simpler because all rays are absorbed.
To proceed, KMT/MK10 made the simplifying assumption that the
H{\small I}-to-H$_2$ transitions are always sharp, such that any
atomic layer has a well-defined length-scale. For the spheres, fully 
atomic and dusty H{\small I} shells are then assumed to surround fully molecular H$_2$ cores.
Rays that cross through the H{\small I} shells are
either unattenuated or are absorbed by dust, or by H$_2$-lines under the assumption
that the molecular fractions in the shells are small. 
Rays that impinge on the H$_2$ cores are fully
attenuated.  MK10 employ a ``variable Eddington factor"
formalism to close the angular moments of the radiative transfer equation for the 
radiation fields in the H{\small I} shells.  For a given total cloud mass and
radiation field intensity a complicated iterative procedure 
then yields the spherical and nested H{\small I}/H$_2$ structures, and associated H$_2$ mass fractions. 

KMT/MK10 implicitly assumed that the H$_2$ lines are always in the fully overlapping regime, 
independent of $Z'$ and $\sigma_g$, and they did not consider the possible reduction of the effective 
dissociation bandwidth by H$_2$-dust. KMT/MK10
did not make the distinction between H{\small I}-dust and H$_2$-dust that we have been emphasizing.

According to KMT/MK10, the spherical H{\small I}/H$_2$ ``shell-core" structures depend on two 
dimensionless parameters, ``$\chi$" and ``$\tau_r$".
In their definition, ``$\chi/f_1$ is the ratio of the number of LW photons
absorbed by dust to the number absorbed by H$_2$",
where $f_1$ is the fraction of gas in atomic form.
As we clarify below, $\chi$ is actually identical to our $\alpha G$ -
in the low-$Z'$, small-$\sigma_g$, ($w=1$) limit where H$_2$-dust absorption is negligible.
The second parameter, $\tau_r\equiv n\sigma_g r$, is the
``dust optical depth associated with the cloud radius $r$", where
$n$ is the hydrogen nucleon density at the cloud surface. 
For a slab (of finite width) the corresponding second 
parameter is the total dust optical depth 
$\tau_z\equiv \sigma_g nz$ through the slab, where $z$ is the linear extent.
The parameters $\tau_r$ and $\tau_z$ will enter into our
discussion of the H$_2$ mass fractions for spheres and slabs (\S 4.2).

We wish to compare our analytic theory for the H{\small I} column densities for 
planar clouds to the KMT/MK10 formulae for spheres.  This will
enable us to also compare results for the H$_2$ mass fractions.
 We will show that any differences between 
corresponding (i.e.,~properly normalized) slabs and spheres
are very small and that
it is therefore advantageous to use our much simpler and fully analytic plane-parallel formalism.

We start (in \S 4.1) by clarifying the relationship between the KMT/MK10 $\chi$ and our $\alpha G$. To do this
we reanalyze the ``transfer-dissociation equation" presented by \citet{Krumholz_2008}. This also
enables a comparison of our formulae for the H{\small I} columns
to KMT/MK10's similar, though not identical, expressions for the atomic columns
for planar geometry.  In \S 4.2.1 we develop a simple expression for $f_{{H_2}}$ 
for optically thick slabs for two-sided irradiation using our analytic expressions for the 
total H{\small I}-dust optical depths for one-sided irradiation of semi-infinite clouds, for isotropic or beamed fields.
In \S 4.2.2 we re-express the MK10 formulae for 
uniform density (isochoric) spheres and (isobaric) ``atomic-molecular complexes",
for (a) the critical H{\small I}-dust optical depths required for the formation of H$_2$ cores
and (b) the H$_2$ mass fractions as functions of the total cloud optical depths (or gas masses).
This will enable a clear comparison to our expressions for slabs.

We find that for corresponding spheres and slabs embedded in 
isotropic fields the predicted H{\small I} columns are very similar. The
differences are no greater than 20\%
for $\alpha G$ ranging from 0.01 to 100. 
For two-phased H{\small I} equilibria for which $\alpha G/2 \sim 1$, 
the differences between spheres and slabs are negligible, and the
H{\small I} columns for spheres and slabs are essentially identical. 
In fact, we show that switching from a beamed to an isotropic field for a slab is
much more significant than switching from a slab to a sphere
for a fixed isotropic field.
We then show that for
proper normalization of the three configurations - uniform sphere, complex, or slab -
the predicted H$_2$ mass fractions as functions of total cloud masses or columns are 
also very similar.

\subsection{$\chi$ and $\alpha G$.}

To clarify the relationship between the KMT/MK10 $\chi$ and our $\alpha G$, it is useful to
consider the ``transfer-dissociation equation" developed by \citet{Krumholz_2008}, here
for beamed radiation into a slab. This discussion will also
provide an alternate derivation of
our fundamental formula (Equation~[\ref{S88}]) for the total H{\small I} column density.

~\\

\subsubsection{Transfer-dissociation Equation}

Let $F_\nu(z)$ be the flux density (photons cm$^{-2}$~s$^{-1}$~Hz$^{-1}$) at frequency $\nu$ of beamed LW radiation at
linear depth $z$ into a slab. The radiation flux (cm$^{-2}$~s$^{-1}$) between frequencies $\nu_1$ to $\nu_2$ is then
\begin{equation}
F(z) \  \equiv \ \int_{{\nu_1}}^{{\nu_2}}  F_{\nu}(z) \ d\nu \ \ \ .
\end{equation}
At the cloud surface $F(0)=F_0/2$, where $F_0$ is the flux-integral (our Equation~[\ref{F0}])
for the free-space radiation field.

The transfer equation for $F(z)$ is
\begin{equation}
\label{tran1}
\frac{dF}{dz} \ = \ -n\sigma_g F
- n_2 \ \int_{{\nu_1}}^{{\nu_2}} \sigma_{\nu} F_\nu \ d\nu \ \ \ .
\end{equation} 
Here $n\equiv n_1+2n_2$ is the total (atomic plus molecular) hydrogen gas volume density, 
and $\sigma_{\nu}$ is the (``complicated") cross-section for the multi-line H$_2$ absorption process
\footnote{
The product ${\bar f}_{\rm diss}$ $\int \sigma_{\nu} \,\, d\nu$ is the total dissociation cross section
$\sigma_d^{\rm tot}$ (Equation[\ref{sigdtot}]), \S \ 2.2.1). 
}.
The first term on the right-hand side of Equation~(\ref{tran1}) is the dust absorption rate of 
the local radiative flux. The second term is the H$_2$-line absorption rate.

The steady-state H$_2$ formation-destruction equation is
\begin{equation}
\label{formdesK}
Rnn_1 \ = \ {\bar f}_{\rm diss} n_2  \int_{{\nu_1}}^{{\nu_2}} \sigma_{\nu} F_\nu(z) \ d\nu \ \ \ ,
\end{equation}
where in this expression, ${\bar f}_{\rm diss}$ (as defined in \S 2) 
is the mean fraction of all H$_2$-line absorptions that lead to 
photodissociation.  Following \citet{Krumholz_2008} and our discussion in \S~2, we assume that 
${\bar f}_{\rm diss}$ is independent of cloud depth.

The ``complicated" integral in Equation~(\ref{tran1}) may be eliminated
to give the transfer-dissociation equation \citep{Krumholz_2008}
\begin{equation}
\label{td1}
\frac{dF}{dz} \ = \ -n\sigma_g F \ - \ 
\frac{Rnn_1}{{\bar f}_{\rm diss}} \ \ \ .
\end{equation}
In dimensionless form
\begin{equation}
\label {td2}
\frac{d {\cal F}}{d\tau} \ = \ -{\cal F} \ - \ \frac{2}{\chi}f_1
\end{equation}
where ${\cal F}\equiv F/F(0)$, is the normalized depth-dependent flux, 
$\tau\equiv n\sigma_g z$ is the dust optical depth, and $f_1(\tau)\equiv n_1/n$ is the
depth-dependent fraction of gas in atomic form.

In Equation~(\ref{td2})
\begin{equation}
\label{KMTc}
\chi \equiv \ \frac{{\bar f}_{\rm diss}\sigma_g F_0}{Rn} \ \ \ .
\end{equation}
This is the KMT/MK10 $\chi$, here introduced for one-sided irradiation of a slab by a beamed field
with incident LW flux $F(0)\equiv F_0/2$
\footnote{When \citet{Krumholz_2008} defined their $\chi$ via the transfer-dissociation equation
they wrote the surface flux $F(0)$ in the numerator, rather than 
the full-$4\pi$ free-space $F_0$. However, when discussing
spherical models, KMT/MK10 {\it redefine} their $\chi$ and replace $F(0)$ with $F_0$ in the numerator.
In this paper we use the 
free-space $F_0$ throughout, including here in our discussion of the transfer-dissociation equation.
With this adjustment a factor of 2 appears in the second term on the right-hand side
of Equation~(\ref{td2}), not there in \citet{Krumholz_2008}.
}.
Comparing to our Equation~(\ref{aG2}) it is already clear that $\chi = \alpha G$, for $w=1$.

Because the product $\sigma_g nF_0$ is the dust
absorption rate per-unit-volume of the surface LW radiation flux, and because
$Rn^2f_1/{\bar f}_{\rm diss}$ is the H$_2$-line absorption rate per-unit-volume,
KMT/MK10 interpret $\chi/f_1$ as the ratio of the number of LW photons
absorbed by dust to the number absorbed by H$_2$-lines
at the optically thin cloud surface
(with one factor of $n$ canceling when doing the division).
In this interpretation, the dust absorption being referred to
in the numerator
is {\it independent of whether the gas is atomic or molecular}.

However, as we now show, if the dust absorption term in the 
transfer-dissociation equation is redefined to refer to H{\small I}-dust absorption only,
we will recover our fundamental formula for the
the H{\small I} column density, including a method for
``renormalizing" the dissociation flux to account for H$_2$-dust absorption.
We demonstrate this by analyzing the transfer-dissociation equation,
in three steps as follows.

\subsubsection{Step 1: Transition Assumed Sharp}

First, we assume - as done by KMT/MK10 - that the atomic-to-molecular
transition is {\it always} sharp. We impose this assumption even though 
we know that it is good only if our
$\alpha G \gg 1$, i.e., only if $F_0/n$ and $\chi$ are sufficiently large.
Nevertheless, in seeking a solution for Equation~(\ref{td2}), 
we assume that for any $\chi$ the atomic fraction $f_1=1$ everywhere ${\cal F}$ is non-zero, 
up to the (unknown) transition point where ${\cal F}$ vanishes. At that point
the gas switches suddenly from H{\small I} to H$_2$.  

Setting $f_1=1$, the solution to Equation~(\ref{td2}) is
\begin{equation}
\label{tdsol}
{\cal F}(\tau) \ = \
\frac{\chi + 2}{\chi}{\rm e}^{-\tau} \ - \ \frac{2}{\chi} \ \ \ .
\end{equation}
It is evident that the flux vanishes for $\tau=\tau_{1,{\rm tot}}$ where
\begin{equation}
\label{tau1}
\tau_{1,{\rm tot}} \equiv {\rm ln}[\frac{\chi}{2} + 1] \ \ \ ,
\end{equation}
and this is then the solution for the transition point \citep{Krumholz_2008}.
The total atomic column density up to the transition point is
\begin{equation}
\label{Ntau1}
N_{1,{\rm tot}} = \frac{1}{\sigma_g}{\rm ln}[\frac{\chi}{2}+1] \ \ \ ,
\end{equation}
because by assumption there is no H$_2$ in the photodissociated layer and
$\tau_1$ is the optical depth due to H{\small I}-dust only.

We immediately recognize Equations~(\ref{tau1}) and
(\ref{Ntau1}). They are identical to our expressions~(\ref{S88}) and (\ref{S88tau}) 
for the total H{\small I}-dust optical depth and H{\small I} column density, except that here the 
argument is $\chi$, as given by Equation~(\ref{KMTc}), rather than our $\alpha G$ as defined by 
the similar but not quite identical Equation~(\ref{aG2}).  Indeed, it is apparent that $\alpha G=w\chi$,
where $w$ is the normalized effective dissociation bandwidth as defined by Equation~(\ref{pdiss}).
(This identity is clarified in {\it Step 3}.)

For $\chi \gg 1$,  the attenuation of the LW flux is dominated by dust absorption, and 
${\cal F}(\tau)\approx{\rm e}^{-\tau}$.  
An exponentially attenuating flux is precisely what produces a sharp transition profile. Therefore,
for $\chi \gg 1$ the assumption that the transition is sharp is consistent. Furthermore, 
the exponential attenuation is due specifically to H{\small I}-dust absorption.
Sharp profiles occur when H{\small I}-dust dominates the attenuation.

For $\chi \ll 1$,  dust absorption is negligible compared to H$_2$-lines, 
$\tau \ll 1$ throughout the atomic layer, and
${\cal F}(\tau )=1-2\tau/\chi$ to first-order in $\tau$.
For small $\chi$, Equation~(\ref{tau1}) gives $\tau_{1,{\rm tot}}=\chi/2$, so that
\begin{equation}
\label{Ntau2}
N_{1,{\rm tot}} \ = \ \frac{1}{\sigma_g}\frac{\chi}{2}
\end{equation}
in this limit.  However, when H$_2$-lines dominate the attenuation the
transition is not sharp, and the assumption that $f_1=1$ is inconsistent.

Nevertheless, as we now show in {\it Step 2}, the above expressions
for  ${\cal F}(\tau)$, $\tau_1$, and $N_1^{\rm tot}$ for large and small $\chi$
are valid for 
{\it any} varying $f_1$, {\it whether or not} the H{\small I}/H$_2$
transition is sharp, provided that H$_2$-dust is always negligible
compared to H{\small I} dust or H$_2$-lines, so that
$\tau$ in Equation~(\ref{td2}) can be redefined as the optical depth associated 
with H{\small I}-dust absorption only\footnote{
Indeed, we know that our Equations~(\ref{S88}) 
and (\ref{S88tau}) are valid for all $\alpha G$, {\it whether or not} the
H{\small I}-to-H$_2$ transition is sharp.  This suggests that although
we assumed $f_1=1$ to derive
Equations~(\ref{tau1}) and (\ref{Ntau1}), they are valid
even when $\chi$ is small and 
the transition is not sharp.
}.

\subsubsection{Step 2: H$_2$-Dust Negligible, and $\tau$ Redefined}

With the neglect of H$_2$-dust absorption, Equation~(\ref{td1}) is
\begin{equation}
\label{td1b}
\frac{dF}{dz} \ = \ -n_1\sigma_g F \ - \ \frac{Rnn_1}{{\bar f}_{\rm diss}} 
\end{equation}
where now $n$ has been replaced by $n_1$ in the first term on the right-hand side.
Or, 
\begin{equation}
\label{td3}
\frac{d{\cal F}}{d\tau} \ = \ -{\cal F} \ - \ \frac{2}{\chi}
\end{equation}
where $\chi$ is again given by Equation~(\ref{KMTc}), but where now
\begin{equation}
\tau \equiv \sigma_g \ \int_0^z n_1 \ dz \ \ \ 
\end{equation}
is defined in advance as the optical depth associated with H{\small I}-dust only.

In Equations~(\ref{td1b}) and (\ref{td3}) a factor $f_1\equiv n_1/n$ does not appear.
The solutions are still given by Equations~(\ref{Ntau1}) and (\ref{Ntau2})
for large and small $\chi$, but now with {\it no} assumptions on the shapes of the transition profiles.

For large $\chi$ the transition is sharp because the radiation flux is attenuated
exponentially by the H{\small I}-dust, as shown in {\it Step 1}. However, for low $\chi$,
for which $\tau_{1,{\rm tot}}$ is small, the transition need not be sharp.
In this limit we again have,
\begin{equation}
\label{str1}
N_{1,{\rm tot}} \ = \ \frac{1}{\sigma_g}\frac{\chi}{2} \ = \ \frac{{\bar f}_{\rm diss}F_0/2}{Rn} \ \ \ .
\end{equation}
But this is just a Str{\" o}mgren relation (as also noted by MK10). Therefore, it must
hold {\it independent} of the shape of the H{\small I}/H$_2$ profile provided only that
the entire incident dissociation flux, ${\bar f}_{\rm diss}F_0/2$, is fully absorbed by H$_2$-lines, which it 
indeed will be for small $\tau_{1,{\rm tot}}$. Again, $RnN_{1,{\rm tot}}$ is the integrated H$_2$ formation
rate per unit area, and in steady-state this equals the fully absorbed dissociation flux ${\bar f}_{\rm diss}F_0/2$
impinging on the surface, independent of profile shape.  In fact, we know that for low $\chi$
(i.e., for small $F_0/n$ and $\alpha G$) the transition profile is gradual,
with most of the H{\small I} column built up in the molecular zone, as we have
shown and discussed in \S 3.

Although \citet{Krumholz_2008} derived their expression~(\ref{tau1}) for the dust optical depth 
assuming a sharp transition, it is actually valid for all $\chi$, large and small, 
for sharp {\it and} gradual atomic-hydrogen profiles, provided
H$_2$-dust absorption is negligible compared to H{\small I}-dust and H$_2$-lines.

\subsubsection{Step 3: H$_2$-Dust and Renormalization of the Dissociation Flux}

We now include H$_2$-dust.
In Equations~(\ref{td1}) and (\ref{td2}), the original
dust absorption term, $n\sigma_g F$, is not necessarily negligible - even when $\chi \ll 1$ - if 
the atomic fraction $f_1$ is sufficiently small.  This is the regime where H$_2$-dust absorption may be
significant, even if H{\small I}-dust is negligible.

Due to H$_2$-dust absorption of radiation ``between the lines", not all of the photons in the 
LW-band are absorbed by the H$_2$, even for a fully molecular slab.
The effective dissociation flux is then smaller than ${\bar f}_{\rm diss} F_0/2$.
The Str{\" o}mgren relation then implies that for $\chi \ll 1$ the total H{\small I}-dust optical depth
$\tau_{1,{\rm tot}}$ must be {\it smaller} than $\chi/2$, rather than equal to $\chi/2$ 
as given by Equation~(\ref{str1}).

If $\tau$ is redefined to refer to H{\small I}-dust only as in {\it Step 2}, the effect of
H$_2$-dust absorption can be included in the transfer-dissociation equation
 by a simple ``renormalization" of the effective dissociation flux.
We replace $F_0/2$ with a suitably
reduced $wF_0/2$ equal to the flux of LW photons
absorbed in H$_2$-lines in a dusty molecular slab, {\it excluding in advance} photons
that are inevitably absorbed by H$_2$-dust. The renormalization factor
can be computed in advance for any given $\sigma_g$, and we immediately
recognize $w$ as 
the (normalized) effective dissociation bandwidth that we 
defined and computed in \S~2 and \S~3.


The transfer-dissociation equation for the renormalized LW flux is
\begin{equation}
\label{td1c}
\frac{d(wF)}{dz} \ = \ -n_1\sigma_g wF - \frac{Rnn_1}{{\bar f}_{\rm diss}}
\end{equation}
where again the dust term on the right-hand-side refers to H{\small I}-dust absorption only. 
Photons absorbed by H$_2$-dust anywhere in the slab are excluded from consideration ``in advance".

In dimensionless form
\begin{equation}
\label{rtd2}
\frac{d{\cal F}}{d\tau} \  = \ -{\cal F} - \frac{2}{\chi'}
\end{equation}
where ${\cal F}\equiv F/F(0)$,
$d\tau \equiv  n_1 \sigma_g z$,
and
\begin{equation}
\chi' \ \equiv w\chi \ = \ \frac{{\bar f}_{\rm diss} \sigma_g wF_0}{Rn} \ \ \ .
\end{equation}
We see that 
$\chi'$ is just our $\alpha G$, (Equation~[\ref{aG2}])
Thus, 
\begin{equation}
\alpha G =w\chi
\end{equation}
as already indicated in {\it Step 1}.

Most importantly, in Equations~(\ref{td1c}) and (\ref{rtd2}) a factor ``$f_1$" 
does not appear at all. The solution is
\begin{equation}
\label{closed}
{\cal F}(\tau) \ = \ \frac{\chi' + 2}{\chi'}{\rm e}^{-\tau} \ - \ \frac{2}{\chi'}
\end{equation}
where $\tau$ is the H{\small I}-dust optical depth, and with {\it no} assumptions
on the shape of the H{\small I}-to-H$_2$ transition profile. The flux vanishes for 
$\tau$=$\tau_1$, where 
\begin{equation}
\label{s88pagain}
\tau_{1, {\rm tot}} \ = \ {\rm ln}[\frac{\chi'}{2}+1] \ = \ {\rm ln}[\frac{\alpha G}{2} + 1] \ \ \ ,
\end{equation}
and the expression for the total atomic column is then
\begin{equation}
\label{N1totagain}
N_{1, {\rm tot}} \ = \ \frac{1}{\sigma_g}{\rm ln}[\frac{\alpha G}{2} + 1] \ \ \ .
\end{equation}
We have thus recovered our original expressions~(\ref{S88p}) and (\ref{S88}) for the
total atomic column and H{\small I}-dust opacity for beamed radiation into a slab. It is again clear that
these expressions are general. They are valid for  
large or small $\alpha G$, (strong- or weak-fields), for the regimes of 
significant H{\small I}-dust and/or H$_2$-dust absorption, and are 
independent of the profile shapes, gradual or sharp.
We also have in Equation~(\ref{closed}) a closed-form expression for the 
depth-dependent LW-band flux as a function of the H{\small I}-dust optical depth.

We again see that $\alpha G$ is the fundamental dimensionless parameter.
Physically, $\alpha G$ is
the ratio of the H{\small I}-dust to H$_2$-line absorption rates
of the effective dissociation flux,
excluding any LW photons that are absorbed by H$_2$-dust.
The relationship between $\chi$ 
and our $\alpha G$ 
is now clear. Specifically, $w\chi \equiv \alpha G$, where, crucially,
our factor $w$ accounts for H$_2$-dust absorption,
and the resulting reduction of the effective dissociation flux.
In the low-$Z'$, small-$\sigma_g$ limit, H$_2$-dust is negligible, $w=1$,
and $\chi=\alpha G$.
For high-$Z'$ and large-$\sigma_g$, $w < 1$ 
(decreasing as $\sigma_g^{-1/2}$ or as $Z'^{-1/2}$),  and then
$\alpha G < \chi$.
KMT/MK10 did not make the distinction between 
H{\small I}-dust and H$_2$-dust, and ignored the effects of H$_2$-dust entirely
in their definition of $\chi$.
As we have discussed in \S~2 and 3, H$_2$-dust can reduce the
effective dissociation bandwidth by a factor $\sim 4$ for the 
(realistic) range of metallicities in galaxies. 

In Figure~\ref{Fig:Schema} we again schematically summarize the four regimes
that are incorporated by our formula for the total H{\small I} column density.  In each quadrant,
large or small $\alpha G$, for large or small $Z'$, we indicate the shape
of the H{\small I}-to-H$_2$ transition (sharp or gradual), the dominant
source of opacity (H{\small I}-dust, H$_2$-lines, or H$_2$-lines plus H$_2$-dust),
and the expression for the H{\small I}-dust opacity in terms of the
parameters $\sigma_g$, $F_0$, $Rn$, and $w$.

\begin{figure}[h!]
\centerline{\includegraphics[width=1.0 \columnwidth]{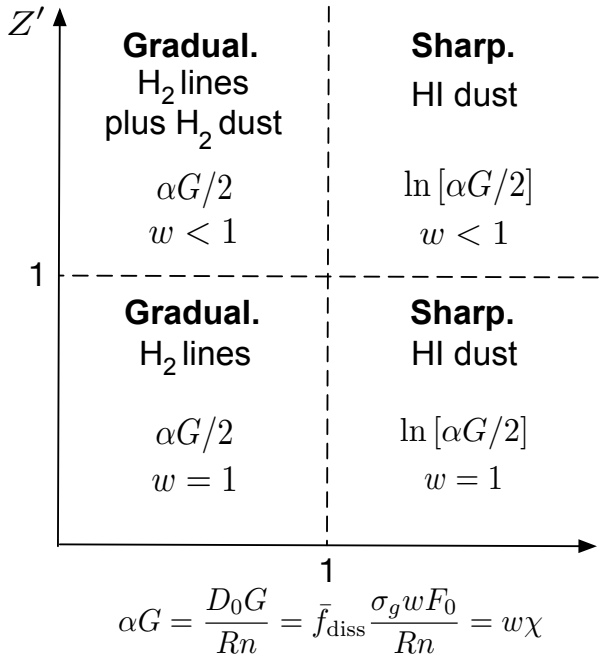}}
\caption{Dominant sources of FUV absorption (H{\small I}-dust, H$_2$-lines and H$_2$-dust),
H{\small I}-to-H$_2$ transition profile shapes (gradual or sharp), and total H{\small I}-dust 
opacities (large or small), in the $\alpha G= D_0 G / (Rn) = {\bar f}_{\rm diss} \sigma_g wF_0 / (Rn) \equiv w\chi$ versus $Z'$ plane.}
\label{Fig:Schema}
\end{figure}

\subsection{Critical H{\small I}-dust Opacities and H$_2$ Mass Fractions: Slabs Versus Spheres}

\subsubsection{Slabs}

Given our expressions (Equations~[\ref{S88}] or [\ref{S88iso}]) for the total H{\small I}-dust 
optical depths in semi-infinite slabs we can write down a simple formula
for the integrated H$_2$ mass fraction, $f_{{H_2}}$, for planar geometry, for a comparison to 
the KMT/MK10 results for spheres.  Our main focus is a comparison of slabs 
and spheres irradiated by isotropic fields, but we also consider beamed radiation for slabs.

For an isotropic field with a given $I_{\rm UV}$ a uniformly illuminated sphere
corresponds to {\it two-sided} irradiation of a slab of finite width, with properly normalized
total optical depths for the sphere and slab.  
 For a slab with total dust thickness
$\tau_z\equiv\sigma_gnz= \sigma_gN$ (where $N$ is the total gas column density in the normal direction)
the H$_2$ mass fraction is simply
\begin{equation}
\label{fH2slab}
f_{{H_2}}^{\rm p} \ = \ 1 - \frac{1}{y} \ \ \ ,
\end{equation}
where $y\equiv \tau_z/\tau_1$, and $\tau_1$ is the H{\small I}-dust depth summed over
both sides of the slab (the superscript ``p" indicates
plane-parallel slab).  By definition, $y \ge 1$. 
We plot $f_{{H_2}}^{\rm p}$ versus $y$ in Figure~\ref{Fig:f_y}.

\begin{figure}[h!]
\centerline{\includegraphics[width=1.0 \columnwidth]{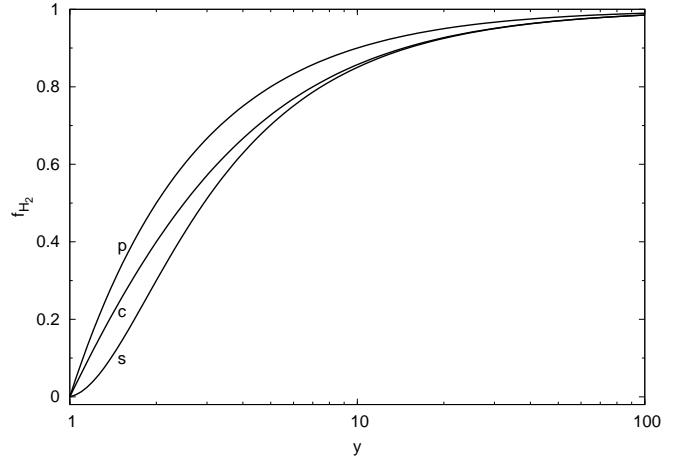}}
\caption{H$_2$ mass fractions as given by Equations~(\ref{fH2slab}), (\ref{fH2sphere}),
and (\ref{fH2complex}), for plane-parallel slabs (``p"), uniform-density spheres (``s"),
and complexes (``c"), as functions of the normalized H{\small I}-dust optical depth $y$.}
\label{Fig:f_y}
\end{figure}

For {\it optically thick} slabs in which the radiation incident on both sides is fully absorbed,
and for illumination by isotropic fields, $\tau_1=\tau_1^{\rm p}$ where
\begin{equation}
\label{t1slab}
\tau_1^{\rm p} \ = \ 2\langle \mu \rangle \ {\rm ln}\bigl[\frac{1}{\langle\mu \rangle}\frac{\alpha G}{4} + 1\bigr] 
\ = \ 1.6 \ {\rm ln}\bigl[\frac{\alpha G}{3.2} + 1\bigr]   \ \ \ .
\end{equation}
This is twice the optical depth given by our Equation~(\ref{S88piso}) for one-sided illumination of a
semi-infinite slab. (We have set $\langle \mu \rangle=0.8$, as found in \S~3.)
For beamed radiation $\tau_1=\tau_1^{\rm p,b}$ where
\begin{equation}
\label{t2slab}
\tau_1^{\rm p,b} \ = \ 2\ {\rm ln}\bigl[\frac{\alpha G}{2} + 1\bigr] \ \ \ ,
\end{equation}
given our Equation~(\ref{S88p}) for one-sided illumination
(the superscript ``b"  is for beamed radiation).
We plot $\tau_1^{\rm p}$ and 
$\tau_1^{\rm p,b}$ versus $\alpha G$ in Figure~\ref{Fig:tau1_aG}.

For two-sided illumination of optically thick slabs, Equation~(\ref{fH2slab}) for 
$f_{{H_2}}$ together with Equations~(\ref{t1slab}) or~(\ref{t2slab}) for the
total H{\small I}-dust optical depths may be used to compute the integral
H$_2$ mass fractions whether the H{\small I}-to-H$_2$ transitions are gradual or sharp.
For $\alpha G \gg 1$, an optically thick slab consists
of a simple H{\small I}-H$_2$-H{\small I} sandwich structure, with two fully atomic layers 
outside an inner H$_2$ zone,
with sharp transitions between the atomic and molecular layers.  For sharp transitions
a ``critical" optically thick slab occurs for $y=1$ for which an H$_2$ layer just appears
at the midplane, and $\tau_1^{\rm p}$ or $\tau_1^{\rm p,b}$ are then the critical H{\small I}-dust
optical depths.  The planar sandwich for sharp transitions corresponds to the spherical core-shell structures
considered by KMT/MK10, as illustrated in Figure~\ref{Fig:SphereBeam}.

\begin{figure}[h!]
\centerline{\includegraphics[width=1.0 \columnwidth]{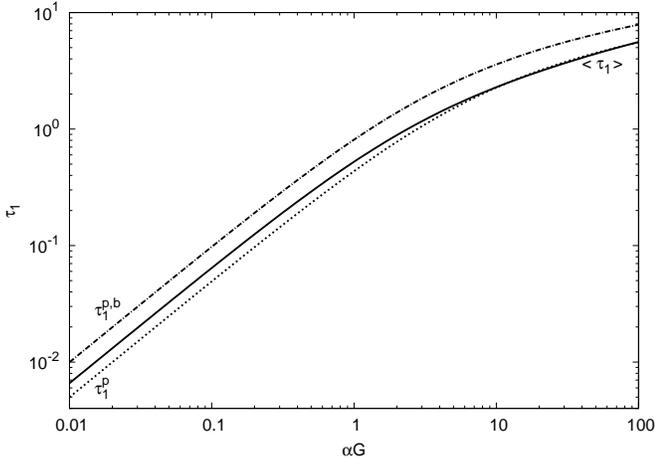}}
\caption{Critical (total) H{\small I}-dust optical depths; $\tau_1^{\rm p,b}$ for beamed fields into a slab (dot-dashed),
$\tau_1^{\rm p}$ for isotropic fields into a slab (dotted), and $\langle \tau_1 \rangle$ 
for critical spheres in isotropic fields (solid), as functions of $\alpha G$, as given by 
Equations~(\ref{t2slab}), (\ref{t1slab}), and (\ref{t1sphere}).}
\label{Fig:tau1_aG}
\end{figure}

\begin{figure*}
\centerline{\includegraphics[width=0.9 \textwidth]{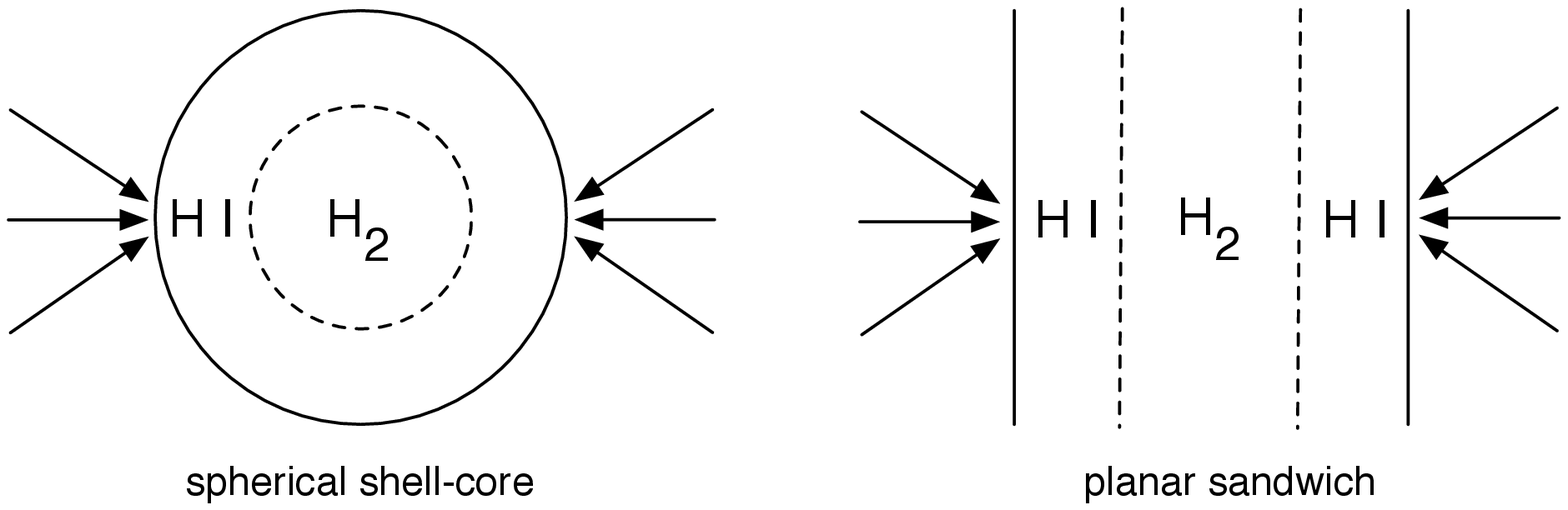}}
\caption{Corresponding shell-core structures for spheres, and sandwich structure for
slabs, for sharp H{\small I}-to-H$_2$ transitions, and illumination by isotropic radiation.}
\label{Fig:SphereBeam}
\end{figure*}

\subsubsection{Spheres}

MK10 carried out their iterative radiative transfer procedure to compute spherical
 H{\small I}-shell H$_2$-core structures for two types of systems. First, they
 considered ``uniform density" spheres for which the density of
 hydrogen nuclei, $n=n_1+2n_2$, is constant through the spheres
 as in our slabs. Second, they considered
 ``atomic-molecular complexes" in which
 the H$_2$ gas density, $n_2$, is ten times the atomic density, $n_1$, 
 for approximate pressure equilibrium between the shells and the cores.
MK10 then presented fitting formulae (their Equations [82] and [93]) for 
$f_{{H_2}}$ in uniform density spheres and complexes,
as functions of their cloud parameters $\chi$ and $\tau_r$. 

We can re-express the MK10 formulae for $f_{{H_2}}$ as simple functions of $y$,
similar to our Equation (\ref{fH2slab}) for slabs. For spheres
$y\equiv\langle \tau \rangle /  \langle \tau_1 \rangle$, where
$\langle \tau \rangle \equiv \sigma_g \langle N \rangle$ is the area-averaged
total dust column density of the sphere,
where $\langle N \rangle~\equiv~M_{gas}/m\pi r^2$ is the average
gas column density, $r$ is the cloud radius, and $m$ is the
mean particle mass per hydrogen nucleon.  
For a uniform density sphere, the total optical depth
$\langle \tau \rangle = \langle \tau_r \rangle \equiv (4/3)\sigma_gnr$.
For a complex, $\langle \tau \rangle \ge (4/3)\sigma_gn_1r\equiv \langle \tau_r \rangle$,
where $n_1$ is the gas density in the atomic shell. The parameter $y$ is defined such that
the total optical depth $\langle \tau \rangle$ is normalized relative
to  the 
critical (area-averaged) dust depth $\langle \tau_1 \rangle$ required for the appearance of an H$_2$ core at $r=0$.
The critical depths $\langle \tau_1 \rangle$
are auxiliary quantities also computed by MK10 for the uniform density spheres and complexes,
assuming irradiation by isotropic radiation fields.  The $\langle \tau_1 \rangle$ for spheres correspond
to our $\tau_1^{\rm p}$ for two-sided illumination of
slabs by isotropic fields. In our terminology, these are the (total) {\it H{\small I}-dust} optical depths 
in spheres and slabs.  For sharp transitions these are critical optical depths.

The MK10 fitting formula (their Eq.~[82]) for the H$_2$ mass fraction in
uniform density spheres may be rewritten as\footnote{
This follows from MK10 Eq.~(82)
for $x_{{H_2}}^3\equiv f_{{H_2}}$, where $x_{{H_2}}$ is the scaled radius of the
H$_2$ core in a uniform density cloud. 
For a critical cloud $x_{{H_2}}=0$, and this occurs when their
$\xi_d=1.944$, where
$\xi_d~\equiv~2.22 {\rm ln}[1 + 0.6\chi + 0.01\chi^2]/\langle \tau_r \rangle$.
This gives our expression (\ref{t1sphere}) for the critical depth 
$\langle \tau_1 \rangle$
where we have replaced $\chi$ with our $\alpha G$.
Our expression (\ref{fH2sphere}) is then MK10 Eq.~(82),
where $y= 2 / \xi_d$.
}
\begin{equation}
\label{fH2sphere}
f_{{H_2}}^{\rm s} \ \simeq \ 1 \ - \ \frac{1.5}{y + 0.5y^{-1.8}} \ \ \ 
\end{equation}
(where the superscript ``s" is for spheres).
Their critical H{\small I}-dust optical depth is
\begin{equation}
\label{t1sphere}
\langle \tau_1 \rangle \ = \ 
1.1\times {\rm ln}[1 + 0.6\alpha G + 0.01(\alpha G)^2] \ \ \  . 
\end{equation}
Again, in Equation~(\ref{fH2sphere}) $y\equiv\langle \tau \rangle /  \langle \tau_1 \rangle$.
In Figures~\ref{Fig:f_y} and \ref{Fig:tau1_aG}
we plot $f_{{H_2}}^{\rm s}$ versus $y$,
and $\langle \tau_1 \rangle$ versus $\alpha G$, 
for comparisons to our expressions for slabs.

For complexes, the MK10 fit formula for the H$_2$ mass fraction (their Eq.~[93]) 
may be rewritten as\footnote{
For MK10 Eq.~(93) for the
complexes their $f_{{H_2}}(s)$ just vanishes for the critical value
$s=2$, where $s\equiv 2.22 {\rm ln}[1 + 0.6\chi + 0.01\chi^2]/\langle \tau_c \rangle$
and $\langle \tau_c \rangle$ is the total area-averaged dust optical depth of the complex
(MK10, Eq.~[91]).  This again gives our expression (\ref{t1sphere}) for the critical depth 
$\langle \tau_1^{\rm c} \rangle$.
Our expression (\ref{fH2complex})
is then MK10 Eq.~(93), where their $s\equiv 2/y$.
}
\begin{equation}
\label{fH2complex}
f_{{H_2}}^{\rm c} \ \simeq \ 1 - \frac{1.5}{y + 0.5} \ \ \ .
\end{equation}
(The superscript ``c" is for complexes.)  
Fully atomic complexes and uniform density spheres are identical, so
the critical H{\small I}-dust depth $\langle \tau_1 \rangle$
for complexes is also given by Equation~(\ref{t1sphere}).
Due to the compression of the molecular gas the functional form for $f_{{H_2}}$
is altered compared to uniform density spheres, and the $y^{-1.8}$ factor
in the denominator of Equation~(\ref{fH2sphere}) is replaced by unity
in Equation~(\ref{fH2complex}).
We also plot $f_{{H_2}}^{\rm c}$
for the complexes in Figure~\ref{Fig:f_y}.

Figure~\ref{Fig:tau1_aG} shows the similarity in the
critical H{\small I}-dust optical depths
for slabs and spheres. For isotropic fields, and for
$\alpha G$ ranging from 0.01 to 100, the differences are no greater than 25\%. 
Furthermore, 
Figure~\ref{Fig:tau1_aG} shows that 
switching from isotropic to corresponding beamed radiation for a slab is in fact much 
more significant than switching from a slab to a sphere for an isotropic field.

For isotropic fields, the H{\small I}-dust optical depths for slabs and complexes are equal
for $\alpha G\approx 12$, so for two-phased HI equilibrium ($\alpha G \sim 2$) the differences
between the HI-dust optical depths for spheres and slabs are negligible.
Any further differences between spheres and slabs arise only because of 
remaining small differences in the functional forms for $f_{{H_2}}$.

Figure~\ref{Fig:f_y} shows the similarity and small differences in $f_{{H_2}}$
for slabs, uniform-density spheres, and complexes, as given by Equations~(\ref{fH2slab}),
(\ref{fH2sphere}), and (\ref{fH2complex}).
For slabs, our formula~(\ref{fH2slab}) for $f_{{H_2}}$ is unaltered if the
molecular gas is assumed to be denser than the atomic gas, and
there is no distinction between uniform-density and isobaric conditions.
The differences are all small. For example, 
for a spherical complex compared to a slab, the percentage difference in 
$f_{{H_2}}(y)$ is at most 40\% at $y=3$.  

\subsubsection{H$_2$ Mass Fractions and Star-Formation Thresholds in Self-Regulated Gas}

If H$_2$ is a requirement for star-formation we may define the cloud gas column 
 at which $f_{{H_2}}=0.5$ 
 as the ``star-formation threshold".  
 This is a plausible
 definition for sharp H{\small I}-to-H$_2$ transitions for which a ``sterile" atomic layer is well defined.
 As given by Equations~(\ref{fH2slab}), (\ref{fH2sphere}), and (\ref{fH2complex}), 
 $f_{{H_2}}=0.5$  for
 $y=2$, 2.93, and 2.5, for slabs, uniform density spheres, and complexes.
 We define $\Sigma_{gas,*}$ as the threshold gas mass surface density for which $f_{{H_2}}=0.5$.
 
For estimates of {\it metallicity-dependent} H$_2$ mass fractions and star-formation thresholds in 
Kennicutt-Schmidt relations for galaxy disks we adopt the KMT/MK10 ``self-regulation" ansatz that the
H{\small I} in star-forming clouds is typically driven to the CNM densities/pressures required for two-phase 
equilibria as set by the stellar FUV radiation fields.  Thus, for any $Z'$ we assume that 
$\alpha G=(\alpha G)_{\rm CNM}(Z')$ (Equation~[\ref{aGcnm}]).  Because 
$(\alpha G)_{\rm CNM} \sim 1$~to~2 for all $Z'$ the H{\small I}-to-H$_2$ transitions
are sharp to a good approximation (as argued by KMT/MK10).
For our slabs we therefore assume H{\small I}-H$_2$-H{\small I} sandwich structures, for
which critical optically thick slabs occur at $y=1$ (as discussed in \S4.2.1).


We present four sets of computations to compare results for spheres versus slabs.

First, in Figure~\ref{Fig:Comparison_McKee}, panel (a), we use Equations~(\ref{fH2complex}) and (\ref{t1sphere})
to reproduce the MK10 results for $f_{{H_2}}^{\rm c}(\Sigma_{gas})$
for spherical atomic-molecular-complexes (Fig.~5 in MK10).
Here, $\Sigma_{gas}\equiv (m/\sigma_g) \langle \tau \rangle$ is
the area-averaged gas mass surface density where $m$ is the mean particle mass per hydrogen nucleus
and $\langle \tau \rangle$ is the area-averaged dust opacity.
Thus, in Equation~(\ref{fH2complex})
$y=\Sigma_{gas}/\Sigma_1$ where $\Sigma_1\equiv (m/\sigma_g) \langle \tau_1 \rangle$,
and where $\langle\tau_1\rangle$ is given by Equation~(\ref{t1sphere}) for each $(\alpha G)_{\rm CNM}(Z')$.
We set $m=2.34\times 10^{-24}$~g as appropriate for a cosmic hydrogen-helium mixture.
KMT/MK10 assume that the dust cross section scales with metallicity as
 $\sigma_g=1.0\times 10^{-21}Z'$~cm$^2$, so we
set our $\phi_g=1/1.9$ (see Equation~[\ref{Sstand}]). 
MK10 also implicitly assume that H$_2$-dust absorption is negligible for all $Z'$ so we 
exclude our H$_2$-dust term, $(2.64\phi_gZ')^{1/2}$, in the denominator of Equation~(\ref{aGcnm}).  
The resulting curves for $f_{{H_2}}$ as functions of $\Sigma_{gas}$ (M$_\odot$~pc$^{-2}$) 
are displayed in Figure~\ref{Fig:Comparison_McKee} for $Z'$ ranging from 0.01 to 10.
They are a precise reproduction of the MK10 results (their Fig.~5).  
For example, with the above assumptions, $Z'=1$ gives $(\alpha G)_{\rm CNM}=3.6$
so that $\langle \tau_1 \rangle=1.3$, and
a molecular core appears ($y=1$) for $\Sigma_{gas}=14.6$~$M_\odot$~pc$^{-2}$.
The $f_{{H_2}}^{\rm c}=0.5$ star-formation threshold ($y=2.5$) is then
$\Sigma_{gas,*}=36.5$~$M_\odot$~pc$^{-2}$.  For smaller (larger) $Z'$ the
curves shift to the right (left) exactly as in Fig.~5 of MK10.

\begin{figure*}
\centerline{\includegraphics[width=0.9 \textwidth]{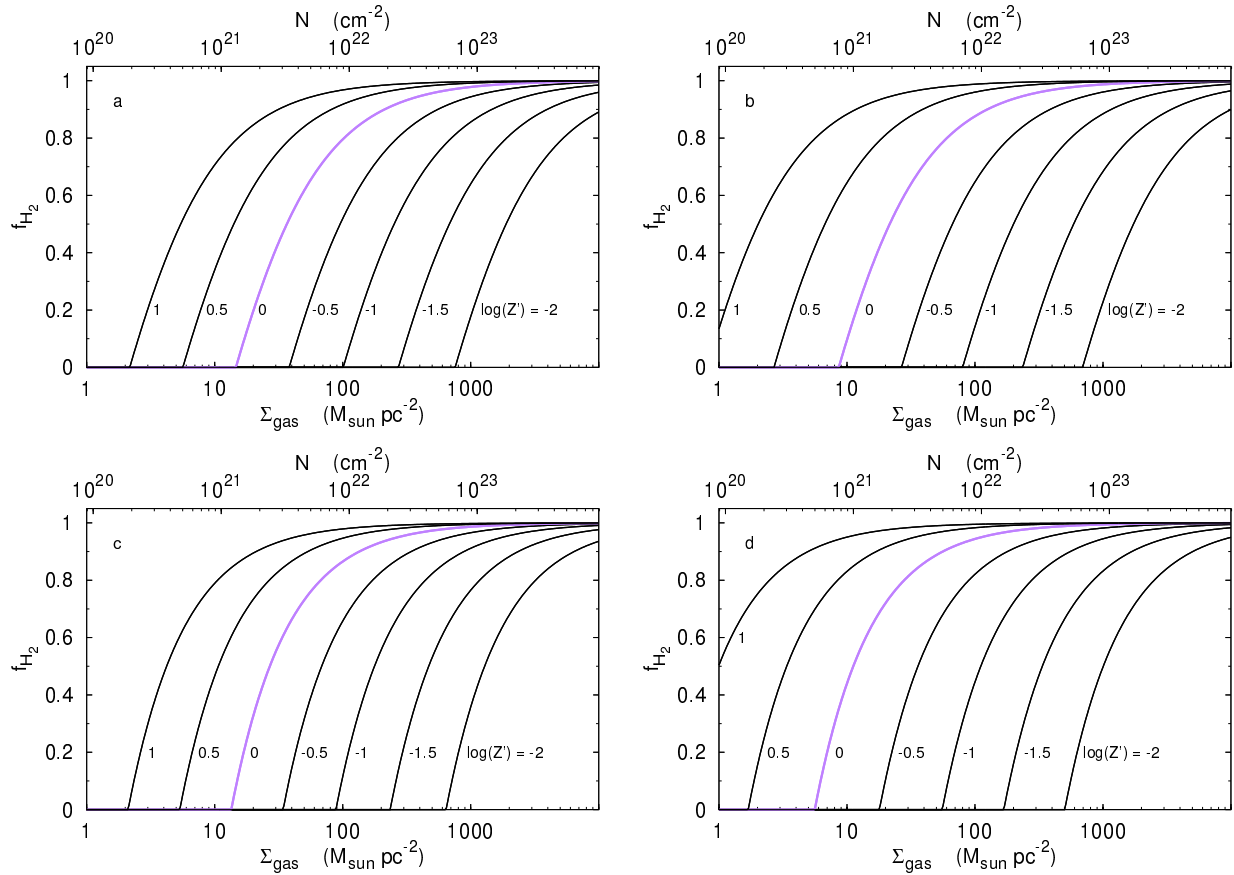}}
\caption{H$_2$ mass fractions as functions of total gas-mass surface density
for metallicities $Z'$ from 0.01 to 10 times solar. Panel (a):  Spherical complex,
with $\sigma_g=1.0\times 10^{-21}Z'$~cm$^2$, and excluding  H$_2$-dust absorption in
setting $(\alpha G)_{\rm CNM}$, as assumed by MK10. Panel (b): Spherical complex
with $\sigma_g=1.0\times 10^{-21}Z'$~cm$^2$, but including H$_2$-dust.
Panel ( c): Plane-parallel slab, with the MK10 relation $\sigma_g=1.0\times 10^{-21}Z'$~cm$^2$,
and excluding H$_2$ dust. 
Panel (d): Slab with $\sigma_g=1.9\times 10^{-21}Z'$~cm$^2$ and including H$_2$-dust.}
\label{Fig:Comparison_McKee}
\end{figure*}

Second, in Figure~\ref{Fig:Comparison_McKee}, panel (b), we show results for spherical complexes, again with
$\phi_g=1/1.9$, but now {\it with} the H$_2$-dust
term included in the estimate for $(\alpha G)_{\rm CNM}(Z')$.   As expected, for very low $Z'$ the $f_{{H_2}}$
curves and the corresponding star-formation thresholds are unaltered,
since this is the complete line-overlap regime for which H$_2$-dust is negligible.  
However, for $Z'\gtrsim 0.3$ the curves start to shift to the left and the thresholds
are reduced to smaller gas mass surface densities.
These shifts are due to the reductions of the effective 
dissociation fluxes by the non-negligible H$_2$ dust opacities.
For example, when H$_2$-dust is included for $Z'=1$ we have that 
$(\alpha G)_{\rm CNM}=1.6$ (for $\phi_g=1/1.9$) so that  
$\langle \tau_1 \rangle=0.77$,
and the H$_2$ core appears at
$\Sigma_{gas}\simeq 8.6$~$M_\odot$~pc$^{-2}$, and 
$\Sigma_{gas,*}=21.5$~$M_\odot$~pc$^{-2}$.  The inclusion of H$_2$-dust 
in the estimate for $(\alpha G)_{\rm CNM}$ modifies the 
MK10 results for solar and super-solar metallicities.

Third, in Figure~\ref{Fig:Comparison_McKee}, panel (c), we again adopt the MK10 assumptions, $\phi_g=1/1.9$ and
negligible H$_2$-dust for any $Z'$, but for slabs instead of spheres, and we 
use Equation~(\ref{fH2slab}) to compute $f_{{H_2}}^{\rm p}(\Sigma_{gas})$.
Here, $\Sigma_{gas}=(m/\sigma_g)\tau_z$ where $\tau_z$ is the total dust depth
of the slab, so that $y=\Sigma_{gas}/\Sigma_1$
where $\Sigma_1=(m/\sigma_g) \tau_1^{\rm p}$, and the critical H{\small I}-dust optical depth $\tau_1^{\rm p}$
is given by Equation~(\ref{t1slab}) for each $(\alpha G)_{\rm CNM}(Z')$.
We have already seen that the functional forms for the mass fractions
are very similar for complexes and slabs (with at most a $40\%$ difference
at $y=3$) and that the critical H{\small I}-dust columns are essentially equal for 
$\alpha G\sim 1$ (as for self-regulated gas).  It is therefore not surprising
that the curves for slabs are almost identical to the MK10
results in Figure~\ref{Fig:Comparison_McKee}.  For $Z'=1$,
we again have that
$\alpha G_{\rm CNM}=3.6$ so that  $\tau_1^{\rm p}=1.2$, so that  H$_2$
appears for $\Sigma_{gas}\simeq13.4$~M$_\odot$~pc$^{-2}$ 
and the star-formation threshold ($y=2$) is at $\Sigma_{gas,*}=26.8$~M$_\odot$~pc$^{-2}$.
For metallicities $Z'\gtrsim 0.3$, the neglect of H$_2$-dust in setting $(\alpha G)_{\rm CNM}$
is more significant than switching from spherical to plane-parallel cloud geometry.

In Figure~\ref{Fig:Comparison_McKee}, panel (d), we again present results for slabs,
but now with our preferred and somewhat larger $\sigma_g=1.9\times 10^{-21}Z'$~cm$^2$
($\phi_g=1$), and with the inclusion of H$_2$-dust absorption in setting 
$(\alpha G)_{\rm CNM}(Z')$.    The $f_{{H_2}}^{\rm p}$ curves and thresholds are 
shifted to lower gas mass surface densities compared to Figure~\ref{Fig:Comparison_McKee}.
For $Z'=1$, we now have $(\alpha G)_{\rm CNM}=2.6$ given that $\tau_1^{\rm p}=0.95$, so that 
H$_2$ appears at $\Sigma_{gas}=5.6$~M$_\odot$~pc$^{-2}$ and the star-formation threshold
surface gas mass density is $\Sigma_{gas,*}=11$~M$_\odot$~pc$^{-2}$.

Finally, in Figure~\ref{Fig:SigmaThreshold} we plot $\Sigma_{gas,*}(Z')$ for the above four model sets. 
At any $Z'$ the star-formation thresholds vary by factors of 2 to 3 for the four model assumptions, for
$Z'$ between 0.01 and $\sim 2$ . 
Again, we are assuming that $\alpha G=(\alpha G)_{\rm CNM}(Z')$ for which the critical 
H{\small I}-dust optical depths are all of order unity. The threshold gas mass surface densities 
therefore scale as $\sim 2m/\sigma_g$, and we have that 
\begin{equation}
\label{Sthresh}
\Sigma_{gas,*}(Z') \ \approx \ \frac{12}{\phi_g Z'} \ \ \ {\rm M}_\odot \ {\rm pc}^{-2} \ \ \ .
\end{equation}
For $Z'\sim 1$, the star-formation thresholds are of order 10~M$_\odot$~pc$^{-2}$.

The predicted H{\small I} columns and star-formation
thresholds appear consistent with observations, at least for $\sim$solar metallicity systems.
For example, in a recent GALFA-H{\small I} study of the H{\small I}-to-H$_2$ transition on sub-parsec scales
in the Perseus giant molecular cloud, \cite{Lee_12} find an almost constant H{\small I}
surface density, of $\Sigma_{HI}\sim 6-8$~M$_\odot$~pc$^{-2}$, in each of the mapped sub-regions.
For $Z'=1$,
this is consistent with the H{\small I} columns expected for $\alpha G$ of order unity,
including conditions appropriate for two-phased equilibrium.
Thus, Perseus is in the H{\small I}-dust limited and strong-field regime.
On larger scales,  \cite{Leroy_08}
find a mean total gas column of 14~M$_\odot$~pc$^{-2}$ at the 
H{\small I}-to-H$_2$ transition radii  (see their Table 5)
in their analysis of the H{\small I} and CO distributions in the 
THINGS and HERACLES galaxy surveys.  The observed threshold column
density of 14~M$_\odot$~pc$^{-2}$ is in harmony with 
our 1D predictions or the spherical KMT/MK10 models (see Figure~\ref{Fig:SigmaThreshold}).
However, in this interpretation there must be typically one primary UV absorbing 
cloud (GMC) per line of sight in the galaxy surveys, with negligible additional 
diffuse H I, since our predicted thresholds are for single cold clouds.

\begin{figure}[h!]
\centerline{\includegraphics[width=1.0 \columnwidth]{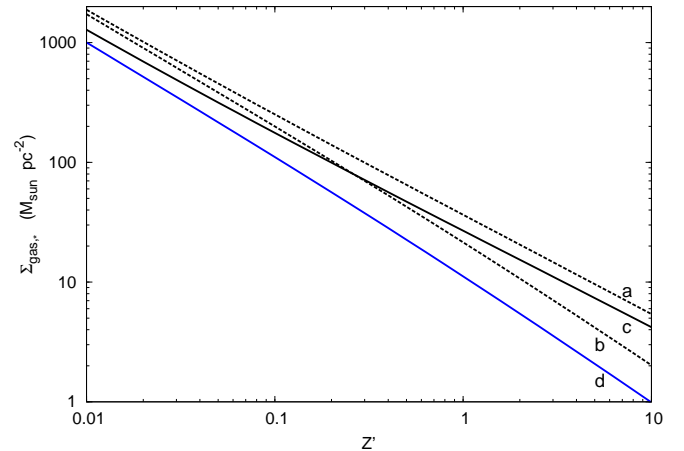}}
\caption{The star-formation threshold gas-mass surface density $\Sigma_{{\rm gas},*}$
at which the H$_2$ mass fraction $f_{H_2}=0.5$, for the four models, (a), (b), (c), and (d)
in Figure~\ref{Fig:Comparison_McKee} (see text).}
\label{Fig:SigmaThreshold}
\end{figure}

\section{Recap and Summary}

In this paper we have presented analytic theory and numerical computations for the 
atomic-to-molecular (H{\small I}-to-H$_2$) transitions, and the build-up of 
atomic hydrogen (H{\small I}) gas columns, in the PDRs of 
optically thick interstellar clouds illuminated by 
photodissociating far-ultraviolet Lyman-Werner (LW) radiation fields.  
We have focussed on simple idealized uniform density (1D) planar slabs that fully 
absorb the incident radiation.
This enables a Str{\" o}mgren type analysis (Sternberg 1988, ``S88") 
for the total steady-state column density of H{\small I} that is maintained
by the incident photodissociating flux.  The behavior is more complicated than
for classical (dust-free) photoionized H{\small II} regions, (a) because H$_2$ photodissociation
is a (multi)-line absorption process, and (b) because of the generally inevitable competition
between H$_2$-line, H$_2$-dust, and H{\small I}-dust absorption of the LW photons.


As discussed in \S~1, the H{\small I}-to-H$_2$ transition in interstellar gas has
been investigated by numerous authors over many years with varying levels of analytic
and numerical sophistication. In this paper we extend the analytic theory 
developed in S88, and consider 1D slabs irradiated by either beamed or isotropic fields.
The beamed configuration is appropriate for clouds near a localized radiation source.
Isotropic irradiation may be more appropriate
for ambient or ``global" conditions in galaxies.  In our theoretical development we make
the important distinction between H{\small I}-dust and H$_2$-dust absorption,
not previously discussed in the literature.
The three-way competition between H{\small I}-dust,
H$_2$-dust and H$_2$-line absorption determines the behavior and dependence
of the atomic column density on the environmental parameters.

Our analytic formulae and results are developed in \S~2, and verified in \S~3
with detailed numerical radiative transfer computations, including for the
H$_2$-dust limited dissociation bandwidth.  In \S~4 we compare
our results to models that assume spherical symmetry.

To conclude this paper we recap our basic results:

The fundamental dimensionless parameter in the problem is 
\begin{equation*}
\alpha G \ = \ \frac{D_0G}{Rn} 
\ = \ \frac{\sigma_g{\bar F}_\nu W_{g,{\rm tot}}}{Rn} 
\ = \ {\bar f}_{\rm diss}\frac{\sigma_g wF_0}{Rn}  \ \ \ 
\end{equation*}
$$
= \ 1.54 \ \frac{I_{\rm UV}}{(n/100~{\rm cm}^{-3})} \ \frac{\phi_g}{1+(2.64\phi_gZ')^{1/2}} \ \ \ .
$$
(See also expressions~[\ref{aG_A}] and [\ref{aG_B}]).
Here, $I_{\rm UV}$ is the free-space FUV intensity (relative to the Draine field),
$D_0$ is the free-space (optically thin) H$_2$ photodissociation rate (s$^{-1}$), $n$ 
is the total hydrogen gas volume density (cm$^{-3}$), $R$ is the H$_2$ formation rate 
coefficient (cm$^3$~s$^{-1}$), $G$ (dimensionless) is
the mean self-shielding factor, 
$\sigma_g$ is the far-UV continuum dust absorption cross section (cm$^2$), 
${\bar F}_\nu$  is the 
mean photon flux density (cm$^{-2}$~s$^{-1}$~Hz$^{-1}$) in the (912-1108 \AA) LW band,
$F_0$ is the total LW photon flux (cm$^{-2}$~s$^{-1}$),
${\bar f}_{\rm diss}$ is the mean dissociation probability,
$Z'$ is the metallicity relative to solar, 
$W_{g,{\rm tot}}$  is the total H$_2$-dust limited dissociation bandwidth (Hz),
$w$ (dimensionless) is the normalized bandwidth,
and $\phi_g$ is a 
factor of order unity depending on the dust-grain absorption properties.
The final dependence of $\alpha G$ on $Z'$ is via the metallicity-dependent
dissociation bandwidth $w$ (see Equation~[\ref{defw}]).

For two-sided illumination of an optically thick slab by beamed radiation, our analytic expression for
the total atomic  (H{\small I}) column density is (see Equation~[\ref{S88}]),
\begin{equation*}
\begin{split}
N_{\rm HI} & \ = \ 2\times \frac{1}{\sigma_g} \ {\rm ln}\bigl[\frac{\alpha G}{2}  + 1\bigr]\\
& \ = \ \frac{1.05\times 10^{21}}{Z'\phi_g} \ 
{\rm ln}\bigl[\frac{\alpha G}{2}  + 1\bigr] \ \ \ \ {\rm cm}^{-2} \ \ \ ,
\end{split}
\end{equation*}
or as a mass surface density (including a factor of 1.4 for helium)
\begin{equation*}
\Sigma_{gas,\textrm{HI}} \ = \ \frac{11.9}{Z'\phi_g} \ 
{\rm ln}\bigl[\frac{\alpha G}{2}  + 1\bigr] \ \ \ \ {\rm M}_\odot \ {\rm pc}^{-2} \ \ \ .
\end{equation*}
For two-sided irradiation by an isotropic field (see Equation~[\ref{S88piso}]),
\begin{equation*}
\begin{split}
N_{\rm HI}&  \ = \ 2\times \frac{\langle \mu \rangle}{\sigma_g}\ 
{\rm ln}\bigl[\frac{1}{{\langle \mu \rangle}}\frac{\alpha G}{4}  + 1\bigr] \\
& \ = \ \frac{8.42\times 10^{20}}{Z'\phi_g} \
{\rm ln}\bigl[\frac{\alpha G}{3.2}  + 1\bigr] \ \ \ \ {\rm cm}^{-2} \ \ \ ,
\end{split}
\end{equation*}
(where ${\langle \mu \rangle}=0.8$), or
\begin{equation*}
\Sigma_{gas,\textrm{HI}} \ = \ \frac{9.5}{Z'\phi_g} \ 
{\rm ln}\bigl[\frac{\alpha G}{3.2}  + 1\bigr] \ \ \ \ {\rm M}_\odot \ {\rm pc}^{-2} \ \ \ .
\end{equation*}
\newline
These expressions are generally valid for all regimes, from the weak- to strong-field
limits (small and large $\alpha G$), for all H{\small I}-to-H$_2$ transition profile
shapes (gradual or sharp), and for arbitrary metallicity $Z'$ (small or large).

For an optically thick cloud with total gas mass surface density $\Sigma_{gas}$, the
integral molecular (H$_2$) gas fraction is
\begin{equation*}
f_{{H_2}}(\Sigma_{gas}) \ = \ 1 - \frac{\Sigma_{HI}}{\Sigma_{gas}}
\end{equation*}
with $\Sigma_{HI}$ as given above for either beamed or isotropic fields.
In Figure~\ref{Fig:Comparison_McKee} (panel [a] versus [c]) we compare our results for
1D slabs to the H$_2$ mass fractions for spheres. The differences are very small.

Our formulae for the H{\small I} columns and H$_2$ mass fractions may be incorporated 
easily into hydrodynamics simulations for galaxy evolution.

For $\alpha G/2 \lesssim 1$ the
H{\small I}-to-H$_2$ transitions are gradual, and
the atomic columns are built up in the
predominantly molecular portions of the cloud.  For $\alpha G/2 \gtrsim 1$
the transitions are sharp, and
the H{\small I} is built up in outer fully atomic layers.
For sharp transitions, the H{\small I}-H$_2$-H{\small I} sandwich structure
for slabs corresponds to H{\small I}-shell/H$_2$-core
structures for spheres. If star-formation requires the conversion to H$_2$, then 
a threshold surface density,
$\Sigma_{gas,*}$, may be defined at which the molecular fraction $f_{{H_2}}=0.5$.
For $\alpha G/2 \gtrsim 1$ and sharp transitions we then have that
$\Sigma_{gas,*}\equiv 2\Sigma_{HI}$.

For self-regulated galaxy disks (ala KMT/MK10) in which the H{\small I} is 
driven to two-phased thermal equilibrium (and heated by FUV photoelectric
emission from dust-grains) the dimensionless parameter $\alpha G$ is restricted to 
a narrow range close to $(\alpha G)_{\rm CNM}$, where
\begin{equation*}
\frac{(\alpha G)_{\rm CNM}}{2} \ \approx \ 1 \ \ \ ,
\end{equation*}
independent of the metallicity $Z'$. For such self-regulated systems, 
our theory then predicts a metallicity dependent 
H{\small I}-to-H$_2$ star-formation threshold for individual clouds
\begin{equation*}
\Sigma_{gas,*}(Z') \ \approx \ \frac{12}{\phi_g Z'} \ \ \ {\rm M}_\odot \ {\rm pc}^{-2} \ \ \ .
\end{equation*}
This appears consistent with Galactic and extragalactic observations, 
at least for solar metallicity systems.

\begin{acknowledgements}
We thank Chris McKee for many discussions and for detailed comments on our manuscript,
including assistance with Equation~(\ref{fH2sphere}). We thank Alex Dalgarno, 
Reinhard Genzel, Avi Loeb, Tsevi Mazeh, David Neufeld, 
Ewine van Dishoeck, and the referee for helpful comments and conversations about this work.
A.S. is supported by the DFG via German-Israeli Project Cooperation grant STE1869/1-1/GE625/15-1,
and by a PBC Israel Science Foundation I-CORE Program, grant 1829/12. 
This work was supported in part by grant SYMPATICO (ANR-11-BS56-0023) 
from the French Agence Nationale de la Recherche and by the French CNRS national program PCMI. 
The early stages of our project was funded by the France-Israel High Council for Science and Technology. 
\end{acknowledgements}

\appendix

\section{Glossary of Symbols}

Table~2  is a glossary of symbols we have used in this paper.
In this listing we also refer to the Equations in which the various
symbols appear and/or are defined.













\LongTables
\begin{table*}[h!]
\begin{longtable}{lll}
\hline
Symbol		 		& Definition	& 	               Units \\						
\tableline
\tableline
${\cal I}_\nu$                & far-UV specific photon intensity.                                              &   photons cm$^{-2}$ s$^{-1}$ Hz$^{-1}$ sr$^{-1}$ 	\\
${\cal I}_\nu^{ISM}$          & free space far-UV specific intensity (Eq.~[2]).                              	&   photons cm$^{-2}$ s$^{-1}$ Hz$^{-1}$ sr$^{-1}$    \\
$I_{\rm UV}$                  & intensity scaling factor relative to the unit free-space Draine field.        	&   ...                                					\\
                              &                                                                               	&                                      					\\
$F_\nu$                       & flux density,  $F_\nu\equiv 4\pi {\cal I}_\nu$.                                	&   photons cm$^{-2}$ s$^{-1}$ Hz$^{-1}$           	\\
${\bar F}_\nu$                & mean flux density in LW band (Eq.~[12]).                                     	&   photons cm$^{-2}$ s$^{-1}$ Hz$^{-1}$           	\\
$F_0$                         & free-space LW band photon flux integral (Eqns. [3] and [4]).                  	&   photons cm$^{-2}$ s$^{-1}$                 			\\
$F_\nu(z)$                    & beamed photon flux density at linear depth $z$ in a slab.                     	&   photons cm$^{-2}$ s$^{-1}$ Hz$^{-1}$           	\\
$F(z)$                        & beamed photon flux at linear depth $z$ in a slab (Eq. [76]).                   &   photons cm$^{-2}$ s$^{-1}$  								\\
                              &                                                                                &                       										\\
$\sigma_{\nu,d}$              & H$_2$ photodissociation cross section at frequency $\nu$ (Eqns. [7] and [77]). &  cm$^2$                 		\\
$\sigma_d$                    & individual H$_2$ line photodissociation cross section (Eq. [9]).               &  cm$^2$ Hz              		\\
$\sigma_d^{\rm tot}$          & H$_2$ photodissociation cross section summed over all lines (Eq. [13]).        &  cm$^2$ Hz              		\\
                              &                                                                                &                       		\\
$D$                           & depth-dependent H$_2$ photodissociation rate.                                  &  s$^{-1}$                 	\\
$D_0$                         & free-space H$_2$ photodissociation rate (Eq. [5] and [15]).                    &  s$^{-1}$                 	\\
$D(0)$                        & H$_2$ photodissociation rate at the surface of a optically thick slab (Eq [6]).&  s$^{-1}$                 	\\
                              &                                                                                &                       		\\
$P_{vj}$                      & LW photon H$_2$ pumping rate out of ro-vibrational level $vj$ (Table 1).        &  s$^{-1}$                 	\\
$\langle f_{\rm diss}\rangle_{vj}$ & mean dissociation probability out of level $vj$ (Table 1).                     & ...                   		\\
${\bar f}_{\rm diss}$         & mean dissociation fraction per absorbed LW photon (Eq. [18]).                  & ...                   		\\
                              &                                                                                &                      			\\
$\sigma_g$                    & far-UV grain absorption cross section per hydrogen nucleon (Eq. [20]).         &  cm$^2$                 		\\
$\phi_g$                      & order-unity grain composition factor.                                          &  ...                  		\\
$Z'$                          & metallicity relative to solar abundances.                                                &  ...                  		\\
                              &                                                                                &                       		\\
$R$                           & grain-surface H$_2$ formation rate coefficient (Eq. [21]).                     &  cm$^3$ s$^{-1}$            \\
$T$                           & gas temperature.                                                               &  K                    		\\
                              &                                                                                &                       		\\
$n_1$                         & atomic hydrogen (HI) volume density.                                           &  cm$^{-3}$            		\\
$n_2$                         & molecular hydrogen (H$_2$) volume density.                                     &  cm$^{-3}$            		\\
$n$                           & total hydrogen gas volume density, $\equiv n_1+2n_2$.                         	&  cm$^{-3}$            		\\
$N_1$                         & HI column density.                                                             &  cm$^{-2}$            		\\
$N_2$                         & H$_2$ column density.                                                          &  cm$^{-2}$            		\\
$N$                           & hydrogen gas column density, $N\equiv N_1+2N_2$.                               &  cm$^{-2}$            		\\
$\tau_g\equiv \sigma_gN$      & dust opacity in normal direction.                                              &  ...                  		\\
                              &                                                                                &                       		\\
$N_{1,\rm tot}$               & total HI column density on one side of an optically thick slab.       			&  cm$^{-2}$             	 	\\
$\tau_{1,{\rm tot}}\equiv \sigma_g$  & HI-dust opacity in normal direction.                            			& ...                   		\\
                              &                                                                                &                       		\\
$W_d(N_2)$                    & H$_2$-line-overlap-limited dissociation bandwidth.                             &  Hz                   		\\
$W_g(N_2)$                    & H$_2$-dust-limited dissociation bandwidth.                                     &  Hz                   		\\
$W_{d,{\rm tot}}$             & total H$_2$-dust-limited dissociation bandwidth.                               &  Hz                   		\\
$w$                           & normalized H$_2$-dust-limited dissociation bandwidth (Eq. [29]).               &  ...                  		\\
$f_{shield}(N_2)$             & H$_2$ self-shielding function (Eqns. [16] and [72]).                           &  ...                  		\\
                              &                                                                                &                       		\\
$<\mu>$                       & mean ray-angle factor for HI-dust opacity (Eqns. [63] and [75]).      			&  ...                  		\\
                              &                                                                                &                       		\\
$\alpha\equiv D_0/(Rn)$       & free-space atomic-to-molecular density ratio (Eqns. [34] and [44]).            &  ...                  		\\
$G(\sigma_g)$                 & mean self-shielding factor (Eqns. [41], [45]-[48]).                            &  ...                  		\\
                              &                                                                                &                       		\\
$\alpha G$                    & fundamental dimensionless parameter for the HI-to-H$_2$ transition (Eqns. [49]-[51]).         &  ...                  		\\
                              &                                                                                &                       		\\
$n_{\rm CNM}$                 & CNM gas density for multi-phased HI equilibrium (Eq. [58]).           			&  cm$^{-3}$            		\\
$(\alpha G)_{\rm CNM}$        & $\alpha G$ for two-phase HI equilibrium (Eq. [59]).                   			&  ...                  		\\
                              &                                                                                &                       		\\
$\Sigma_{{H_2}}$              & H$_2$ mass surface density in plane-parallel slab.                             &  M$_\odot$ pc$^{-2}$      	\\
$\Sigma_{gas}$            & total gas mass surface density in plane-parallel slab.                         &  M$_\odot$ pc$^{-2}$      	\\
$M_{{H_2}}$                   & H$_2$ gas mass within sphere.                                                  &  M$_\odot$            		\\
$M_{gas}$                 & total gas mass within sphere.                                                  &  M$_\odot$            		\\
                              &                                                                                &                       		\\
$\tau_z\equiv n\sigma_gz$     & total dust optical depth through slab of finite linear width z.                &  ...                  		\\
$\tau_r\equiv n\sigma_gr$     & dust optical depth along radius $r$ of a sphere.                               &  ...                  		\\
                              &                                                                                &                       		\\
$y$                           & total dust optical depth normalized to the HI-dust optical depth (Eq. [97]).   &  ...                  		\\
                              &                                                                                &                       		\\
$\tau_1^{\rm p}$              & HI-dust optical depth, finite slab, two-sided-irradiation, isotropic (Eq. [98]). & ...                 		\\
$\tau_1^{\rm p,b}$            & HI-dust optical depth, finite slab, two-sided-radiation, beamed (Eq. [99]).    & ...                 			\\
$\langle \tau_1 \rangle$      & mean HI-dust optical depth, sphere, isotropic irradiation (Eq. [101]).         & ...                 			\\
                              &                                                                                &                     			\\
$f_{{H_2}}^{\rm p}$           & H$_2$ mass-fraction for slab (Eq. [97]).                                       & ...                 			\\
$f_{{H_2}}^{\rm s}$           & H$_2$ mass fraction for uniform density sphere (Eq. [100]).                    & ...                 			\\
$f_{{H_2}}^{\rm c}$           & H$_2$ mass fraction for spherical atomic-molecular complex (Eq. [102]).        & ...                 			\\
                              &                                                                                &                     			\\
$\Sigma_{gas,*}$              & Star-formation threshold gas surface density at which $f_{H_2}=0.5$.           & M$_\odot$~pc$^{-2}$     		\\
\tableline
\end{longtable}
\end{table*}
\clearpage

\bibliography{Arxiv_Sternberg14}

\begin{thebibliography}{201}
\expandafter\ifx\csname natexlab\endcsname\relax\def\natexlab#1{#1}\fi

\bibitem[{{Aaronson} {et~al.}(1974){Aaronson}, {Black}, \&
  {McKee}}]{Aaronson_74}
{Aaronson}, M., {Black}, J.~H., \& {McKee}, C.~F. 1974, \apjl, 191, L53

\bibitem[{{Abel} {et~al.}(1997){Abel}, {Anninos}, {Zhang}, \&
  {Norman}}]{Abel_97}
{Abel}, T., {Anninos}, P., {Zhang}, Y., \& {Norman}, M.~L. 1997, \nat, 2, 181

\bibitem[{{Abgrall} {et~al.}(1992){Abgrall}, {Le Bourlot}, {Pineau Des Forets},
  {Roueff}, {Flower}, \& {Heck}}]{Abgrall_92}
{Abgrall}, H., {Le Bourlot}, J., {Pineau Des Forets}, G., {et~al.} 1992, \aap,
  253, 525

\bibitem[{{Abgrall} {et~al.}(2000){Abgrall}, {Roueff}, \&
  {Drira}}]{Abgrall_2000}
{Abgrall}, H., {Roueff}, E., \& {Drira}, I. 2000, \aaps, 141, 297

\bibitem[{{Abgrall} {et~al.}(1993{\natexlab{a}}){Abgrall}, {Roueff}, {Launay},
  {Roncin}, \& {Subtil}}]{Abgrall_1993_b}
{Abgrall}, H., {Roueff}, E., {Launay}, F., {Roncin}, J.~Y., \& {Subtil}, J.~L.
  1993{\natexlab{a}}, \aaps, 101, 273

\bibitem[{{Abgrall} {et~al.}(1993{\natexlab{b}}){Abgrall}, {Roueff}, {Launay},
  {Roncin}, \& {Subtil}}]{Abgrall_1993_a}
{Abgrall}, H., {Roueff}, E., {Launay}, F., {Roncin}, J.~Y., \& {Subtil}, J.~L.
  1993{\natexlab{b}}, \aaps, 101, 323

\bibitem[{{Ahn} {et~al.}(2009){Ahn}, {Shapiro}, {Iliev}, {Mellema}, \&
  {Pen}}]{Ahn_09}
{Ahn}, K., {Shapiro}, P.~R., {Iliev}, I.~T., {Mellema}, G., \& {Pen}, U.-L.
  2009, \apj, 695, 1430

\bibitem[{{Albornoz V{\'a}squez} {et~al.}(2014){Albornoz V{\'a}squez},
  {Rahmani}, {Noterdaeme}, {Petitjean}, {Srianand}, \& {Ledoux}}]{Albornoz_14}
{Albornoz V{\'a}squez}, D., {Rahmani}, H., {Noterdaeme}, P., {et~al.} 2014,
  \aap, 562, A88

\bibitem[{{Allen} {et~al.}(1986){Allen}, {Atherton}, \& {Tilanus}}]{Allen_86}
{Allen}, R.~J., {Atherton}, P.~D., \& {Tilanus}, R.~P.~J. 1986, \nat, 319, 296

\bibitem[{{Allen} {et~al.}(1997){Allen}, {Knapen}, {Bohlin}, \&
  {Stecher}}]{Allen_97}
{Allen}, R.~J., {Knapen}, J.~H., {Bohlin}, R., \& {Stecher}, T.~P. 1997, \apj,
  487, 171

\bibitem[{{Andersson} {et~al.}(1992){Andersson}, {Roger}, \&
  {Wannier}}]{Andersson_92}
{Andersson}, B.-G., {Roger}, R.~S., \& {Wannier}, P.~G. 1992, \aap, 260, 355

\bibitem[{{Andersson} \& {Wannier}(1993)}]{Andersson_93}
{Andersson}, B.-G. \& {Wannier}, P.~G. 1993, \apj, 402, 585

\bibitem[{{Barlow} \& {Silk}(1976)}]{Barlow_76}
{Barlow}, M.~J. \& {Silk}, J. 1976, \apj, 207, 131

\bibitem[{{Bigiel} {et~al.}(2008){Bigiel}, {Leroy}, {Walter}, {Brinks}, {de
  Blok}, {Madore}, \& {Thornley}}]{Bigiel_08}
{Bigiel}, F., {Leroy}, A., {Walter}, F., {et~al.} 2008, \aj, 136, 2846

\bibitem[{{Bisbas} {et~al.}(2012){Bisbas}, {Bell}, {Viti}, {Yates}, \&
  {Barlow}}]{Bisbas_12}
{Bisbas}, T.~G., {Bell}, T.~A., {Viti}, S., {Yates}, J., \& {Barlow}, M.~J.
  2012, \mnras, 427, 2100

\bibitem[{{Black} \& {Dalgarno}(1976)}]{Black_76}
{Black}, J.~H. \& {Dalgarno}, A. 1976, \apj, 203, 132

\bibitem[{{Black} \& {Dalgarno}(1977)}]{Black_77}
{Black}, J.~H. \& {Dalgarno}, A. 1977, \apjs, 34, 405

\bibitem[{{Black} \& {van Dishoeck}(1987)}]{Black_87}
{Black}, J.~H. \& {van Dishoeck}, E.~F. 1987, \apj, 322, 412

\bibitem[{{Blitz} \& {Rosolowsky}(2004)}]{Blitz_04}
{Blitz}, L. \& {Rosolowsky}, E. 2004, \apjl, 612, L29

\bibitem[{{Blitz} \& {Rosolowsky}(2006)}]{Blitz_06}
{Blitz}, L. \& {Rosolowsky}, E. 2006, \apj, 650, 933

\bibitem[{{Bohlin} {et~al.}(1978){Bohlin}, {Savage}, \& {Drake}}]{Bohlin_78}
{Bohlin}, R.~C., {Savage}, B.~D., \& {Drake}, J.~F. 1978, \apj, 224, 132

\bibitem[{{Bok} {et~al.}(1955){Bok}, {Lawrence}, \& {Menon}}]{Bok_55}
{Bok}, B.~J., {Lawrence}, R.~S., \& {Menon}, T.~K. 1955, \pasp, 67, 108

\bibitem[{{B{\"o}ker} {et~al.}(2003){B{\"o}ker}, {Lisenfeld}, \&
  {Schinnerer}}]{Boker_03}
{B{\"o}ker}, T., {Lisenfeld}, U., \& {Schinnerer}, E. 2003, \aap, 406, 87

\bibitem[{{Bolatto} {et~al.}(2011){Bolatto}, {Leroy}, {Jameson}, {Ostriker},
  {Gordon}, {Lawton}, {Stanimirovi{\'c}}, {Israel}, {Madden}, {Hony},
  {Sandstrom}, {Bot}, {Rubio}, {Winkler}, {Roman-Duval}, {van Loon},
  {Oliveira}, \& {Indebetouw}}]{Bolatto_11}
{Bolatto}, A.~D., {Leroy}, A.~K., {Jameson}, K., {et~al.} 2011, \apj, 741, 12

\bibitem[{{Bromm} {et~al.}(2009){Bromm}, {Yoshida}, {Hernquist}, \&
  {McKee}}]{Bromm_09}
{Bromm}, V., {Yoshida}, N., {Hernquist}, L., \& {McKee}, C.~F. 2009, \nat, 459,
  49

\bibitem[{{Browning} {et~al.}(2003){Browning}, {Tumlinson}, \&
  {Shull}}]{Browning_03}
{Browning}, M.~K., {Tumlinson}, J., \& {Shull}, J.~M. 2003, \apj, 582, 810

\bibitem[{{Burton} {et~al.}(1990){Burton}, {Hollenbach}, \&
  {Tielens}}]{Burton_90}
{Burton}, M.~G., {Hollenbach}, D.~J., \& {Tielens}, A.~G.~G.~M. 1990, \apj,
  365, 620

\bibitem[{{Burton} {et~al.}(1978){Burton}, {Liszt}, \& {Baker}}]{Burton_78}
{Burton}, W.~B., {Liszt}, H.~S., \& {Baker}, P.~L. 1978, \apjl, 219, L67

\bibitem[{{Cardelli} {et~al.}(1989){Cardelli}, {Clayton}, \&
  {Mathis}}]{Cardelli_1989}
{Cardelli}, J.~A., {Clayton}, G.~C., \& {Mathis}, J.~S. 1989, \apj, 345, 245

\bibitem[{{Carruthers}(1970)}]{Carruthers_70}
{Carruthers}, G.~R. 1970, \apjl, 161, L81

\bibitem[{{Cazaux} \& {Tielens}(2002)}]{Cazaux_02}
{Cazaux}, S. \& {Tielens}, A.~G.~G.~M. 2002, \apjl, 575, L29

\bibitem[{{Christensen} {et~al.}(2012){Christensen}, {Quinn}, {Governato},
  {Stilp}, {Shen}, \& {Wadsley}}]{Christensen_12}
{Christensen}, C., {Quinn}, T., {Governato}, F., {et~al.} 2012, \mnras, 425,
  3058

\bibitem[{{Ciardi} {et~al.}(2000){Ciardi}, {Ferrara}, \& {Abel}}]{Ciardi_00}
{Ciardi}, B., {Ferrara}, A., \& {Abel}, T. 2000, \apj, 533, 594

\bibitem[{{Crighton} {et~al.}(2013){Crighton}, {Bechtold}, {Carswell},
  {Dav{\'e}}, {Foltz}, {Jannuzi}, {Morris}, {O'Meara}, {Prochaska}, {Schaye},
  \& {Tejos}}]{Crighton_13}
{Crighton}, N.~H.~M., {Bechtold}, J., {Carswell}, R.~F., {et~al.} 2013, \mnras,
  433, 178

\bibitem[{{Cui} {et~al.}(2005){Cui}, {Bechtold}, {Ge}, \& {Meyer}}]{Cui_05}
{Cui}, J., {Bechtold}, J., {Ge}, J., \& {Meyer}, D.~M. 2005, \apj, 633, 649

\bibitem[{{Dalgarno}(2006)}]{Dalgarno_06}
{Dalgarno}, A. 2006, Proceedings of the National Academy of Science, 103, 12269

\bibitem[{{Dav{\'e}} {et~al.}(2013){Dav{\'e}}, {Katz}, {Oppenheimer},
  {Kollmeier}, \& {Weinberg}}]{Dave_13}
{Dav{\'e}}, R., {Katz}, N., {Oppenheimer}, B.~D., {Kollmeier}, J.~A., \&
  {Weinberg}, D.~H. 2013, \mnras, 434, 2645

\bibitem[{{de Jong}(1972)}]{deJong_72}
{de Jong}, T. 1972, \aap, 20, 263

\bibitem[{{de Jong} {et~al.}(1980){de Jong}, {Boland}, \&
  {Dalgarno}}]{deJong_80}
{de Jong}, T., {Boland}, W., \& {Dalgarno}, A. 1980, \aap, 91, 68

\bibitem[{{Diaz-Miller} {et~al.}(1998){Diaz-Miller}, {Franco}, \&
  {Shore}}]{Diaz_98}
{Diaz-Miller}, R.~I., {Franco}, J., \& {Shore}, S.~N. 1998, \apj, 501, 192

\bibitem[{{Dijkstra} {et~al.}(2008){Dijkstra}, {Haiman}, {Mesinger}, \&
  {Wyithe}}]{Dijkstra_08}
{Dijkstra}, M., {Haiman}, Z., {Mesinger}, A., \& {Wyithe}, J.~S.~B. 2008,
  \mnras, 391, 1961

\bibitem[{{Draine}(1978)}]{Draine_1978}
{Draine}, B.~T. 1978, \apjs, 36, 595

\bibitem[{{Draine}(2003)}]{Draine_2003}
{Draine}, B.~T. 2003, \araa, 41, 241

\bibitem[{{Draine}(2011)}]{Draine_2011}
{Draine}, B.~T. 2011, {Physics of the Interstellar and Intergalactic medium}
  (Princeton University Press)

\bibitem[{{Draine} \& {Bertoldi}(1996)}]{Draine_1996}
{Draine}, B.~T. \& {Bertoldi}, F. 1996, \apj, 468, 269

\bibitem[{{Elmegreen}(1993)}]{Elmegreen_93}
{Elmegreen}, B.~G. 1993, \apj, 411, 170

\bibitem[{{Elmegreen} \& {Elmegreen}(1987)}]{Elmegreen_87}
{Elmegreen}, B.~G. \& {Elmegreen}, D.~M. 1987, \apj, 320, 182

\bibitem[{{Faucher-Gigu{\`e}re} {et~al.}(2013){Faucher-Gigu{\`e}re},
  {Quataert}, \& {Hopkins}}]{Faucher_13}
{Faucher-Gigu{\`e}re}, C.-A., {Quataert}, E., \& {Hopkins}, P.~F. 2013, \mnras,
  433, 1970

\bibitem[{{Federman} {et~al.}(1979){Federman}, {Glassgold}, \&
  {Kwan}}]{Federman_79}
{Federman}, S.~R., {Glassgold}, A.~E., \& {Kwan}, J. 1979, \apj, 227, 466

\bibitem[{{Feldmann} {et~al.}(2012){Feldmann}, {Hernandez}, \&
  {Gnedin}}]{Feldmann_12}
{Feldmann}, R., {Hernandez}, J., \& {Gnedin}, N.~Y. 2012, \apj, 761, 167

\bibitem[{{Fialkov} {et~al.}(2012){Fialkov}, {Barkana}, {Tseliakhovich}, \&
  {Hirata}}]{Fialkov_12}
{Fialkov}, A., {Barkana}, R., {Tseliakhovich}, D., \& {Hirata}, C.~M. 2012,
  \mnras, 424, 1335

\bibitem[{{Field} {et~al.}(1969){Field}, {Goldsmith}, \& {Habing}}]{Field_69}
{Field}, G.~B., {Goldsmith}, D.~W., \& {Habing}, H.~J. 1969, \apjl, 155, L149

\bibitem[{{Field} {et~al.}(1966){Field}, {Somerville}, \&
  {Dressler}}]{Field_66}
{Field}, G.~B., {Somerville}, W.~B., \& {Dressler}, K. 1966, \araa, 4, 207

\bibitem[{{Fitzpatrick}(1999)}]{Fitzpatrick_1999}
{Fitzpatrick}, E.~L. 1999, \pasp, 111, 63

\bibitem[{{Flannery} {et~al.}(1980){Flannery}, {Roberge}, \&
  {Rybicki}}]{Flannery_1980}
{Flannery}, B.~P., {Roberge}, W., \& {Rybicki}, G.~B. 1980, \apj, 236, 598

\bibitem[{{Foltz} {et~al.}(1988){Foltz}, {Chaffee}, \& {Black}}]{Foltz_88}
{Foltz}, C.~B., {Chaffee}, Jr., F.~H., \& {Black}, J.~H. 1988, \apj, 324, 267

\bibitem[{{France} {et~al.}(2013){France}, {Nell}, {Kane}, {Burgh}, {Beasley},
  \& {Green}}]{France_13}
{France}, K., {Nell}, N., {Kane}, R., {et~al.} 2013, \apjl, 772, L9

\bibitem[{{Fu} {et~al.}(2010){Fu}, {Guo}, {Kauffmann}, \& {Krumholz}}]{Fu_10}
{Fu}, J., {Guo}, Q., {Kauffmann}, G., \& {Krumholz}, M.~R. 2010, \mnras, 409,
  515

\bibitem[{{Fukui} {et~al.}(2014){Fukui}, {Okamoto}, {Kaji}, {Yamamoto},
  {Torii}, {Hayakawa}, {Tachihara}, {Dickey}, {Okuda}, {Ohama}, {Kuroda}, \&
  {Kuwahara}}]{Fukui_14}
{Fukui}, Y., {Okamoto}, R., {Kaji}, R., {et~al.} 2014, \apj, 796, 59

\bibitem[{{Fumagalli} {et~al.}(2010){Fumagalli}, {Krumholz}, \&
  {Hunt}}]{Fumagalli_10}
{Fumagalli}, M., {Krumholz}, M.~R., \& {Hunt}, L.~K. 2010, \apj, 722, 919

\bibitem[{{Ge} \& {Bechtold}(1997)}]{Ge_97}
{Ge}, J. \& {Bechtold}, J. 1997, \apjl, 477, L73

\bibitem[{{Genzel} {et~al.}(2012){Genzel}, {Tacconi}, {Combes}, {Bolatto},
  {Neri}, {Sternberg}, {Cooper}, {Bouch{\'e}}, {Bournaud}, {Burkert},
  {Comerford}, {Cox}, {Davis}, {F{\"o}rster Schreiber}, {Garcia-Burillo},
  {Gracia-Carpio}, {Lutz}, {Naab}, {Newman}, {Saintonge}, {Shapiro}, {Shapley},
  \& {Weiner}}]{Genzel_12}
{Genzel}, R., {Tacconi}, L.~J., {Combes}, F., {et~al.} 2012, \apj, 746, 69

\bibitem[{{Genzel} {et~al.}(2013){Genzel}, {Tacconi}, {Kurk}, {Wuyts},
  {Combes}, {Freundlich}, {Bolatto}, {Cooper}, {Neri}, {Nordon}, {Bournaud},
  {Burkert}, {Comerford}, {Cox}, {Davis}, {F{\"o}rster Schreiber},
  {Garc{\'{\i}}a-Burillo}, {Gracia-Carpio}, {Lutz}, {Naab}, {Newman},
  {Saintonge}, {Shapiro Griffin}, {Shapley}, {Sternberg}, \&
  {Weiner}}]{Genzel_13}
{Genzel}, R., {Tacconi}, L.~J., {Kurk}, J., {et~al.} 2013, \apj, 773, 68

\bibitem[{{Gillmon} \& {Shull}(2006)}]{Gillmon_06a}
{Gillmon}, K. \& {Shull}, J.~M. 2006, \apj, 636, 908

\bibitem[{{Gillmon} {et~al.}(2006){Gillmon}, {Shull}, {Tumlinson}, \&
  {Danforth}}]{Gillmon_06b}
{Gillmon}, K., {Shull}, J.~M., {Tumlinson}, J., \& {Danforth}, C. 2006, \apj,
  636, 891

\bibitem[{{Gir} {et~al.}(1994){Gir}, {Blitz}, \& {Magnani}}]{Gir_94}
{Gir}, B.-Y., {Blitz}, L., \& {Magnani}, L. 1994, \apj, 434, 162

\bibitem[{{Glassgold} \& {Langer}(1974)}]{Glassgold_74}
{Glassgold}, A.~E. \& {Langer}, W.~D. 1974, \apj, 193, 73

\bibitem[{{Glover} \& {Brand}(2003)}]{Glover_03}
{Glover}, S.~C.~O. \& {Brand}, P.~W.~J.~L. 2003, \mnras, 340, 210

\bibitem[{{Glover} {et~al.}(2010){Glover}, {Federrath}, {Mac Low}, \&
  {Klessen}}]{Glover_10}
{Glover}, S.~C.~O., {Federrath}, C., {Mac Low}, M.-M., \& {Klessen}, R.~S.
  2010, \mnras, 404, 2

\bibitem[{{Gnedin} {et~al.}(2009){Gnedin}, {Tassis}, \& {Kravtsov}}]{Gnedin_09}
{Gnedin}, N.~Y., {Tassis}, K., \& {Kravtsov}, A.~V. 2009, \apj, 697, 55

\bibitem[{{Goicoechea} \& {Le Bourlot}(2007)}]{Goicoechea_2007}
{Goicoechea}, J.~R. \& {Le Bourlot}, J. 2007, \aap, 467, 1

\bibitem[{{Goldshmidt} \& {Sternberg}(1995)}]{Goldshmidt_95}
{Goldshmidt}, O. \& {Sternberg}, A. 1995, \apj, 439, 256

\bibitem[{{Goldsmith} \& {Li}(2005)}]{Goldsmith_05}
{Goldsmith}, P.~F. \& {Li}, D. 2005, \apj, 622, 938

\bibitem[{{Goldsmith} {et~al.}(2007){Goldsmith}, {Li}, \& {Kr{\v
  c}o}}]{Goldsmith_07}
{Goldsmith}, P.~F., {Li}, D., \& {Kr{\v c}o}, M. 2007, \apj, 654, 273

\bibitem[{{Gomez} {et~al.}(1998){Gomez}, {Lebron}, {Rodriguez}, {Garay},
  {Lizano}, {Escalante}, \& {Canto}}]{Gomez_98}
{Gomez}, Y., {Lebron}, M., {Rodriguez}, L.~F., {et~al.} 1998, \apj, 503, 297

\bibitem[{{Gould} \& {Harwit}(1963)}]{GouldHarwit_63}
{Gould}, R.~J. \& {Harwit}, M. 1963, \apj, 137, 694

\bibitem[{{Gould} \& {Salpeter}(1963)}]{Gould_63}
{Gould}, R.~J. \& {Salpeter}, E.~E. 1963, \apj, 138, 393

\bibitem[{{Habart} {et~al.}(2003){Habart}, {Boulanger}, {Verstraete}, {Pineau
  des For{\^e}ts}, {Falgarone}, \& {Abergel}}]{Habart_03}
{Habart}, E., {Boulanger}, F., {Verstraete}, L., {et~al.} 2003, \aap, 397, 623

\bibitem[{{Habart} {et~al.}(2004){Habart}, {Boulanger}, {Verstraete},
  {Walmsley}, \& {Pineau des For{\^e}ts}}]{Habart_04}
{Habart}, E., {Boulanger}, F., {Verstraete}, L., {Walmsley}, C.~M., \& {Pineau
  des For{\^e}ts}, G. 2004, \aap, 414, 531

\bibitem[{{Habing}(1968)}]{Habing_1968}
{Habing}, H.~J. 1968, \bain, 19, 421

\bibitem[{{Haiman} {et~al.}(1996){Haiman}, {Rees}, \& {Loeb}}]{Haiman_96}
{Haiman}, Z., {Rees}, M.~J., \& {Loeb}, A. 1996, \apj, 467, 522

\bibitem[{{Haiman} {et~al.}(1997){Haiman}, {Rees}, \& {Loeb}}]{Haiman_97}
{Haiman}, Z., {Rees}, M.~J., \& {Loeb}, A. 1997, \apj, 476, 458

\bibitem[{{Heiles}(1969)}]{Heiles_69}
{Heiles}, C. 1969, \apj, 156, 493

\bibitem[{{Heiner} {et~al.}(2009){Heiner}, {Allen}, \& {van der
  Kruit}}]{Heiner_09}
{Heiner}, J.~S., {Allen}, R.~J., \& {van der Kruit}, P.~C. 2009, \apj, 700, 545

\bibitem[{{Heiner} {et~al.}(2011){Heiner}, {Allen}, \& {van der
  Kruit}}]{Heiner_11}
{Heiner}, J.~S., {Allen}, R.~J., \& {van der Kruit}, P.~C. 2011, \mnras, 416, 2

\bibitem[{{Heyer} {et~al.}(2004){Heyer}, {Corbelli}, {Schneider}, \&
  {Young}}]{Heyer_04}
{Heyer}, M.~H., {Corbelli}, E., {Schneider}, S.~E., \& {Young}, J.~S. 2004,
  \apj, 602, 723

\bibitem[{{Hill} \& {Hollenbach}(1978)}]{Hill_78}
{Hill}, J.~K. \& {Hollenbach}, D.~J. 1978, \apj, 225, 390

\bibitem[{{Hirashita} \& {Ferrara}(2005)}]{Hirashita_05}
{Hirashita}, H. \& {Ferrara}, A. 2005, \mnras, 356, 1529

\bibitem[{{Hollenbach} \& {Natta}(1995)}]{Hollenbach_95}
{Hollenbach}, D. \& {Natta}, A. 1995, \apj, 455, 133

\bibitem[{{Hollenbach} {et~al.}(1971){Hollenbach}, {Werner}, \&
  {Salpeter}}]{Hollenbach_71}
{Hollenbach}, D.~J., {Werner}, M.~W., \& {Salpeter}, E.~E. 1971, \apj, 163, 165

\bibitem[{{Holzbauer} \& {Furlanetto}(2012)}]{Holzbauer_12}
{Holzbauer}, L.~N. \& {Furlanetto}, S.~R. 2012, \mnras, 419, 718

\bibitem[{{Jura}(1974)}]{Jura_74}
{Jura}, M. 1974, \apj, 191, 375

\bibitem[{{Kaufman} {et~al.}(1999){Kaufman}, {Wolfire}, {Hollenbach}, \&
  {Luhman}}]{Kaufman_99}
{Kaufman}, M.~J., {Wolfire}, M.~G., {Hollenbach}, D.~J., \& {Luhman}, M.~L.
  1999, \apj, 527, 795

\bibitem[{{Kim} {et~al.}(2013){Kim}, {Krumholz}, {Wise}, {Turk}, {Goldbaum}, \&
  {Abel}}]{Kim_13}
{Kim}, J.-h., {Krumholz}, M.~R., {Wise}, J.~H., {et~al.} 2013, \apj, 775, 109

\bibitem[{{Knapen} {et~al.}(2006){Knapen}, {Allen}, {Heaton}, {Kuno}, \&
  {Nakai}}]{Knapen_06}
{Knapen}, J.~H., {Allen}, R.~J., {Heaton}, H.~I., {Kuno}, N., \& {Nakai}, N.
  2006, \aap, 455, 897

\bibitem[{{Knapp}(1974)}]{Knapp_74}
{Knapp}, G.~R. 1974, \aj, 79, 527

\bibitem[{{Krumholz} {et~al.}(2008){Krumholz}, {McKee}, \&
  {Tumlinson}}]{Krumholz_2008}
{Krumholz}, M.~R., {McKee}, C.~F., \& {Tumlinson}, J. 2008, \apj, 689, 865

\bibitem[{{Krumholz} {et~al.}(2009){Krumholz}, {McKee}, \&
  {Tumlinson}}]{Krumholz_2009}
{Krumholz}, M.~R., {McKee}, C.~F., \& {Tumlinson}, J. 2009, \apj, 693, 216

\bibitem[{{Kr{\v c}o} \& {Goldsmith}(2010)}]{Krco_10}
{Kr{\v c}o}, M. \& {Goldsmith}, P.~F. 2010, \apj, 724, 1402

\bibitem[{{Kuhlen} {et~al.}(2013){Kuhlen}, {Madau}, \& {Krumholz}}]{Kuhlen_13}
{Kuhlen}, M., {Madau}, P., \& {Krumholz}, M.~R. 2013, \apj, 776, 34

\bibitem[{{Lagos} {et~al.}(2011){Lagos}, {Baugh}, {Lacey}, {Benson}, {Kim}, \&
  {Power}}]{Lagos_11}
{Lagos}, C.~D.~P., {Baugh}, C.~M., {Lacey}, C.~G., {et~al.} 2011, \mnras, 418,
  1649

\bibitem[{{Le Bourlot} {et~al.}(1999){Le Bourlot}, {Pineau des For{\^e}ts}, \&
  {Flower}}]{LeBourlot_99}
{Le Bourlot}, J., {Pineau des For{\^e}ts}, G., \& {Flower}, D.~R. 1999, \mnras,
  305, 802

\bibitem[{{Le Petit} {et~al.}(2006){Le Petit}, {Nehm{\'e}}, {Le Bourlot}, \&
  {Roueff}}]{LePetit_2006}
{Le Petit}, F., {Nehm{\'e}}, C., {Le Bourlot}, J., \& {Roueff}, E. 2006, \apjs,
  164, 506

\bibitem[{{Ledoux} {et~al.}(2006){Ledoux}, {Petitjean}, \&
  {Srianand}}]{Ledoux_06}
{Ledoux}, C., {Petitjean}, P., \& {Srianand}, R. 2006, \apjl, 640, L25

\bibitem[{{Lee} {et~al.}(2007){Lee}, {Pak}, {Dixon}, \& {van
  Dishoeck}}]{Lee_07}
{Lee}, D.-H., {Pak}, S., {Dixon}, W.~V.~D., \& {van Dishoeck}, E.~F. 2007,
  \apj, 655, 940

\bibitem[{{Lee} {et~al.}(1996){Lee}, {Herbst}, {Pineau des Forets}, {Roueff},
  \& {Le Bourlot}}]{Lee_96}
{Lee}, H.-H., {Herbst}, E., {Pineau des Forets}, G., {Roueff}, E., \& {Le
  Bourlot}, J. 1996, \aap, 311, 690

\bibitem[{{Lee} {et~al.}(2012){Lee}, {Stanimirovi{\'c}}, {Douglas}, {Knee}, {Di
  Francesco}, {Gibson}, {Begum}, {Grcevich}, {Heiles}, {Korpela}, {Leroy},
  {Peek}, {Pingel}, {Putman}, \& {Saul}}]{Lee_12}
{Lee}, M.-Y., {Stanimirovi{\'c}}, S., {Douglas}, K.~A., {et~al.} 2012, \apj,
  748, 75

\bibitem[{{Lee} {et~al.}(2014){Lee}, {Stanimirovi{\'c}}, {Wolfire}, {Shetty},
  {Glover}, {Molina}, \& {Klessen}}]{Lee_14}
{Lee}, M.-Y., {Stanimirovi{\'c}}, S., {Wolfire}, M.~G., {et~al.} 2014, \apj,
  784, 80

\bibitem[{{Leitch-Devlin} \& {Williams}(1985)}]{LeitchDevlin_85}
{Leitch-Devlin}, M.~A. \& {Williams}, D.~A. 1985, \mnras, 213, 295

\bibitem[{{Lepp} \& {Shull}(1984)}]{Lepp_1984}
{Lepp}, S. \& {Shull}, J.~M. 1984, \apj, 280, 465

\bibitem[{{Leroy} {et~al.}(2008){Leroy}, {Walter}, {Brinks}, {Bigiel}, {de
  Blok}, {Madore}, \& {Thornley}}]{Leroy_08}
{Leroy}, A.~K., {Walter}, F., {Brinks}, E., {et~al.} 2008, \aj, 136, 2782

\bibitem[{{Levshakov} \& {Varshalovich}(1985)}]{Levshakov_85}
{Levshakov}, S.~A. \& {Varshalovich}, D.~A. 1985, \mnras, 212, 517

\bibitem[{{Li} \& {Goldsmith}(2003)}]{Li_03}
{Li}, D. \& {Goldsmith}, P.~F. 2003, \apj, 585, 823

\bibitem[{{Liszt} \& {Lucas}(2002)}]{Liszt_02}
{Liszt}, H. \& {Lucas}, R. 2002, \aap, 391, 693

\bibitem[{{Liszt}(2007)}]{Liszt_07}
{Liszt}, H.~S. 2007, \aap, 461, 205

\bibitem[{{Liszt} \& {Burton}(1979)}]{Liszt_79}
{Liszt}, H.~S. \& {Burton}, W.~B. 1979, \apj, 228, 105

\bibitem[{{London}(1978)}]{London_78}
{London}, R. 1978, \apj, 225, 405

\bibitem[{{Mac Low} \& {Glover}(2012)}]{MacLow_12}
{Mac Low}, M.-M. \& {Glover}, S.~C.~O. 2012, \apj, 746, 135

\bibitem[{{Madden} {et~al.}(1993){Madden}, {Geis}, {Genzel}, {Herrmann},
  {Jackson}, {Poglitsch}, {Stacey}, \& {Townes}}]{Madden_93}
{Madden}, S.~C., {Geis}, N., {Genzel}, R., {et~al.} 1993, \apj, 407, 579

\bibitem[{{Matthews} {et~al.}(2003){Matthews}, {Purton}, {Roger}, {Dewdney}, \&
  {Mitchell}}]{Matthews_03}
{Matthews}, H.~E., {Purton}, C.~R., {Roger}, R.~S., {Dewdney}, P.~E., \&
  {Mitchell}, G.~F. 2003, \apj, 592, 176

\bibitem[{{McCutcheon} {et~al.}(1978){McCutcheon}, {Shuter}, \&
  {Booth}}]{McCutcheon_78}
{McCutcheon}, W.~H., {Shuter}, W.~L.~H., \& {Booth}, R.~S. 1978, \mnras, 185,
  755

\bibitem[{{McKee} \& {Krumholz}(2010)}]{McKee_2010}
{McKee}, C.~F. \& {Krumholz}, M.~R. 2010, \apj, 709, 308

\bibitem[{{Mebold} {et~al.}(1982){Mebold}, {Winnberg}, {Kalberla}, \&
  {Goss}}]{Mebold_82}
{Mebold}, U., {Winnberg}, A., {Kalberla}, P.~M.~W., \& {Goss}, W.~M. 1982,
  \aap, 115, 223

\bibitem[{{M{\'e}sz{\'a}ros}(1968)}]{Meszaros_68}
{M{\'e}sz{\'a}ros}, P. 1968, \apss, 2, 510

\bibitem[{{Miyake} {et~al.}(2010){Miyake}, {Stancil}, {Sadeghpour}, {Dalgarno},
  {McLaughlin}, \& {Forrey}}]{Miyake_10}
{Miyake}, S., {Stancil}, P.~C., {Sadeghpour}, H.~R., {et~al.} 2010, \apjl, 709,
  L168

\bibitem[{{Myers} {et~al.}(1978){Myers}, {Ho}, {Schneps}, {Chin}, {Pankonin},
  \& {Winnberg}}]{Myers_78}
{Myers}, P.~C., {Ho}, P.~T.~P., {Schneps}, M.~H., {et~al.} 1978, \apj, 220, 864

\bibitem[{{Neufeld} \& {Spaans}(1996)}]{Neufeld_96}
{Neufeld}, D.~A. \& {Spaans}, M. 1996, \apj, 473, 894

\bibitem[{{Noterdaeme} {et~al.}(2010){Noterdaeme}, {Petitjean}, {Ledoux},
  {L{\'o}pez}, {Srianand}, \& {Vergani}}]{Noterdaeme_10}
{Noterdaeme}, P., {Petitjean}, P., {Ledoux}, C., {et~al.} 2010, \aap, 523, A80

\bibitem[{{Offner} {et~al.}(2013){Offner}, {Bisbas}, {Viti}, \&
  {Bell}}]{Offner_13}
{Offner}, S.~S.~R., {Bisbas}, T.~G., {Viti}, S., \& {Bell}, T.~A. 2013, \apj,
  770, 49

\bibitem[{{Ostriker} {et~al.}(2010){Ostriker}, {McKee}, \&
  {Leroy}}]{Ostriker_10}
{Ostriker}, E.~C., {McKee}, C.~F., \& {Leroy}, A.~K. 2010, \apj, 721, 975

\bibitem[{{Palla} {et~al.}(1983){Palla}, {Salpeter}, \& {Stahler}}]{Palla_83}
{Palla}, F., {Salpeter}, E.~E., \& {Stahler}, S.~W. 1983, \apj, 271, 632

\bibitem[{{Pirronello} {et~al.}(1997){Pirronello}, {Liu}, {Shen}, \&
  {Vidali}}]{Pirronello_97}
{Pirronello}, V., {Liu}, C., {Shen}, L., \& {Vidali}, G. 1997, \apjl, 475, L69

\bibitem[{{Popping} {et~al.}(2014){Popping}, {Somerville}, \&
  {Trager}}]{Popping_13}
{Popping}, G., {Somerville}, R.~S., \& {Trager}, S.~C. 2014, \mnras, 442, 2398

\bibitem[{{Rachford} {et~al.}(2009){Rachford}, {Snow}, {Destree}, {Ross},
  {Ferlet}, {Friedman}, {Gry}, {Jenkins}, {Morton}, {Savage}, {Shull},
  {Sonnentrucker}, {Tumlinson}, {Vidal-Madjar}, {Welty}, \&
  {York}}]{Rachford_09}
{Rachford}, B.~L., {Snow}, T.~P., {Destree}, J.~D., {et~al.} 2009, \apjs, 180,
  125

\bibitem[{{Rachford} {et~al.}(2002){Rachford}, {Snow}, {Tumlinson}, {Shull},
  {Blair}, {Ferlet}, {Friedman}, {Gry}, {Jenkins}, {Morton}, {Savage},
  {Sonnentrucker}, {Vidal-Madjar}, {Welty}, \& {York}}]{Rachford_02}
{Rachford}, B.~L., {Snow}, T.~P., {Tumlinson}, J., {et~al.} 2002, \apj, 577,
  221

\bibitem[{{Rand} {et~al.}(1992){Rand}, {Kulkarni}, \& {Rice}}]{Rand_92}
{Rand}, R.~J., {Kulkarni}, S.~R., \& {Rice}, W. 1992, \apj, 390, 66

\bibitem[{{Rank} {et~al.}(1971){Rank}, {Townes}, \& {Welch}}]{Rank_71}
{Rank}, D.~M., {Townes}, C.~H., \& {Welch}, W.~J. 1971, Science, 174, 1083

\bibitem[{{Reach} {et~al.}(1994){Reach}, {Koo}, \& {Heiles}}]{Reach_94}
{Reach}, W.~T., {Koo}, B.-C., \& {Heiles}, C. 1994, \apj, 429, 672

\bibitem[{{Read}(1981)}]{Read_81}
{Read}, P.~L. 1981, \mnras, 194, 863

\bibitem[{{Richter} {et~al.}(2001){Richter}, {Sembach}, {Wakker}, \&
  {Savage}}]{Richter_01}
{Richter}, P., {Sembach}, K.~R., {Wakker}, B.~P., \& {Savage}, B.~D. 2001,
  \apjl, 562, L181

\bibitem[{{Robertson} \& {Kravtsov}(2008)}]{Robertson_08}
{Robertson}, B.~E. \& {Kravtsov}, A.~V. 2008, \apj, 680, 1083

\bibitem[{{Roger} \& {Dewdney}(1992)}]{Roger_92}
{Roger}, R.~S. \& {Dewdney}, P.~E. 1992, \apj, 385, 536

\bibitem[{{Roger} {et~al.}(2004){Roger}, {McCutcheon}, {Purton}, \&
  {Dewdney}}]{Roger_04}
{Roger}, R.~S., {McCutcheon}, W.~H., {Purton}, C.~R., \& {Dewdney}, P.~E. 2004,
  \aap, 425, 553

\bibitem[{{Roger} \& {Pedlar}(1981)}]{Roger_81}
{Roger}, R.~S. \& {Pedlar}, A. 1981, \aap, 94, 238

\bibitem[{{R{\"o}hser} {et~al.}(2014){R{\"o}hser}, {Kerp}, {Winkel},
  {Boulanger}, \& {Lagache}}]{Rohser_14}
{R{\"o}hser}, T., {Kerp}, J., {Winkel}, B., {Boulanger}, F., \& {Lagache}, G.
  2014, \aap, 564, A71

\bibitem[{{Safranek-Shrader} {et~al.}(2012){Safranek-Shrader}, {Agarwal},
  {Federrath}, {Dubey}, {Milosavljevi{\'c}}, \& {Bromm}}]{Safranek_12}
{Safranek-Shrader}, C., {Agarwal}, M., {Federrath}, C., {et~al.} 2012, \mnras,
  426, 1159

\bibitem[{{Sancisi} {et~al.}(1974){Sancisi}, {Goss}, {Anderson}, {Johansson},
  \& {Winnberg}}]{Sancisi_74}
{Sancisi}, R., {Goss}, W.~M., {Anderson}, C., {Johansson}, L.~E.~B., \&
  {Winnberg}, A. 1974, \aap, 35, 445

\bibitem[{{Savage} {et~al.}(1977){Savage}, {Bohlin}, {Drake}, \&
  {Budich}}]{Savage_77}
{Savage}, B.~D., {Bohlin}, R.~C., {Drake}, J.~F., \& {Budich}, W. 1977, \apj,
  216, 291

\bibitem[{{Schruba} {et~al.}(2011){Schruba}, {Leroy}, {Walter}, {Bigiel},
  {Brinks}, {de Blok}, {Dumas}, {Kramer}, {Rosolowsky}, {Sandstrom},
  {Schuster}, {Usero}, {Weiss}, \& {Wiesemeyer}}]{Schruba_11}
{Schruba}, A., {Leroy}, A.~K., {Walter}, F., {et~al.} 2011, \aj, 142, 37

\bibitem[{{Schuster} {et~al.}(2007){Schuster}, {Kramer}, {Hitschfeld},
  {Garcia-Burillo}, \& {Mookerjea}}]{Schuster_07}
{Schuster}, K.~F., {Kramer}, C., {Hitschfeld}, M., {Garcia-Burillo}, S., \&
  {Mookerjea}, B. 2007, \aap, 461, 143

\bibitem[{{Shaw} {et~al.}(2005){Shaw}, {Ferland}, {Abel}, {Stancil}, \& {van
  Hoof}}]{Shaw_05}
{Shaw}, G., {Ferland}, G.~J., {Abel}, N.~P., {Stancil}, P.~C., \& {van Hoof},
  P.~A.~M. 2005, \apj, 624, 794

\bibitem[{{Shaya} \& {Federman}(1987)}]{Shaya_87}
{Shaya}, E.~J. \& {Federman}, S.~R. 1987, \apj, 319, 76

\bibitem[{{Shull}(1978)}]{Shull_1978}
{Shull}, J.~M. 1978, \apj, 219, 877

\bibitem[{{Smith} {et~al.}(2000){Smith}, {Allen}, {Bohlin}, {Nicholson}, \&
  {Stecher}}]{Smith_00}
{Smith}, D.~A., {Allen}, R.~J., {Bohlin}, R.~C., {Nicholson}, N., \& {Stecher},
  T.~P. 2000, \apj, 538, 608

\bibitem[{{Solomon} \& {Werner}(1971)}]{Solomon_71}
{Solomon}, P.~M. \& {Werner}, M.~W. 1971, \apj, 165, 41

\bibitem[{{Spaans} \& {Neufeld}(1997)}]{SpaansNeufeld_97}
{Spaans}, M. \& {Neufeld}, D.~A. 1997, \apj, 484, 785

\bibitem[{{Spaans} {et~al.}(1994){Spaans}, {Tielens}, {van Dishoeck}, \&
  {Bakes}}]{Spaans_94}
{Spaans}, M., {Tielens}, A.~G.~G.~M., {van Dishoeck}, E.~F., \& {Bakes},
  E.~L.~O. 1994, \apj, 437, 270

\bibitem[{{Spaans} \& {van Dishoeck}(1997)}]{Spaans_97}
{Spaans}, M. \& {van Dishoeck}, E.~F. 1997, \aap, 323, 953

\bibitem[{{Spitzer} {et~al.}(1973){Spitzer}, {Drake}, {Jenkins}, {Morton},
  {Rogerson}, \& {York}}]{Spitzer_73}
{Spitzer}, L., {Drake}, J.~F., {Jenkins}, E.~B., {et~al.} 1973, \apjl, 181,
  L116

\bibitem[{{Spitzer}(1948)}]{Spitzer_48}
{Spitzer}, Jr., L. 1948, \apj, 107, 6

\bibitem[{{Spitzer} \& {Tomasko}(1968)}]{Spitzer_68}
{Spitzer}, Jr., L. \& {Tomasko}, M.~G. 1968, \apj, 152, 971

\bibitem[{{Stecher} \& {Williams}(1967)}]{Stecher_67}
{Stecher}, T.~P. \& {Williams}, D.~A. 1967, \apjl, 149, L29

\bibitem[{{Stephens} \& {Dalgarno}(1972)}]{Stephens_72}
{Stephens}, T.~L. \& {Dalgarno}, A. 1972, \jqsrt, 12, 569

\bibitem[{{Sternberg}(1988)}]{Sternberg_1988}
{Sternberg}, A. 1988, \apj, 332, 400

\bibitem[{{Sternberg} \& {Dalgarno}(1989)}]{Sternberg_1989}
{Sternberg}, A. \& {Dalgarno}, A. 1989, \apj, 338, 197

\bibitem[{{Sternberg} \& {Neufeld}(1999)}]{Sternberg_99}
{Sternberg}, A. \& {Neufeld}, D.~A. 1999, \apj, 516, 371

\bibitem[{{Stoerzer} {et~al.}(1996){Stoerzer}, {Stutzki}, \&
  {Sternberg}}]{Stoerzer_1996}
{Stoerzer}, H., {Stutzki}, J., \& {Sternberg}, A. 1996, \aap, 310, 592

\bibitem[{{Str{\"o}mgren}(1939)}]{Stromgren_39}
{Str{\"o}mgren}, B. 1939, \apj, 89, 526

\bibitem[{{Tacconi} {et~al.}(2010){Tacconi}, {Genzel}, {Neri}, {Cox}, {Cooper},
  {Shapiro}, {Bolatto}, {Bouch{\'e}}, {Bournaud}, {Burkert}, {Combes},
  {Comerford}, {Davis}, {Schreiber}, {Garcia-Burillo}, {Gracia-Carpio}, {Lutz},
  {Naab}, {Omont}, {Shapley}, {Sternberg}, \& {Weiner}}]{Tacconi_10}
{Tacconi}, L.~J., {Genzel}, R., {Neri}, R., {et~al.} 2010, \nat, 463, 781

\bibitem[{{Tacconi} {et~al.}(2013){Tacconi}, {Neri}, {Genzel}, {Combes},
  {Bolatto}, {Cooper}, {Wuyts}, {Bournaud}, {Burkert}, {Comerford}, {Cox},
  {Davis}, {F{\"o}rster Schreiber}, {Garc{\'{\i}}a-Burillo}, {Gracia-Carpio},
  {Lutz}, {Naab}, {Newman}, {Omont}, {Saintonge}, {Shapiro Griffin}, {Shapley},
  {Sternberg}, \& {Weiner}}]{Tacconi_13}
{Tacconi}, L.~J., {Neri}, R., {Genzel}, R., {et~al.} 2013, \apj, 768, 74

\bibitem[{{Takahashi} {et~al.}(1999){Takahashi}, {Masuda}, \&
  {Nagaoka}}]{Takahashi_99}
{Takahashi}, J., {Masuda}, K., \& {Nagaoka}, M. 1999, \mnras, 306, 22

\bibitem[{{Thompson} {et~al.}(2014){Thompson}, {Nagamine}, {Jaacks}, \&
  {Choi}}]{Thompson_14}
{Thompson}, R., {Nagamine}, K., {Jaacks}, J., \& {Choi}, J.-H. 2014, \apj, 780,
  145

\bibitem[{{Tielens} \& {Hollenbach}(1985)}]{Tielens_1985}
{Tielens}, A.~G.~G.~M. \& {Hollenbach}, D. 1985, \apj, 291, 722

\bibitem[{{van der Werf} \& {Goss}(1989)}]{vanderwerf_89}
{van der Werf}, P.~P. \& {Goss}, W.~M. 1989, \aap, 224, 209

\bibitem[{{van der Werf} {et~al.}(2013){van der Werf}, {Goss}, \&
  {O'Dell}}]{vanderwerf_13}
{van der Werf}, P.~P., {Goss}, W.~M., \& {O'Dell}, C.~R. 2013, \apj, 762, 101

\bibitem[{{van der Werf} {et~al.}(1988){van der Werf}, {Goss}, \& {Vanden
  Bout}}]{vanderwerf_88}
{van der Werf}, P.~P., {Goss}, W.~M., \& {Vanden Bout}, P.~A. 1988, \aap, 201,
  311

\bibitem[{{van Dishoeck} \& {Black}(1986)}]{vanDishoeck_1986}
{van Dishoeck}, E.~F. \& {Black}, J.~H. 1986, \apjs, 62, 109

\bibitem[{{van Dishoeck} \& {Black}(1988)}]{vanDishoeck_88}
{van Dishoeck}, E.~F. \& {Black}, J.~H. 1988, \apj, 334, 771

\bibitem[{{van Dishoeck} \& {Black}(1990)}]{vanDishoeck_90}
{van Dishoeck}, E.~F. \& {Black}, J.~H. 1990, \apj, 360, 313

\bibitem[{{van Dishoeck} {et~al.}(2006){van Dishoeck}, {Jonkheid}, \& {van
  Hemert}}]{vanDishoeck_06}
{van Dishoeck}, E.~F., {Jonkheid}, B., \& {van Hemert}, M.~C. 2006, Faraday
  Discussions, 133, 231

\bibitem[{{Viala}(1986)}]{Viala_86}
{Viala}, Y.~P. 1986, \aaps, 64, 391

\bibitem[{{Viala} {et~al.}(1988){Viala}, {Roueff}, \& {Abgrall}}]{Viala_88}
{Viala}, Y.~P., {Roueff}, E., \& {Abgrall}, H. 1988, \aap, 190, 215

\bibitem[{{Visbal} {et~al.}(2014){Visbal}, {Haiman}, {Terrazas}, {Bryan}, \&
  {Barkana}}]{Visbal_14}
{Visbal}, E., {Haiman}, Z., {Terrazas}, B., {Bryan}, G.~L., \& {Barkana}, R.
  2014, \mnras, 445, 107

\bibitem[{{Wakker}(2006)}]{Wakker_06}
{Wakker}, B.~P. 2006, \apjs, 163, 282

\bibitem[{{Wannier} {et~al.}(1991){Wannier}, {Lichten}, {Andersson}, \&
  {Morris}}]{Wannier_91}
{Wannier}, P.~G., {Lichten}, S.~M., {Andersson}, B.-G., \& {Morris}, M. 1991,
  \apjs, 75, 987

\bibitem[{{Wannier} {et~al.}(1983){Wannier}, {Lichten}, \&
  {Morris}}]{Wannier_83}
{Wannier}, P.~G., {Lichten}, S.~M., \& {Morris}, M. 1983, \apj, 268, 727

\bibitem[{{Webber}(1998)}]{Webber_98}
{Webber}, W.~R. 1998, \apj, 506, 329

\bibitem[{{Welty} {et~al.}(2012){Welty}, {Xue}, \& {Wong}}]{Welty_12}
{Welty}, D.~E., {Xue}, R., \& {Wong}, T. 2012, \apj, 745, 173

\bibitem[{{Williams} \& {Maddalena}(1996)}]{Williams_96}
{Williams}, J.~P. \& {Maddalena}, R.~J. 1996, \apj, 464, 247

\bibitem[{{Wilson} {et~al.}(1970){Wilson}, {Jefferts}, \&
  {Penzias}}]{Wilson_70}
{Wilson}, R.~W., {Jefferts}, K.~B., \& {Penzias}, A.~A. 1970, \apjl, 161, L43

\bibitem[{{Wise} \& {Abel}(2007)}]{Wise_07}
{Wise}, J.~H. \& {Abel}, T. 2007, \apj, 671, 1559

\bibitem[{{Wolcott-Green} {et~al.}(2011){Wolcott-Green}, {Haiman}, \&
  {Bryan}}]{Wolcott_11}
{Wolcott-Green}, J., {Haiman}, Z., \& {Bryan}, G.~L. 2011, \mnras, 418, 838

\bibitem[{{Wolfire} {et~al.}(2010){Wolfire}, {Hollenbach}, \&
  {McKee}}]{Wolfire_10}
{Wolfire}, M.~G., {Hollenbach}, D., \& {McKee}, C.~F. 2010, \apj, 716, 1191

\bibitem[{{Wolfire} {et~al.}(2003){Wolfire}, {McKee}, {Hollenbach}, \&
  {Tielens}}]{Wolfire_2003}
{Wolfire}, M.~G., {McKee}, C.~F., {Hollenbach}, D., \& {Tielens}, A.~G.~G.~M.
  2003, \apj, 587, 278

\bibitem[{{Wolfire} {et~al.}(2008){Wolfire}, {Tielens}, {Hollenbach}, \&
  {Kaufman}}]{Wolfire_08}
{Wolfire}, M.~G., {Tielens}, A.~G.~G.~M., {Hollenbach}, D., \& {Kaufman}, M.~J.
  2008, \apj, 680, 384

\bibitem[{{Wolniewicz} {et~al.}(1998){Wolniewicz}, {Simbotin}, \&
  {Dalgarno}}]{Wolniewicz_1998}
{Wolniewicz}, L., {Simbotin}, I., \& {Dalgarno}, A. 1998, \apjs, 115, 293

\bibitem[{{Wong} \& {Blitz}(2002)}]{Wong_02}
{Wong}, T. \& {Blitz}, L. 2002, \apj, 569, 157

\bibitem[{{Wrathmall} {et~al.}(2007){Wrathmall}, {Gusdorf}, \&
  {Flower}}]{Wrathmall07}
{Wrathmall}, S.~A., {Gusdorf}, A., \& {Flower}, D.~R. 2007, \mnras, 382, 133

\bibitem[{{Yoshida} {et~al.}(2003){Yoshida}, {Abel}, {Hernquist}, \&
  {Sugiyama}}]{Yoshida_03}
{Yoshida}, N., {Abel}, T., {Hernquist}, L., \& {Sugiyama}, N. 2003, \apj, 592,
  645

\end{thebibliography}
\bibliographystyle{aa}

\end{document}